\documentclass[final,5p,11pt,authoryear]{elsarticle}

\usepackage[dvipsnames]{xcolor}
\usepackage[nopatch]{microtype}
\usepackage{booktabs}
\usepackage{amsmath,amssymb,amsfonts}
\usepackage{algorithmic}
\usepackage{graphicx}
\usepackage{textcomp}
\usepackage{url}
\usepackage{tabularx}
\usepackage{nameref}
\usepackage{subcaption}
\usepackage{wrapfig}
\usepackage{amsthm}
\usepackage{lineno}
\usepackage[inkscapearea=page]{svg}
\usepackage{longtable}
\usepackage{pdflscape}  
\usepackage{booktabs} 
\usepackage{caption}  
\usepackage{tikz}
\usetikzlibrary{chains, positioning, shapes.misc, arrows.meta, backgrounds}
\usetikzlibrary{shapes, fit}
\usepackage{lipsum}
\usepackage{colortbl}

\emergencystretch=\maxdimen
\usepackage{array}
\usepackage{siunitx} 

\definecolor{processManual}{HTML}{D9D9D9}    
\definecolor{processAutomated}{HTML}{DAE8FC}  
\definecolor{processGuided}{HTML}{FFE6CC}    

\definecolor{natureImplicit}{HTML}{FFE6CC} 
\definecolor{natureExplicit}{HTML}{D5E8D4} 

\usepackage{float} 

\newcommand{\greyline}{\vspace{1.0em}{\color{lightgray}\hrule height 0.5pt}\vspace{1.0em}}

\journal{-}

\begin{document}

\begin{frontmatter}


\title{Attention is also needed for form design}

\author{B. Sankar\fnref{label1}\corref{corauth}}
\ead{sankarb@iisc.ac.in}
\ead[url]{https://www.sankar.studio}
\cortext[corauth]{Corresponding Author}

\affiliation[label1]{
            organization={Indian Institute of Science (IISc)},
            addressline={Department of Mechanical Engineering}, 
            city={CV Raman Road, Bangalore},
            postcode={560012}, 
            state={Karnataka},
            country={India}}

\author{Dibakar Sen\fnref{label2}}
\ead{dibakar@iisc.ac.in}
\affiliation[label2]{
           organization={Indian Institute of Science (IISc)},
           addressline={Department of Design and Manufacturing}, 
           city={CV Raman Road, Bangalore},
           postcode={560012}, 
           state={Karnataka},
           country={India}}

\begin{abstract}
Conventional product form design is a cognitively demanding process, fundamentally limited by its time-consuming nature, reliance on the subjective expertise of the designer, and the opaque translation of inspiration into tangible concepts. This research introduces a novel, attention-aware framework that systematically restructures this creative workflow by integrating two synergistic systems: EUPHORIA, an immersive Virtual Reality environment that uses eye-tracking to implicitly capture a designer's aesthetic preferences, and RETINA, an agentic AI pipeline that translates these implicit preferences into concrete design outputs. The foundational principles of this approach were validated through a two-part study. Initial experiments confirmed a significant positive correlation between a user's gaze duration and their explicit preference, establishing implicit attention as a reliable data source. Subsequent experiments demonstrated that this attention could be effectively directed, with external emotional stimuli inducing a strong convergence in participants' visual selections and thematic interpretations. A comparative study in which four designers solved challenging design problems using four distinct workflows, ranging from a fully manual conventional process to an end-to-end automated pipeline. The results showed that the integrated EUPHORIA-RETINA workflow was over four times more time-efficient than the conventional method. Furthermore, a panel of 50 design experts evaluated the 16 final renderings, and the designs generated by the fully automated system consistently received the highest Worthiness (calculated by an inverse Plackett-Luce model based on gradient descent optimization) and Design Effectiveness scores, indicating superior quality in terms of 8 different criteria namely,  novelty, visual appeal, emotional resonance, clarity of purpose, distinctiveness of silhouette, implied materiality, proportional balance and adherence to the brief. This research presents a validated paradigm shift from traditional Computer-Assisted Design (CAD) to a more collaborative model of Designer-Assisting Computers (DAC). By automating logistical and skill-dependent generative tasks, the proposed framework elevates the designer's role to that of a creative director, synergizing human intuition with the analytical and generative power of agentic AI to produce higher-quality designs more efficiently.
\end{abstract}

\begin{graphicalabstract}
\includegraphics[width=\textwidth, height=\textheight, keepaspectratio]{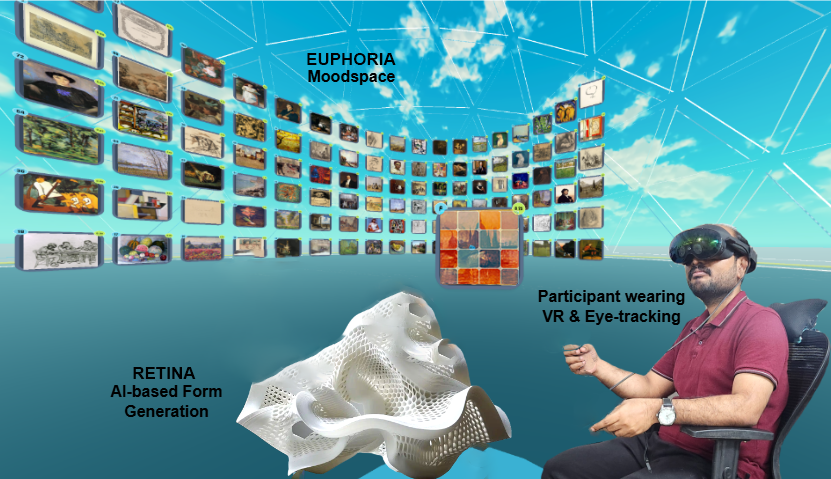}
\end{graphicalabstract}

\begin{highlights}
    \item A novel attention-aware framework, EUPHORIA-RETINA, is proposed, integrating Virtual Reality, Eye-Tracking, and Agentic AI to systematize the form design process.

    \item Implicit visual attention is empirically validated as a reliable, objective proxy for a designer's subjective aesthetic preferences within an immersive moodspace.

    \item Pre-conditioning the mind with thematic stimuli is shown to effectively channel the visual attention, leading to highly convergent and emotionally coherent visual selections among different users.

    \item The proposed automated workflow is demonstrated to be over four times more time-efficient than conventional methods, producing final designs of significantly higher quality as judged by a panel of 50 experts.

    \item A paradigm shift from Computer-Assisted Design (CAD) to Designer-Assisting Computers (DAC) is presented, redefining the designer's role as a strategic curator of AI-driven creative generation.
\end{highlights}

\begin{keyword}
Form Design \sep Agentic AI \sep Virtual Reality (VR) \sep Eye-Tracking (ET) \sep Moodboard \sep Implicit Preference \sep Human-Computer Interaction \sep Generative AI \sep Large Language Model (LLM) \sep Large Image Model (LIM) \sep Attention



\end{keyword}

\end{frontmatter}



\section{Introduction}
\label{sec:introduction}
In product design, the aesthetic and emotional appeal of a product plays a pivotal role in shaping user perception and influencing market success \cite{Lee_Delft_Harada, Norman_2007}. While functionality remains essential, it is often the form of a product—particularly its shape, colour, texture that forms the basis of a consumer’s first impression and drives their engagement \cite{Chang_Lai_Chang_2006, Katicic_Häfner_Ovtcharova_2015}. Form design, therefore, is not a peripheral task but a central phase of the product development process. The design of form not only enhances visual appeal but also communicates brand identity, conveys meaning, and evokes specific emotional responses from users. As such, form design constitutes a rich, cognitively demanding, and creative endeavour for designers, particularly the novice.\cite{Xie_2023}.

\subsection{Moodboarding in Form Design}
\label{subsec:intro_moodboarding}
One of the most widely used methods for initiating form ideation is the creation of moodboards to draw inspiration and set the mood for the design \cite{Mcdonagh_Storer_2004, Endrissat_Islam_Noppeney_2016, Robinson_Visualizing}. Moodboards are curated visual collages that capture the aesthetic essence, thematic direction, or emotional tone of a design. Designers traditionally assemble these boards using a collection of images, materials, textures, and colours that align with the product's conceptual direction \cite{Spawforth-Jones_2021, Gentes_Valentin_Brulé}. These collections serve as visual references that inspire the generation of concept sketches. Prior to moodboarding, the form design process begins with identifying abstract keywords or themes that reflect the product’s intended personality. Subsequently, the designers curate visual materials corresponding to these themes. In earlier days, the designers gathered physical items such as textile swatches, magazine cutouts, paint samples, etc. Integrating digital platforms has modernized the moodboarding process and offers novice designers broader access to visual resources \cite{Brevi_Celi_Gaetani_2019, Chipambwa_Chikwanya_2022}. In contemporary digital workflows, designers typically search for and download images from the internet, organize them in folders, and arrange them inside graphic design software. While this digitization has enhanced the convenience of image collection and allowed for more efficient visual arrangement \cite{Koch_Taffin_Lucero_Mackay_2020}, it has not fundamentally altered the required cognitive burden and expertise. Designers must still exercise considerable discretion in determining which images are relevant, how they relate to one another, and which features may eventually be distilled into product forms, all of which largely depend on the individual designer's experience \cite{Freeman_Marcketti_Karpova_2017}. Therefore, the fundamental process of moodboarding remains largely manual, unstructured, and reliant on subjective judgment.

The moodboarding process as described above, though creatively rich, presents several limitations as follows. First, it is time-intensive, often requiring days or weeks to collect, sort, and synthesize visual references. Second, it is experience-dependent; the quality of the moodboard and the insights derived from it depend heavily on the designer’s ability to intuitively perceive and organize aesthetic patterns and extract feature maps\cite{Casakin_Levy_2023, Kavakli}. Third, it lacks a formal structure, with no standardized procedures for evaluating a moodboard's quality or coherence or for ensuring it captures the end product's emotional and functional needs \cite{Reis_Merino_2021}. The outcome is thus highly subjective, often residing solely in the designer’s cognitive space and untranslatable into objective parameters that others can critique, replicate, or learn from \cite{Kahneman_Sibony_Sunstein_2021, Kim_Ryu_2014}.

These drawbacks point toward the need for systematizing the moodboarding process, particularly in an era where design speed, collaborative workflows, and AI-based creativity are advancing the field of design. Moreover, existing digital tools, while visually versatile, suffer from the inherent limitations of two-dimensional interfaces \cite{Rieuf_Bouchard_Aoussat_2015}. Designers can only view a limited number of images at once on a flat screen, must manually navigate through collections, and have to rely on their memory for comparison of multiple images at once. This imposes a cognitive load and restricts serendipitous discovery \cite{Das_Wu_Skrjanec_Feit_2024}, especially for novice designers who may lack the intuitive fluency to spot visual commonalities or extract design-relevant features from unstructured image sets \cite{Rieuf_etal_2017, Ye_etal_2006}.

\subsection{Relevance of Attention Theory}
\label{subsec:attention_theory}
From a cognitive science perspective, the underlying mechanism that guides the moodboarding process is visual attention \cite{Wickens_2021, Johnson_Proctor_2004a}. Attention, the cognitive process of selectively focusing on certain elements of the visual field while filtering out others, plays a critical role in determining which images a designer engages with, which features they notice, and ultimately, which aesthetic elements make their way into the final product \cite{Itti_Koch_2001, Borji_Itti_2013, Kowler_2011}. In this sense, moodboarding is fundamentally an attentional act of interplay between perception, memory, and preference. Yet, despite its centrality, very few research works focus on attention in form design workflows, particularly moodboarding.

Notably, certain visual features are consistently associated with emotional responses \cite{Lim_Mountstephens_Teo_2020, Brom_Stárková_Lukavský_Javora_Bromová_2016}. For instance, rounded shapes and soft textures often evoke feelings of comfort, while sharp angles may elicit tension \cite{bar2006a}. Similarly, bright, saturated colours are frequently linked to happiness, whereas dark, desaturated tones can correspond to sadness. These mappings, grounded in perceptual biases and affective conditioning, make the manipulation of such features a powerful tool for emotion-driven design, which is the foundational aspect required during moodboarding \cite{Gere_Kókai_Sipos_2017, Tuszynska-Bogucka_etal_2020}.

\subsection{Research Gap}
\label{subsec:intro_researchgap}
Despite this rich psychological foundation, the current moodboarding process does not actively leverage insights from attention science, nor does it incorporate tools that can objectively capture and analyze visual attention in real-time \cite{Park_2016, Duchowski_2002, Novak_etal_2024}. This disconnect represents a research gap at the intersection of design methodology and cognitive science. Specifically, there is a need for systems that can: 
\begin{enumerate}
    \item objectively measure visual attention during form design ideation;
    \item use these measurements to implicitly guide moodboard construction;
    \item support novice designers in identifying emotionally resonant features; and 
    \item translate attention-driven image selection into tangible form elements for concept sketches.
\end{enumerate}

\subsection{Proposition}
\label{subsec:proposition}
To address this need, this research proposes a novel approach: the development of an immersive system called Empathizing User Preferences in a Holodeck Using Real-time Implicit Attention (EUPHORIA) and an Agentic-AI framework called Responsive Embodiment Through Iterative visioN-guided Agentic AI (RETINA). EUPHORIA reimagines the moodboarding process not as a static 2D collage but as a dynamic, interactive \textit{"moodspace"} (MS) situated within a virtual reality (VR) environment. By surrounding the user with a floating constellation of images and tracking their gaze in real-time using eye-tracking technology, EUPHORIA captures implicit attention as participants explore visual stimuli. Rather than relying solely on manual selection, the system uses eye-tracking technology to infer user preferences and automate the construction of personalized, emotionally coherent moodboards. The output of this system is sent to RETINA, an agentic AI framework that uses compound LLMs and specialized agents to systematically extract the visual features upon which the designer spent the most attention, which are then used to automatically generate novel concept sketches and photorealistic renderings.

This research unfolds across three experimental phases, each designed to validate a specific component of the attention model as applicable to form design. Phase 1 investigates the correlation between gaze duration and preference ratings, establishing attention as a proxy for visual liking. Phase 2 explores how attentional patterns shift in response to predefined emotional stimuli, highlighting the role of preconditioning the mind in guiding perception. Phase 3 examines how attention-driven moodboards can enhance the ideation process for designers, particularly novices, by supporting more effective feature extraction and AI-mediated concept sketching. Across all phases, gaze data is analyzed to generate heatmaps, identify salient features, and visualize patterns of visual engagement.

By integrating VR and real-time eye-tracking, EUPHORIA enhances the immersiveness and scalability of the moodboarding experience, offering a structured, data-driven approach to form ideation. It aids designers to work more intuitively while capturing their attentional behaviours in ways that can be analyzed and optimized. Moreover, by making the early stages of moodboarding implicit—turning fleeting gaze patterns into actionable design inputs—the system helps to democratize design creativity, supporting experts and novices alike.

This research contributes a new theoretical and methodological workflow for attention-aware, emotionally resonant, and systematically structured form design. It builds on insights from psychology, design practice, and human-computer interaction to transform moodboards into moodspaces. The following sections detail the conventional stages of moodboarding, the architecture of the EUPHORIA and RETINA system, and the empirical studies that validate its effectiveness in transforming how designers engage with inspiration, emotion, and form.


\greyline  

\section{Form Design Fundamentals}
\label{app:form_design_fundamentals}
To appreciate the need for a systematic, attention-driven framework, it is first essential to understand the conventional process of form design. This process, while creatively fertile, is a multi-stage system that has evolved from traditional hands-on methods to contemporary digital workflows. Despite technological advancements, the core challenges related to subjectivity, cognitive load, and time investment have persisted. The following sections deconstruct this pipeline, from initial user understanding to final product rendering, to highlight the inherent complexities and motivations for reimagining the form design workflow.

\subsection{The Conventional Form Design Process}

The design of a product's form is not an arbitrary act but a structured journey that translates abstract user needs into tangible aesthetic expressions. While individual practices may vary, the conventional process generally follows a sequence of distinct, interlinked stages (Table~\ref{tab:conventional_process_revised}).

\begin{figure*}[htbp]
    \centering 

    \begin{subfigure}[b]{0.54\textwidth}
        \centering
        \includegraphics[width=\textwidth]{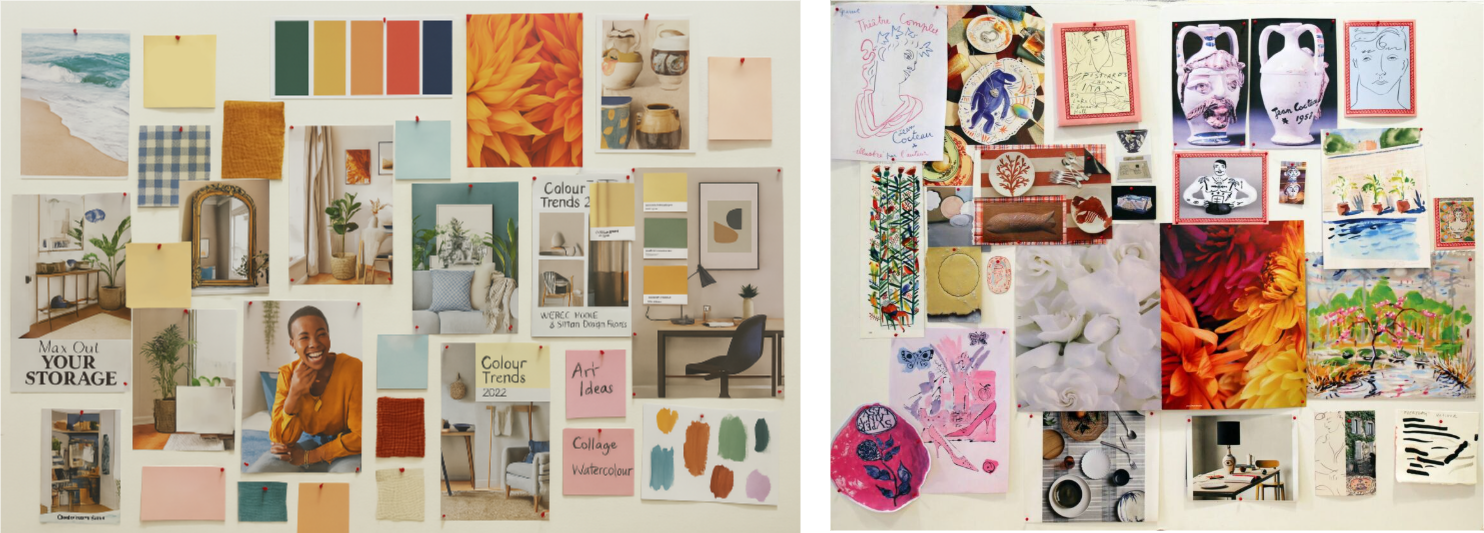}
        \caption{Examples of traditional, physical moodboards. These tactile collages are assembled using a variety of materials, including printed images, magazine cutouts, fabric swatches, and paint samples, all pinned to a physical board to establish a design's theme and feel.}
        \label{fig:traditional_moodboard}
    \end{subfigure}
    \hfill 
    \begin{subfigure}[b]{0.44\textwidth}
        \centering
        \includegraphics[width=\textwidth]{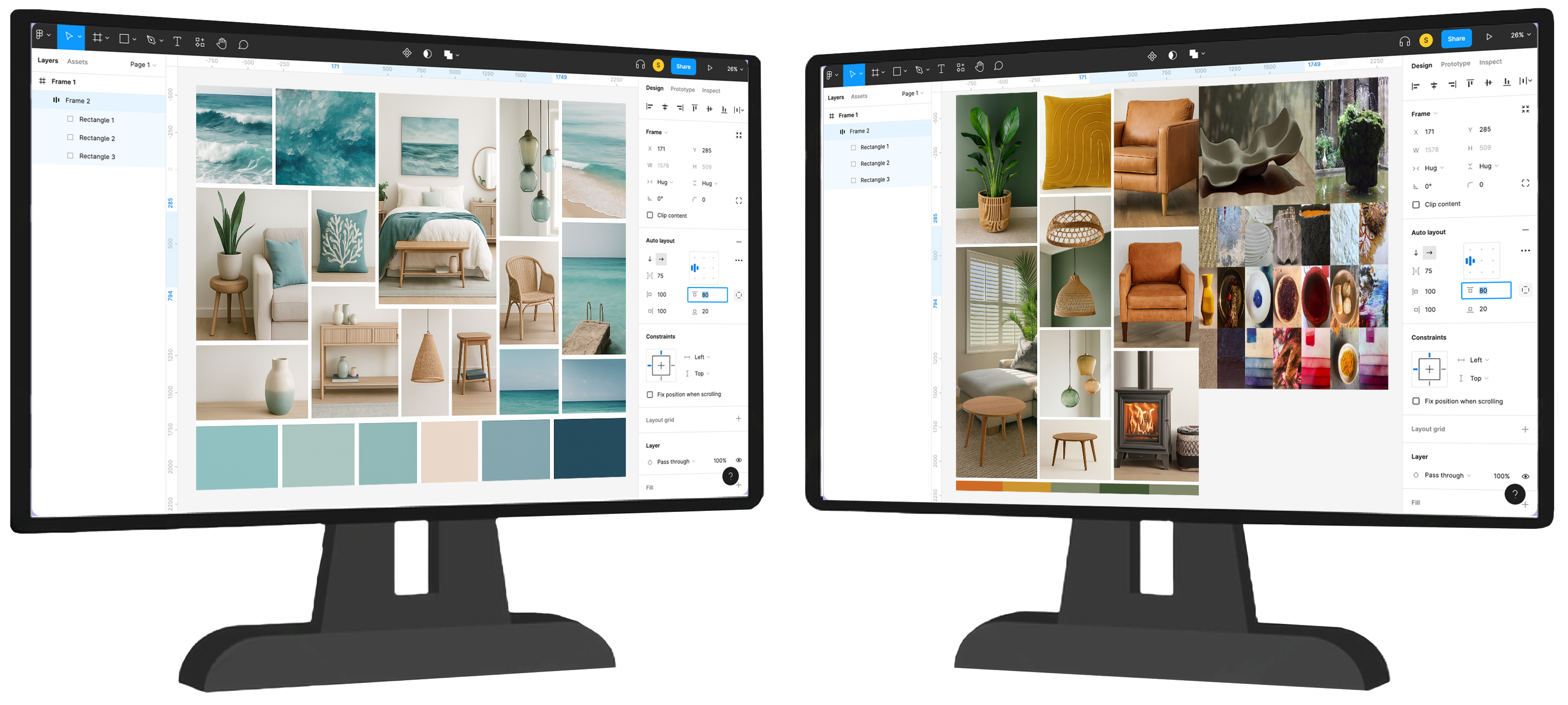}
        \caption{Examples of modern, digital moodboards created within a software interface and displayed on a monitor. This contemporary approach involves curating and arranging digital images in a structured grid on a 2D screen to define the product's aesthetic direction.}
        \label{fig:modern_moodboard}
    \end{subfigure}
    
    \caption{A comparison between traditional and modern moodboarding techniques. (a) Traditional moodboarding involves creating physical collages from tangible materials like printouts, fabric, and paint swatches to build a tactile and visual reference. (b) Modern moodboarding utilises digital software on 2D screens to curate and arrange images sourced online, offering convenience but being confined to a non-tactile, digital space.}
    \label{fig:moodboarding}
\end{figure*}

\subsubsection{Understanding the User \& Keyword Identification}

The form design process commences with understanding the target user group. Designers invest time in researching user needs, tastes, lifestyle aspirations, and the socio-cultural context in which the product will exist. This initial phase of task clarification culminates in the creation of a form design problem statement. The output of this stage is the curation of a list of keywords that encapsulate the desired visual and emotional attributes of the product. These style attributes, captured in textual form—for instance, 'sleek', 'rugged', 'minimalist', or 'playful', etc., serve as the conceptual anchors for the entire design journey. Tools such as the Semantic Differential Scale, Affinity Diagrams, User Journey Maps, etc., are often employed to articulate and prioritize these keywords, ensuring they accurately reflect the intended user experience.

\subsubsection{Moodboarding: Traditional vs. Modern}

With a guiding set of keywords, the designer proceeds to the most crucial and cognitively intensive stage: moodboarding. This is a process of visual exploration intended to inspire and set the emotional tone, i.e., the 'mood' for the design.

In the \textit{traditional approach}, classical designers would painstakingly collect physical materials (Figure~\ref{fig:traditional_moodboard}). This involved gathering snippets from magazine articles, taking photographs of inspirational scenes or objects, and collecting physical samples like fabric swatches or pieces of wood to capture desired textures and colours. These visual elements were then physically pinned on a large whiteboard or wall, grouped by similarities in shape, colour, and texture to embody the different keywords. This method, while tactile and immersive, was time-consuming, often taking days or even weeks.

The advent of the internet and digital tools gave rise to \textit{modern moodboarding}. Contemporary designers now browse image-hosting websites like Pinterest, Unsplash, Behance, etc., to find relevant images for their keywords (Figure~\ref{fig:modern_moodboard}). The collection process involves downloading a large repository of images, which are then curated and selected. Following this, designers use digital whiteboard software such as Figma, Miro, etc., to compose the selected images. However, this modern approach is constrained by the limitations of a 2D computer screen, which can only display a small subset of images at any given time. While digital tools have reduced the time for collection, the fundamental process remains a logistical challenge. In both traditional and modern moodboarding, the success of the outcome relies almost entirely on the designer's skill, experience, and intuition to extract a coherent design language from the visual chaos.

\subsubsection{Geometric Shape Abstraction}

\begin{figure}[htbp]
    \centering
    \includegraphics[width=1\linewidth]{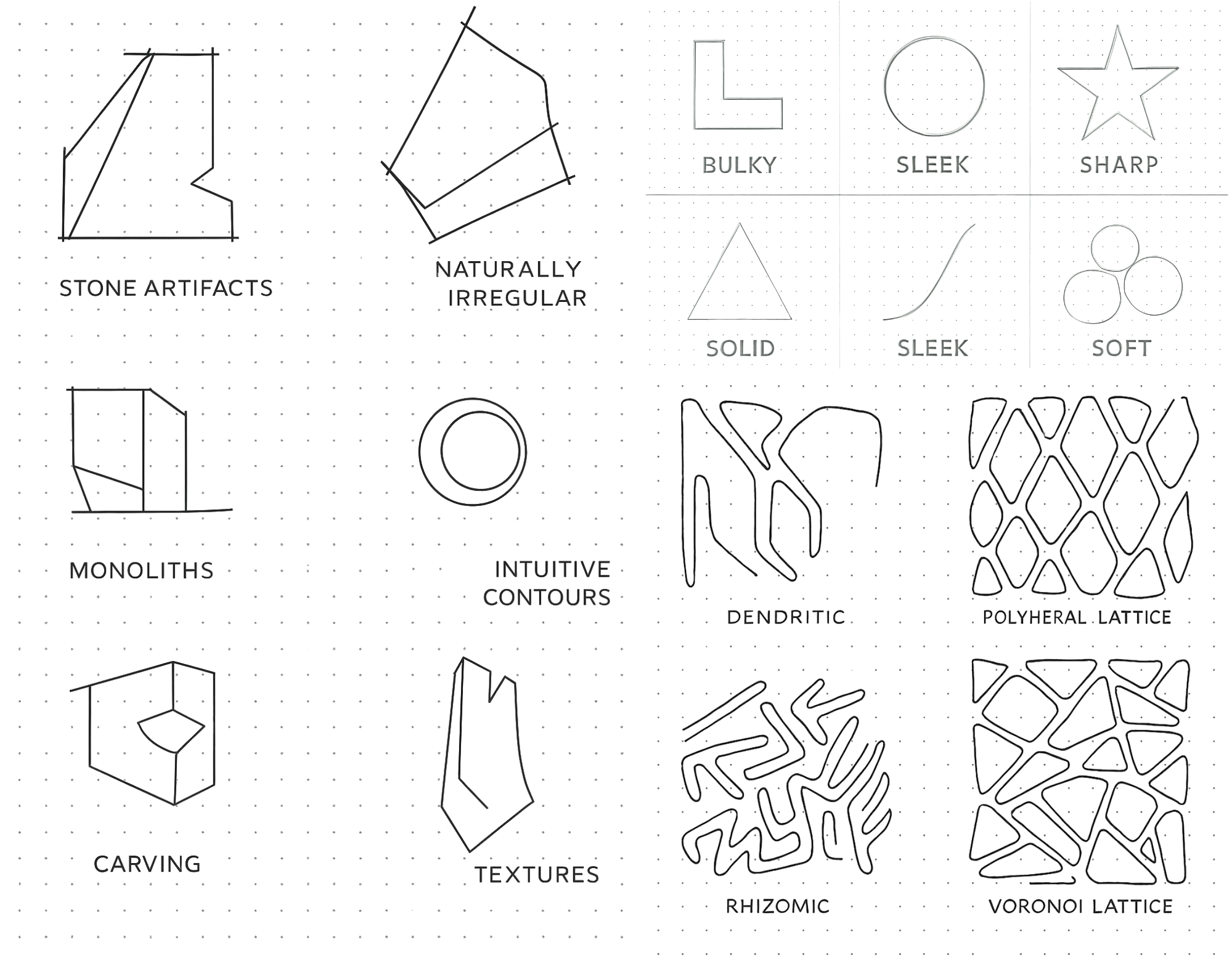}
    \caption{Different types of Geometric Shape Abstractions representing complex visual themes and mental feature maps derived from the moodboard into a vocabulary of simple, tangible forms. The figure showcases a variety of such abstractions, from raw interpretations of concepts like 'stone artifacts' and 'monoliths' (left) to more refined geometric primitives and patterns labeled with descriptive keywords like 'sleek', 'dendritic', or 'Voronoi lattice' (right). These abstracted shapes serve as the fundamental building blocks or 'design DNA' for the subsequent exploration of product forms during the rough sketching phase.}
    \label{fig:geometric_abstraction}
\end{figure}

Once the moodboard is composed, the designer engages in a period of reflection, mentally absorbing the visual information to extract common features and underlying patterns. This process translates the visual styles from the images into mental \textit{feature maps}. The next step is to make these intangible ideas tangible through geometric shape abstraction. Using pen and paper, the designer quickly sketches simple geometric shapes—lines, curves, and basic forms—that represent the core essence of the features they have mentally extracted (Figure~\ref{fig:geometric_abstraction}). This stage marks the first translation of the inspired mood into concrete, albeit simplified, visual entities.

\subsubsection{Exploration through Scribbles}

\begin{figure}[htbp]
    \centering
    \includegraphics[width=1.0\linewidth]{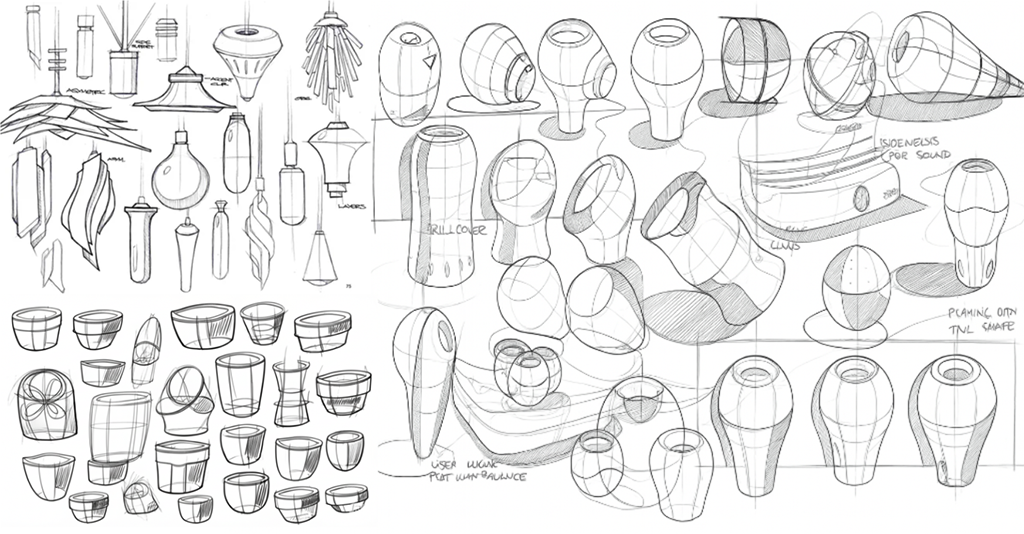}
    \caption{Exploration of Form through Rough Scribbles, a divergent ideation phase where designers mix and match the simple geometric forms derived from the previous abstraction stage. The figure displays a multitude of rapid, multi-stroke sketches, showcasing the fluid process of combining, overlapping, and manipulating these foundational shapes to explore a wide variety of potential product forms.}
    \label{fig:rough_sketch_scribbles}
\end{figure}

The abstracted geometric shapes become the foundational building blocks for form exploration. In this stage, the designer engages in rapid ideation by creating numerous rough sketches termed as \textit{scribbles}. By combining, overlapping, and manipulating the geometric shapes, they explore a wide array of potential product forms (Figure~\ref{fig:rough_sketch_scribbles}). This is a fluid and dynamic process of discovery. Many experienced designers still prefer the tactile feedback of pen on paper over digital tablets for this task, citing that the physical act better facilitates the translation of thought into form.

\subsubsection{Creation of Concept Sketches}

\begin{figure}[htbp]
    \centering
    \includegraphics[width=1\linewidth]{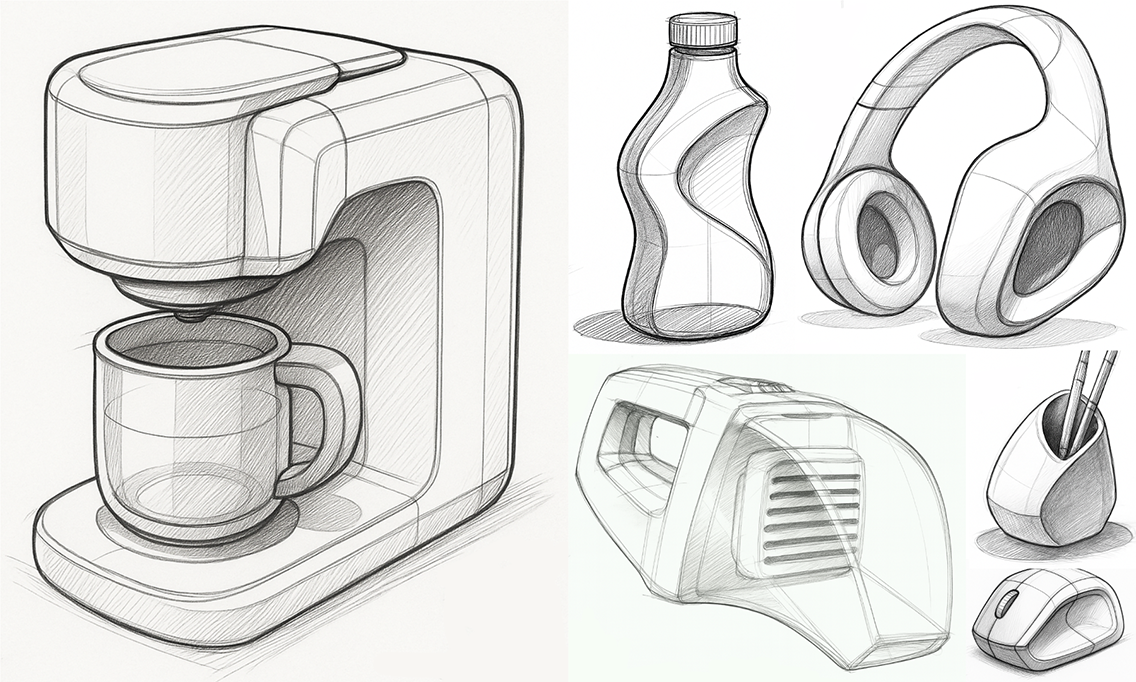}
    \caption{A collection of refined concept sketches of various products. This stage involves developing the most promising ideas from the rough sketch exploration phase into detailed drawings that define the finalized product form before rendering.}
    \label{fig:concept_sketch}
\end{figure}

From the multitude of rough scribbles, the designer begins to converge on promising ideas. The most compelling elements from the exploratory sketches are identified, refined, and integrated to develop detailed product concept sketches (Figure~\ref{fig:concept_sketch}). These sketches are no longer abstract explorations but are coherent representations of the product's potential final form. They clearly define the shape and contours of the product, embodying the form characteristics derived from the initial keywords and the moodboard.

\subsubsection{Photorealistic Form Rendering}

\begin{figure}[htbp]
    \centering
    \includegraphics[width=1\linewidth]{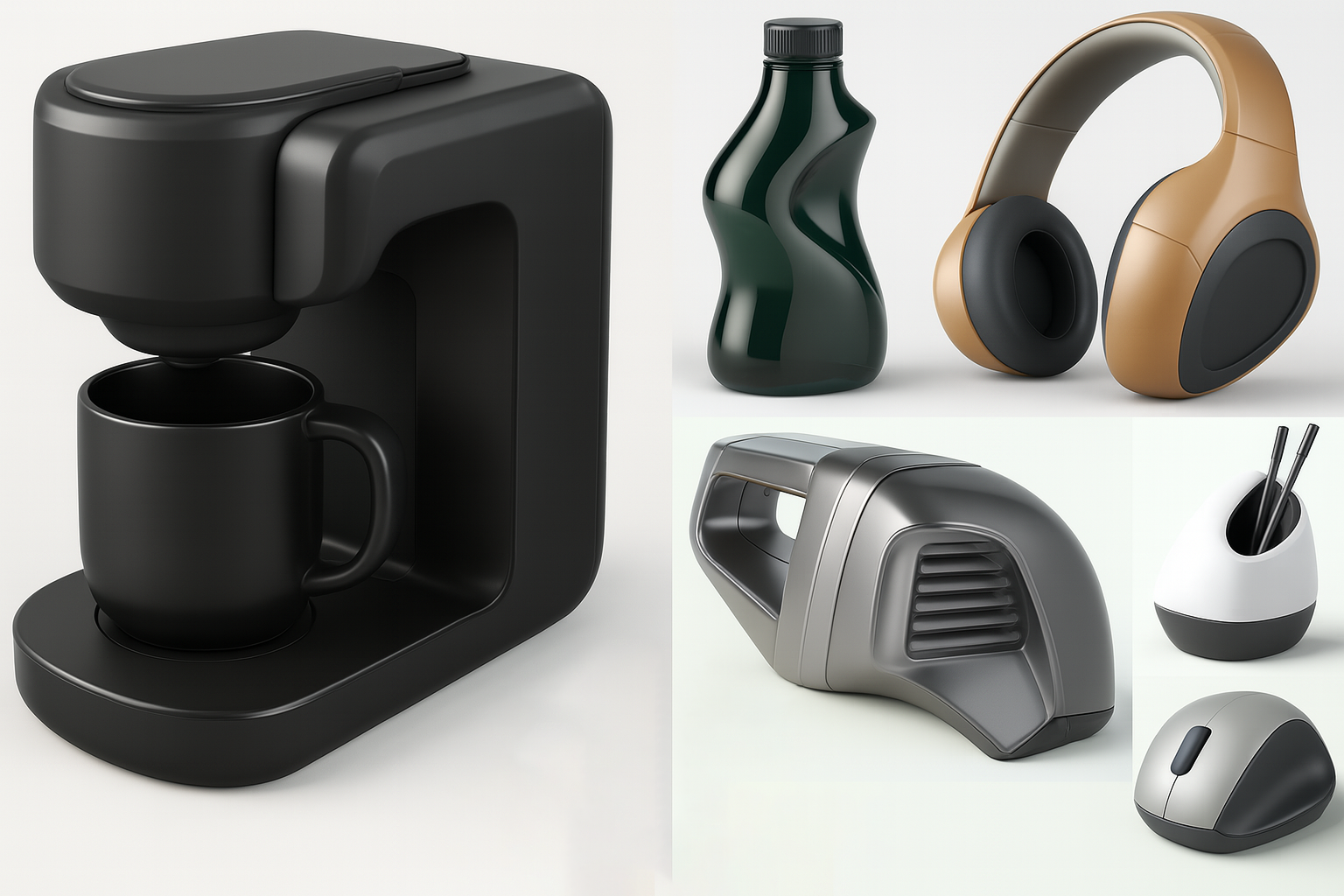}
    \caption{A collection of photorealistic product renderings. This final stage visually realizes the concept sketches by applying colour, texture, and materials to create a lifelike representation of the product's complete form and aesthetic intent.}
    \label{fig:photorealistic_rendering}
\end{figure}

The final stage of the conventional process is rendering, where the elements of colour and texture are applied to the concept sketches to bring the form to life. Traditionally, this was a manual art, with designers using tools like alcohol-based colour markers to meticulously paint over their hand-drawn sketches. Today, this stage is predominantly digital. Designers often scan their sketches and use software like Adobe Photoshop to apply colour, materials, and lighting. More recently, the emergence of AI-powered tools like Vizcom AI has further transformed this step. These platforms can take a hand-drawn concept sketch and a textual prompt describing the desired colour and texture, and automatically generate photorealistic renderings. This final output, an example of which is shown in Figure~\ref{fig:photorealistic_rendering} synthesizes the shape, colour, and texture, presenting a complete visualization of the product's final form design.

\subsubsection{Embodiment and Testing}
\begin{figure}[htbp]
    \centering
    \includegraphics[width=1\linewidth]{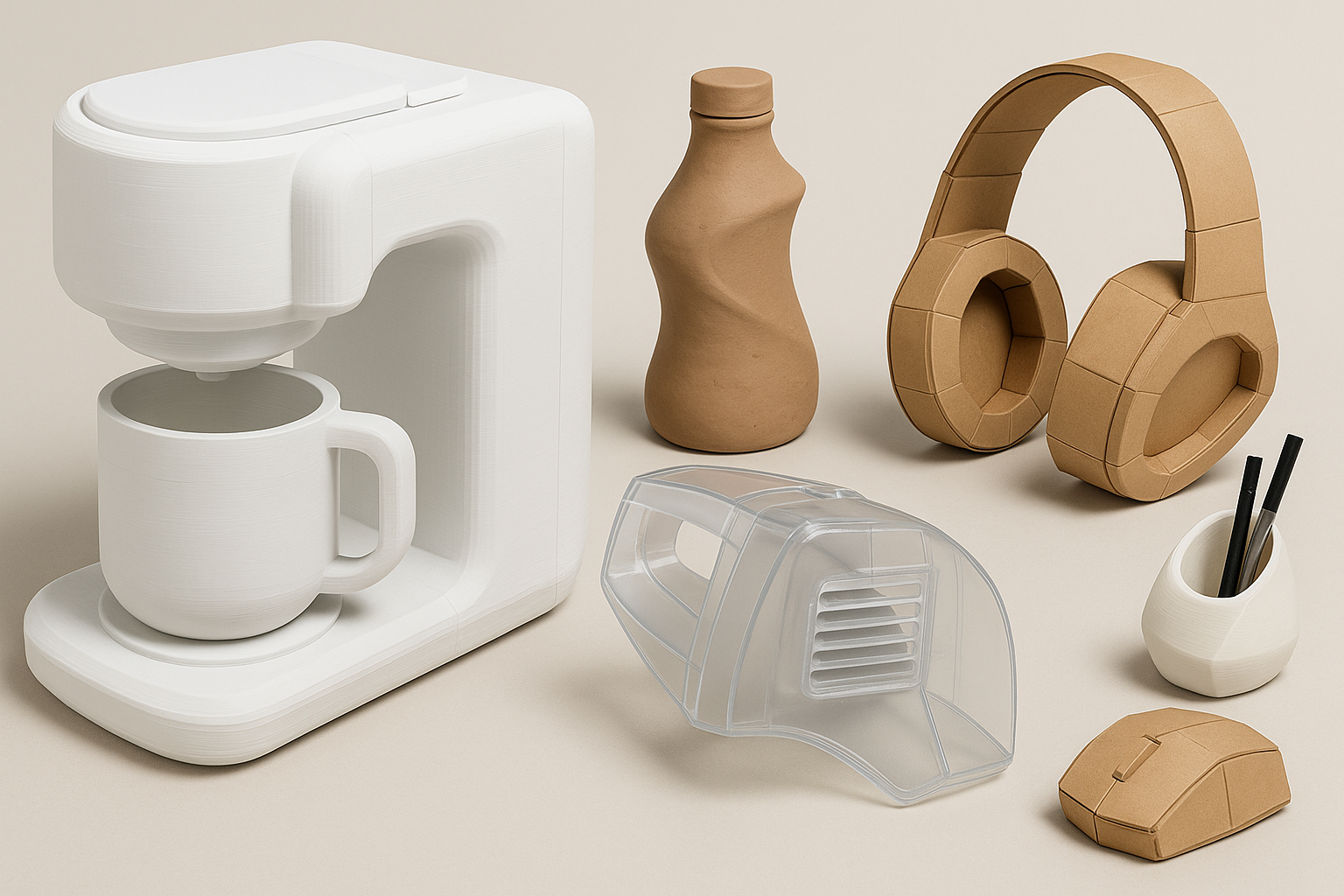}
    \caption{Examples of physical prototypes and mockups, representing the design embodiment phase that follows digital rendering. This stage involves translating the finalized form into tangible objects using techniques such as clay modelling, 3D printing, and high-fidelity prototyping. These physical models are crucial for evaluating the design's real-world qualities, including its tactile feel, ergonomics, scale, and the interplay of light on its surfaces.}
    \label{fig:prototype_embodiment}
\end{figure}

While the rendering often concludes the form design phase, the process typically continues towards physical embodiment (Figure~\ref{fig:prototype_embodiment}). Subsequent stages may involve creating detailed CAD models, mockups, clay models, or 3D-printed prototypes. These steps are essential for evaluating the design's tangible qualities, such as its tactile feel, scale, ergonomics, and the interplay of light on its surfaces. This phase marks a critical transition where purely aesthetic form design converges with the more technical domains of engineering, material science, and physical user experience evaluation.

\begin{table*}[htbp]
\centering
\caption{Breakdown of the Stages in Conventional Form Design Process}
\label{tab:conventional_process_revised}
\begin{tabular}{|p{3.0cm}|p{6.0cm}|p{3.5cm}|p{4cm}|}
\toprule
\textbf{Stage} & \textbf{Inputs \& Outputs} & \textbf{Tools Used} & \textbf{Activity Involved} \\
\midrule
\textbf{Understanding User \& Keyword Identification} & 
\textbf{Input:} User research data, market trends, product brief, stakeholder requirements. \newline \newline
\textbf{Output:} A defined form design problem statement; A curated list of 10-20 descriptive keywords.
&
Semantic Differential Scale, Affinity Diagrams, user interview notes, market analysis reports, user journey maps.
&
Research analysis, user empathy mapping, brainstorming, keyword curation, and prioritization. \\
\hline
\textbf{Moodboarding} &
\textbf{Input:} The curated list of keywords from the previous stage. \newline \newline
\textbf{Output:} A composed visual collage (physical or digital); An implicit mental map of relevant features in the designer's mind.
&
\textbf{Traditional:} Magazines, cameras, physical boards. \newline \textbf{Modern:} Pinterest, Figma, Miro.
&
Collecting, selecting, and composing visual materials; Reflecting to extract common themes and features. \\
\hline
\textbf{Geometric Shape Abstraction} &
\textbf{Input:} The composed moodboard and the designer's mental feature maps derived from it. \newline \newline
\textbf{Output:} A sheet of simple geometric shapes (lines, curves) representing the extracted visual features.
&
Pen, paper, grid layout sheets.
&
Translating abstract mental concepts into tangible, simplified geometric entities through quick sketching. \\
\hline
\textbf{Exploration through Rough Scribbles} &
\textbf{Input:} The set of abstracted geometric shapes. \newline \newline
\textbf{Output:} Multiple rough, multi-stroke concept sketches exploring a wide variety of product forms.
&
Pen, paper, or digital tablets with sketching software like Sketchbook.
&
Rapid ideation by combining and manipulating shapes to quickly explore diverse form variations. \\
\hline
\textbf{Creation of Concept Sketches} &
\textbf{Input:} The collection of rough scribbles and the most promising ideas identified from them. \newline \newline
\textbf{Output:} One or more refined, detailed final concept sketches that clearly define the product's form language.
&
Pen, paper, fine-line sketching tools, or digital sketching software.
&
Refining rough ideas, integrating the best elements, and creating a single, coherent, and detailed line drawing of the product form. \\
\hline
\textbf{Product Form Rendering} &
\textbf{Input:} The final, detailed concept sketch(es). \newline \newline
\textbf{Output:} A photorealistic or stylized visual rendering of the product, showcasing its shape, colour, texture, and materials.
&
\textbf{Traditional:} Alcohol-based markers, paint. \newline \textbf{Modern:} Adobe Photoshop; AI rendering tools like Vizcom AI.
&
Applying colour, texture, lighting, and material finishes to the line sketch to create a lifelike visual representation. \\
\bottomrule
\end{tabular}
\vspace{2.0em}
\end{table*}


\subsection{Pitfalls of Traditional and Modern Moodboarding}

While moodboarding is an indispensable tool for creative inspiration in form design, its conventional execution—both traditional and modern—is fraught with significant challenges that can impede efficiency, objectivity, and creative outcomes. These pitfalls are not merely procedural inconveniences; they represent fundamental structural weaknesses in the process that necessitate a critical re-evaluation.

\subsubsection{Extreme Subjectivity and Experience-Dependency}
The most significant drawback of conventional moodboarding is its profound reliance on the designer's personal subjectivity, intuition, and prior experience. The selection of images, the perception of thematic connections, and the extraction of a design language are all filtered through the individual designer's unique cognitive lens. Consequently, the quality and relevance of a moodboard are linked to the skill level of its creator. An experienced designer may intuitively identify subtle patterns and create a powerful, coherent board, whereas a novice may struggle to move beyond superficial choices, resulting in a weak or disjointed visual narrative. This dependency makes the process difficult to standardize, teach, or replicate consistently across a design team. The outcome remains a highly personal artefact, making objective critique and collaborative refinement challenging.

\subsubsection{Intensive Labour and Time Consumption}
The moodboarding process is highly time-consuming and labour-intensive. The initial stages of collection and selection involve a great deal of logistical work. In the modern context, this translates to hours spent browsing numerous websites, downloading hundreds of images, organizing them into folders, and then manually importing and arranging them in a digital tool. This significant investment in logistical tasks diverts the designer's valuable time and cognitive energy away from the core creative activity: the extraction of features and synthesis of ideas. The entire process, from initial search to final composition, can easily span several days or even weeks, creating a bottleneck in the fast-paced product design and development cycle.

\subsubsection{Cognitive and Environmental Constraints of 2D Interfaces}
Modern digital moodboarding is fundamentally constrained by the medium it employs, i.e., the two-dimensional computer screen. Designers are forced to view a vast collection of images through a small window, able to see only a handful of them at any given time. This sequential viewing process imposes a high cognitive load, as the designer must hold the context of previously seen images in their working memory while evaluating new ones. It hampers the ability to perceive the collection holistically and restricts the potential for serendipitous discovery of connections between disparate images. Furthermore, the typical computer work environment is rife with potential distractions—notifications, other applications, and ambient activity—which can disrupt the focused, immersive state of mind required for deep creative reflection. The very purpose of setting a specific "mood" is compromised when the designer's own mood is susceptible to constant external interference.

\subsubsection{The Opaque Nature of Feature Extraction}
In the conventional workflow, the most critical stages—reflection and extraction—are almost entirely implicit (as detailed in the following section). The process by which a designer observes a collection of images and distills a coherent set of shapes, colour palettes, and textural qualities happens within the confines of their own mind. This "black box" phenomenon means the resulting design language is not explicitly traceable to its visual origins. It is difficult for other team members, stakeholders, or even the designers themselves to articulate precisely why certain features were chosen over others. This lack of transparency hinders collaboration, makes the design rationale difficult to defend, and complicates the process of iterating or building upon the initial concept.

\subsubsection{The Influence of the Designer's Transitory Mood}
Another subtle pitfall is the influence of the designer's own emotional state on the moodboarding process. The objective is to create a moodboard that reflects the visual and emotional needs of the target user. However, the designer's personal mood on any given day can subconsciously bias their selection of images. If the designer is feeling subdued, they might gravitate towards more muted or sombre imagery, even if the brief calls for a vibrant and energetic aesthetic. This introduces an unintended and often unrecognized variable into the design process, risking a misalignment between the final product form and its intended emotional impact.

Collectively, these challenges highlight a need for a new approach that can structure the unstructured, objectify the subjective, and create an immersive and interactive environment conducive to focused creativity.

\greyline

\section{Potential of Attention Theory}
\label{sec:attention_theory}
The field of psychology has long studied the taxonomy and dynamics of attention, offering valuable frameworks for understanding how humans engage with visual material \cite{Bundesen_Habekost_Kyllingsbæk_2005, Hessels_Nuthmann_Nyström_Andersson_Niehorster_Hooge_2024}. Key distinctions include external versus internal attention, stimulus-driven (exogenous) versus goal-driven (endogenous) attention, and implicit versus explicit attention. External attention refers to the sensory processing of environmental stimuli, while internal attention involves focusing on internal thoughts or memories \cite{Solstad_Kaspersen_Eggen_2025}. Stimulus-driven attention is reactive and automatic, whereas goal-driven attention is deliberate and aligned with task objectives. Importantly, implicit attention operates below conscious awareness (at a subconscious level), while explicit attention is consciously directed and controlled.

Theories of attention further illuminate how information is filtered and prioritized. The Spotlight Theory \cite{posner1980orienting} posits that attention moves like a spotlight, enhancing the processing of stimuli in a selected region. The Feature Integration Theory \cite{treisman1980feature} suggests that visual features like colour and shape are processed in parallel before being integrated into unified percepts through focused attention. Furthermore, the Load Theory of Attention \cite{lavie2004a} proposes that attentional capacity is limited, with cognitive load determining whether irrelevant stimuli are filtered out. These theories highlight attention as a selective mechanism governing perception and action. In form design, visual attention determines not only what a designer sees but also what they choose to retain and transform. Empirical studies in cognitive psychology have demonstrated a strong correlation between gaze duration and preference, with individuals tending to gaze longer at items they subsequently rate as preferable \cite{Shimojo_etal_2003, Orquin_MuellerLoose_2013, Saito_etal_2017}. Fixation duration has emerged as a reliable proxy for implicit preference in user experience design and other fields \cite{Wolf_etal_2018}, opening the possibility of using real-time gaze tracking to quantify attentional processes in design ideation.

\greyline

\section{Deconstructing a Moodboard: Attentional Underpinnings}
\label{sec:deconstructing_moodboard}

The design of any physical, tangible product typically unfolds along two interrelated but distinct trajectories: functional design and form design. Functional design addresses the pragmatic aspects of product development—how the product works, how it solves user problems, and how it is manufactured—progressing through well-established stages such as problem identification, concept development, embodiment design, and detailed engineering. In contrast, form design is concerned with the aesthetic, emotive, and perceptual dimensions of the product—how it looks, how it feels, and how it communicates identity and intention. While functional design is rooted in logic and constraint satisfaction, form design is inherently more subjective, context-driven, and emotionally laden. As previously discussed, the moodboarding stage is the most influential phase in the form design process, as it defines the aesthetic trajectory towards which the product evolves.

\subsection{Phases of Moodboard Creation}
\label{subsec:phases_moodboard}

The process of moodboarding typically unfolds in five interlinked sub-stages: \textit{collection, selection, composition, reflection, and extraction}. These stages are often performed iteratively, with feedback loops informing progressive refinement. Each stage is governed by a combination of perceptual mechanisms, cognitive judgments, and physical interactions, which—though rarely articulated explicitly—can be mapped onto distinct types of attention mechanisms as recognized in cognitive psychology. The details of each stage are collectively given in Table~\ref{tab:moodboard_procedural}.

\begin{table*}[htbp]
\centering
\caption{Procedural and Outcome-Oriented Breakdown of Moodboarding Stages}
\label{tab:moodboard_procedural}
\begin{tabular}{|p{2.3cm}|p{5.0cm}|p{2.75cm}|p{3.5cm}|p{2.5cm}|}
\toprule
\textbf{Stage} & \textbf{Description} & \textbf{Cognitive Activity} & \textbf{Physical Action} & \textbf{Outcome} \\
\midrule
\textbf{Collection} & 
\textbf{Traditional:} Gathering physical materials (magazine clippings, photos, fabric swatches) that resonate with keywords. \newline \newline
\textbf{Modern:} Searching and downloading digital images from online sources (Pinterest, Unsplash, Pixabay, etc.) based on keywords.
& Rapid decision-making, affective judgment & 
\textbf{Traditional:} Cutting clippings, taking photographs, gathering physical materials. \newline \newline
\textbf{Modern:} Web navigation, mouse-clicking, downloading files, file organizing.
& Raw visual repository (Physical items vs. Digital files) \\
\hline
\textbf{Selection} & 
\textbf{Traditional:} Often overlaps with collection; involves sorting and prioritizing physical items. \newline \newline
\textbf{Modern:} A distinct step of filtering the downloaded digital image repository.
& Comparative analysis, preference filtering, assessing relevance and resonance &
\textbf{Traditional:} Sorting items into piles, discarding less relevant materials. \newline \newline
\textbf{Modern:} Opening and viewing digital files, deleting or moving files between folders.
& Refined subset of materials (Physical vs. Digital) \\
\hline
\textbf{Composition} & 
\textbf{Traditional:} Arranging and grouping physical materials on a large whiteboard or wall. \newline \newline
\textbf{Modern:} Arranging and grouping digital images within a software application (Figma, Miro).
& Pattern recognition, thematic grouping, establishing visual hierarchy &
\textbf{Traditional:} Pinning, gluing, or taping items onto a physical board. \newline \newline
\textbf{Modern:} Resizing, dragging, layering, and positioning digital images using a mouse and keyboard.
& Structured moodboard layout (Physical board vs. Digital file) \\
\hline
\textbf{Reflection} & 
Observing the composed layout (physical or digital) to absorb visual information and identify latent themes and patterns.
& Deep immersion, conceptual linkage, identifying emotional triggers &
Minimal physical interaction; prolonged staring and cognitive engagement with the composed board.
& Mental mapping of emotional and visual patterns \\
\hline
\textbf{Extraction} &
Distilling implicit patterns into explicit features (shapes, colours, textures) to inform concept sketching.
& Visual-to-form translation, synthesis of abstract ideas into concrete elements &
Sketching on paper, digital clipping of image regions, creating colour palettes, writing down notes.
& Actionable feature map (colour scale, textures, shape cues) \\
\bottomrule
\end{tabular}
\end{table*}

\subsubsection{Stage 1: Collection of Visual Material}

Moodboarding begins with the collection of reference materials (often images) that represent the designer’s initial interpretation of the desired product direction. This is often guided by a curated set of keywords emergent from the previous stage. The number of keywords can vary, typically ranging between 10 to 20. Once keywords are established, designers scour the internet for relevant visual material. Platforms like Pinterest, Unsplash, Behance, Pixabay, Instagram, etc., serve as rich repositories. In traditional settings, this stage involved physical cutouts from magazines or fabric swatches, but contemporary designers predominantly rely on digital image collection. The act of collecting images is explicit, requiring the designer to make quick yet nuanced decisions about whether a particular image resonates with the thematic direction. This decision is guided by a blend of perceptual saliency and prior experience. The collection process is cognitively demanding and relies on rapid affective judgments rather than fine-grained reasoning. It typically spans from a few hours to several days, resulting in a curated visual repository of 50 to 100 images for digital moodboards.

\subsubsection{Stage 2: Selection of Relevant Images}

After accumulating a substantial image pool, designers proceed to select a refined subset that aligns more precisely with the design brief. This stage involves a deeper level of inspection and is governed by both internal criteria (designer intent) and external features (image content). Here, the designer engages in comparative judgment—evaluating how well each image fits the emotional or functional direction of the intended product. This selection process is also explicit and predominantly cognitive, involving focused attention on individual images to assess form relevance, material expressiveness, and emotional resonance. This stage generally takes one to two days, and the selected set typically constitutes one-third to one-half of the original image pool.

\subsubsection{Stage 3: Composition of images}

Having identified a refined set of images, the next step involves their spatial arrangement and grouping based on visual and thematic similarity. Designers arrange images into clusters based on shared attributes—such as colour palettes, surface textures, or geometric forms. This stage enables the designer to perceive patterns and commonalities that may not be evident when images are viewed in isolation. The composition phase is both a cognitive and physical activity, where designers use digital tools to resize, reposition, and layer images in a visually expressive layout. Images are often arranged in uneven grids, with some scaled up to emphasize dominance and others reduced to indicate secondary influence. This phase demands alternating attention and usually takes two to three days, resulting in a visually structured representation of the mood direction.

\subsubsection{Stage 4: Reflection on the images}
Once the moodboard is spatially organized, designers enter a more reflective phase. They closely observe the moodboard as a whole, identifying aesthetic patterns, conceptual themes, or emotional triggers embedded within the image clusters. This phase is largely internal and is governed by sustained attention, as designers spend prolonged periods engaging with the imagery to derive latent insights. Unlike earlier stages, this phase is less about manipulation and more about cognitive immersion, where the designer makes mental notes of recurring elements that evoke the intended emotional response. This highly subjective stage typically spans over a day.

\subsubsection{Stage 5: Extraction of features}
The final stage involves distilling the implicit visual patterns into explicit, actionable features that can inform the generation of form sketches. The designer identifies specific elements—such as a particular curvature, material texture, or colour gradient—that can be translated into physical form. These features may then be diagrammed, annotated, or digitally clipped for reuse. Here, the designer moves from reflective internal attention to a blend of internal and external attention, combining memory-based recognition with fresh perceptual inspection. The extracted features form the seed for concept sketching and typically last a day, often continuing in parallel with subsequent refinement activities.


\subsection{Attention Taxonomy in Moodboarding}
\label{subsec:attention_taxonomy}
This multi-stage, highly cognitive process reveals the latent complexity and attentional richness of moodboarding. Each stage involves distinct attentional demands. Importantly, these attentional types do not occur in isolation; rather, they are interwoven, with designers oscillating between implicit impressions and explicit decisions. Furthermore, recognizing which stages are explicit and experience-driven opens up avenues for leveraging real-time eye-tracking to provide scaffolding for novice designers. Understanding how attention functions in moodboarding makes it possible to design tools that support, enhance, or automate specific stages. To better formalize the role of attention in moodboarding, Table~\ref{tab:moodboard_cognitive} presents a structured overview of the five stages along multiple dimensions: task description, attentional mechanism, explicitness, cognitive and physical demands, typical duration, and tangible outcomes.

\begin{table*}[!ht]
\centering
\caption{Mapping of Attention to Moodboarding Phases}
\label{tab:moodboard_cognitive}
\begin{tabular}{|p{3.0cm}|p{2.5cm}|p{3cm}|p{2.5cm}|p{4.5cm}|}
\toprule
\textbf{Phase} & \textbf{Attention Type} & \textbf{Attention Mechanism} & \textbf{Designer's Role} & \textbf{Key Challenges for Novice Designers} \\
\midrule
\textbf{Collection} & Explicit & Pre-attentive & Gather and Explore & Information overload; Difficulty in assessing long-term relevance of initial finds. \\
\hline
\textbf{Selection} & Explicit & Selective & Filter and Curate & Overcoming subjective bias; Cognitive load of comparing numerous items; Fear of discarding potentially useful material. \\
\hline
\textbf{Composition} & Explicit & Alternating, Divided & Architecture and Story-telling & Creating a coherent visual narrative from disparate elements; Balancing details with the overall composition. \\
\hline
\textbf{Reflection} & Implicit & Focused, Stimulus-driven & Analyze and Contemplate & Opaque mental process (the "black box"); Cognitive fatigue from prolonged concentration; Difficulty in articulating latent insights. \\
\hline
\textbf{Extraction} & Implicit & Sustained, Goal-driven & Synthesize and Translate & Translating abstract feelings and patterns into concrete, actionable design features; Lack of traceability between inspiration and output. \\
\bottomrule
\end{tabular}
\end{table*}


\subsection{Need for a Attention-Aware Approach}
\label{subsec:research_gap}

The deconstruction of the conventional moodboarding process, juxtaposed with its inherent pitfalls, reveals a significant and compelling research gap at the intersection of design methodology, cognitive science, and human-computer interaction. The current paradigm, whether traditional or modern, is fundamentally misaligned with the cognitive processes it seeks to support. It burdens the designer with explicit, logistical tasks (collection, selection, composition) that consume time and mental resources, while leaving the most crucial creative steps (reflection, extraction) as an implicit, opaque, and unsupported mental exercise. This workflow is not only inefficient but also erects significant barriers for novice designers and hinders objective collaboration within teams.

The core of this disconnect lies in the failure of existing tools and methods to acknowledge, measure, or leverage the designer's \textit{visual attention}. While cognitive psychology provides robust theories linking attention to preference, emotion, and feature perception, these principles have not been translated into practical tools for form design. The process remains reliant on the designer's conscious, explicit choices, ignoring the rich stream of subconscious, implicit data generated by their gaze as they interact with visual stimuli. This oversight represents a missed opportunity to create a more intuitive, data-driven, and human-centric design process.

This analysis points to the necessity of \textit{inverting the conventional workflow}. A new paradigm is required where the laborious, explicit tasks of collection and selection are made implicit and automated, freeing the designer to concentrate their cognitive faculties on the creative acts of reflection and extraction, which, in turn, must be made more explicit and transparent.

Addressing this gap requires the development of a novel system founded on a new set of principles. Such a system must satisfy the following requirements:

\paragraph{\textbf{Requirements}} 

\begin{enumerate}
    \item \textbf{Supporting Large Stimuli Set:} The system should provide a vast number of visual stimuli to facilitate comprehensive exploration.
    
    \item \textbf{Reduced Physical Effort:} 
    The system should not require large-scale movement of material or physically demanding action from the designer. 

    \item \textbf{Reduced Cognitive Effort:} The system should enable the designer's natural cognitive processes for segregating visual assets. Since the implicit attention mechanism bears less cognitive load, it is preferred.
    
    \item \textbf{Elicitation of Selection Rationale:} The system should capture the subconscious rationale driving the designer's preferences to support feature extraction.
    
    \item \textbf{Increased Efficiency:} The system should enhance productivity by significantly reducing the time required for the form design process.
\end{enumerate}

\paragraph{\textbf{System Features}}

\begin{enumerate}
    \item \textbf{Provide an Immersive and Scalable Environment:} To overcome the cognitive and spatial limitations of 2D screens, the system must offer a distraction-free, immersive environment. This space should be capable of displaying a large number of visual stimuli simultaneously, facilitating holistic perception and serendipitous discovery in a way that mimics a real-world gallery rather than a digital folder.

    \item \textbf{Enable Implicit and Objective Preference Capture:} To reduce subjectivity and the burden of manual selection, the system must be able to objectively capture a user's preferences implicitly. By leveraging real-time eye-tracking technology, it can use attentional metrics, such as gaze fixation duration, as a reliable proxy for interest and liking, thereby automating the selection process based on the designer's subconscious cues.

    \item \textbf{Facilitate a Structured and Guided Workflow:} To support novice designers and streamline the process for experts, the system must impose a systematic and automated structure on the ideation process. It should guide the user through distinct phases of exploration, making the creative journey more manageable and repeatable.

    \item \textbf{Support Explicit and Data-Driven Feature Extraction:} To make the opaque process of creative synthesis transparent, the system must be able to translate implicit attentional data into explicit, actionable design inputs. This involves automatically identifying and extracting salient visual features (such as dominant shapes, colours, and textures) from the regions of an image that captured the designer's attention.

    \item \textbf{Democratize the Design Process:} By automating logistical tasks and providing cognitive support, the system should empower designers of all skill levels. It should lower the barrier to entry for novices while providing experts with a powerful tool to augment their intuition and accelerate their workflow.
\end{enumerate}

A design system that holistically integrates an immersive virtual environment with implicit attention-sensing is required to systematically restructure the form design process. The following section of this paper introduce EUPHORIA, a system designed and built to address this very gap by transforming the conventional moodboard into a responsive, immersive, and interactive \textit{moodspace} (MS).


\greyline

\section{Translating Moodboard to Moodspace}
\label{sec:euphoria_design}

To address the limitations inherent in conventional form design and bridge the identified research gap, the authors introduce \textbf{EUPHORIA} (Empathizing User Preferences on a Holodeck using Real-time Implicit Attention). EUPHORIA is a novel system that reimagines the foundational process of moodboarding by fundamentally restructuring its workflow and interaction paradigm. It moves beyond the static, two-dimensional \textit{moodboard} to create a dynamic, responsive, immersive, and intelligent \textit{moodspace} (MS). This transformation is achieved through the synergistic integration of core technologies: Virtual Reality (VR) and Eye-Tracking (ET). The system is architecturally designed not merely to digitize the existing process, but to invert it, aligning the workflow with the natural cognitive and attentional processes of the designer.

\subsection{The Core Philosophical Shift: Inverting the Design Workflow}
The central innovation of the EUPHORIA system lies in its radical departure from the conventional moodboarding sequence. As established, the traditional process burdens the designer with explicit, labour-intensive tasks of \textit{collection}, \textit{selection}, and \textit{composition}, while the most creatively vital stages of \textit{reflection} and \textit{extraction} remain implicit and mentally taxing. EUPHORIA systematically inverts this workflow.

By leveraging a designer's innate visual attention as the primary input, the system renders the stages of \textit{collection, selection, and composition implicit and automated}. The designer no longer needs to manually search, download, click, or drag images. Instead, their natural gaze behaviour within the immersive environment implicitly signals their preferences, which the system captures and uses to automatically curate and compose relevant visual stimuli. This frees the designer's cognitive resources from logistical overhead such as clicking, dragging, resizing, etc., allowing them to focus entirely on the higher-order creative tasks of \textit{reflection and extraction}. These stages, in turn, are transformed from an opaque mental process into an \textit{explicit, data-driven activity}, supported by system-generated outputs that visualize the designer's attentional patterns and extracted visual features. This philosophical shift from a manual-explicit to an implicit-automated paradigm is the cornerstone of the EUPHORIA architecture.

Furthermore, EUPHORIA replaces the two-dimensional confines of digital screens with a spatial, multi-modal virtual environment. This enables designers to view and interact with hundreds of floating images at once in a three-dimensional layout—an experience akin to standing in a gallery rather than scrolling through a screen. This architectural shift from moodboard to moodspace represents a core conceptual advancement in the system.

\begin{figure*}[!ht]
    \centering
    \includegraphics[width=0.9\linewidth]{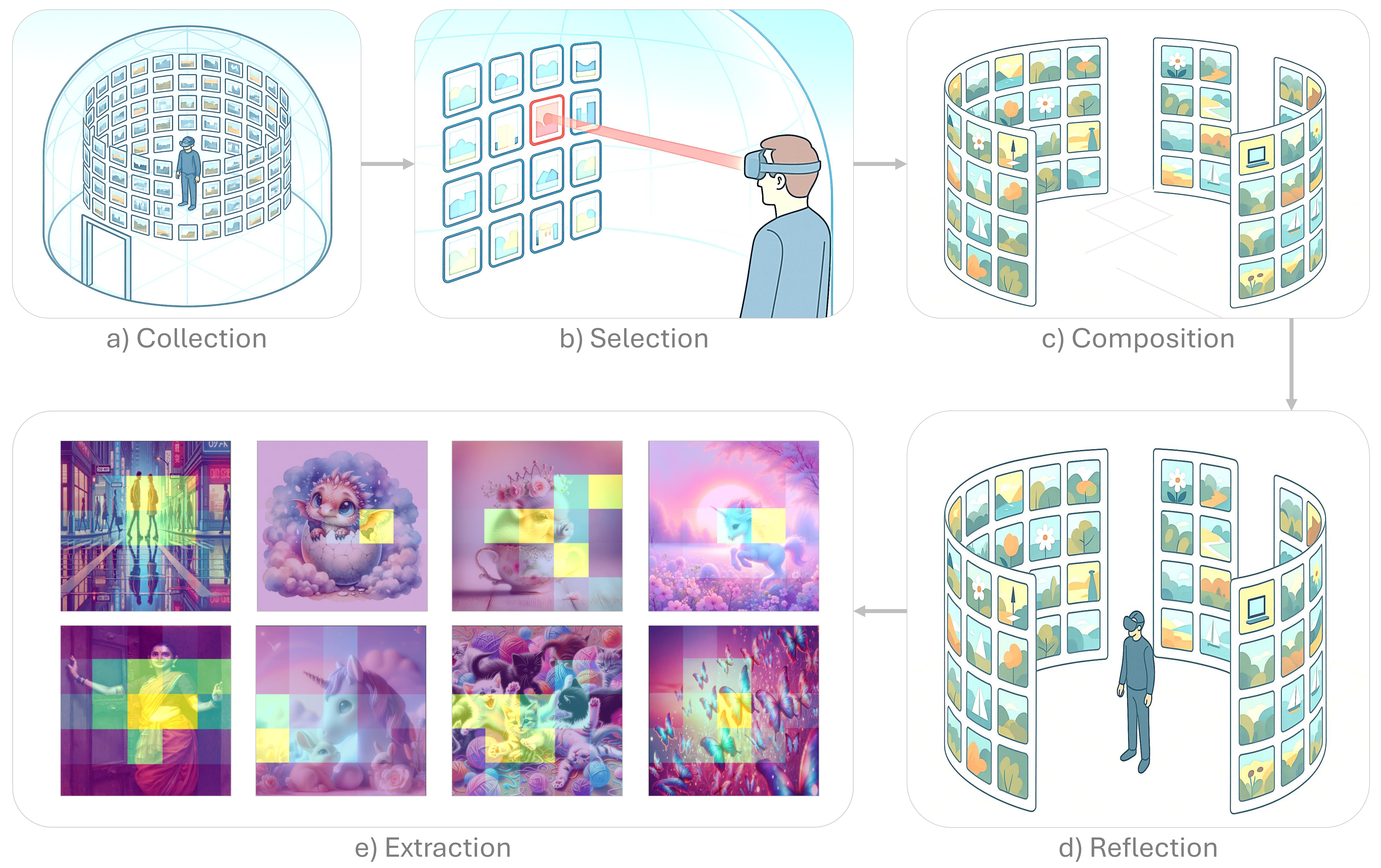}
    \caption{The workflow of the EUPHORIA system, illustrating how the conventional moodboarding stages are transformed into an implicit, attention-driven process within the immersive moodspace. The process unfolds as follows: 
    \textbf{(a) Collection:} The designer is situated at the center of a large constellation of visual stimuli, enabling a comprehensive and immersive initial exploration. 
    \textbf{(b) Selection:} Instead of manual clicking, the system uses real-time eye-tracking to implicitly select images that capture the designer's sustained gaze fixation. 
    \textbf{(c) Composition:} The attention-selected images are automatically composed into a refined, more focused moodspace. 
    \textbf{(d) Reflection:} The designer then reflects upon this curated set of images. 
    \textbf{(e) Extraction:} The system makes the reflection process explicit by generating attention heatmaps over the selected images. These heatmaps highlight the specific Regions of Interest (ROIs) that subconsciously attracted the designer's attention, providing a tangible, data-driven output for the subsequent feature extraction and form generation stages.}
    \label{fig:euphoria_workflow}
\end{figure*}

\subsection{Architectural Components}
The EUPHORIA system is built upon a carefully integrated triad of technologies, with each component addressing specific requirements identified in the research gap. The detailed flow of EUPHORIA system is given in Figure~\ref{fig:euphoria_workflow}.

\subsubsection{Virtual Reality (VR): The Foundation of the Immersive Moodspace}
To overcome the spatial and cognitive constraints of 2D screens, EUPHORIA is built as a standalone application in a VR environment using the Unity 3D engine. The system creates a vast, open, and distraction-free moodspace that replaces the conventional flat canvas. When a user enters the application, they are situated at the centre of a large, translucent glass dome, under an open sky. This design choice minimizes external distractions and fosters a state of focused immersion.

\begin{figure}[!ht]
    \centering
    \includegraphics[width=1\linewidth]{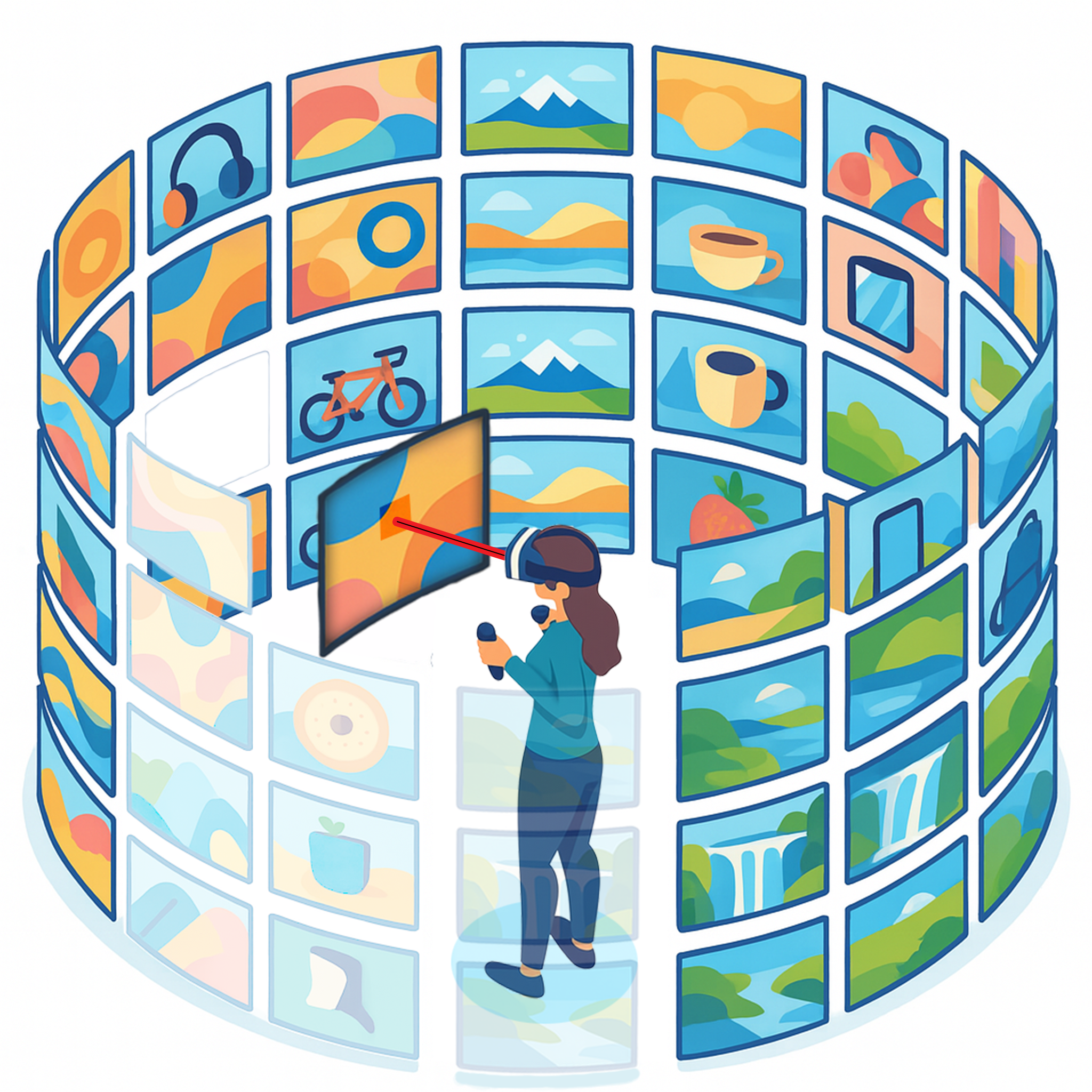}
    \caption{A schematic representation of the EUPHORIA system in action, illustrating the immersive 'moodspace' paradigm. A designer, equipped with a VR headset, stands at the center of a cylindrical canvas populated by a constellation of visual stimuli. The system utilizes integrated eye-tracking to monitor the designer's gaze, represented by the ray tracing a path to a specific image region. Images that capture the designer's sustained attention automatically move closer for detailed inspection, enabling an intuitive, hands-free selection process. Additionally, the designer can use a handheld joystick controller to rotate the entire image canvas, allowing for seamless browsing of the extensive visual library.}
    \label{fig:euphoria_system_in_action}
\end{figure}

The visual stimuli (images) are not confined to a small window but are populated in a large-scale, three-dimensional arrangement. They are displayed in an evenly spaced grid along the circumference of a vast, invisible cylinder that encircles the user. This spatial layout allows the designer to be surrounded by hundreds of images simultaneously (the system currently supports between 300 to 500 images at once, with the capacity for more), enabling holistic perception and facilitating the discovery of serendipitous connections that would be impossible on a flat screen. The user can load images either from a local directory or by using an integrated search feature that makes API calls to image-hosting websites based on keywords, populating the moodspace in real-time. This immersive presentation transforms the act of browsing from a tedious scrolling exercise into a natural, exploratory experience akin to walking through a gallery, thereby directly addressing the need for an immersive and scalable environment. A schematic representation of the same is shown in Figure~\ref{fig:euphoria_system_in_action}.

\subsubsection{Eye-Tracking (ET): The Engine for Implicit Preference Capture}
The second critical component of EUPHORIA is the integration of real-time eye-tracking, facilitated by the built-in ET system of the Meta Quest Pro headset. This technology serves as the engine for capturing the designer's implicit preferences objectively and non-intrusively. Drawing from established principles in cognitive psychology that correlate gaze duration with interest and preference, the system continuously monitors where the user is looking within the moodspace.

In the background, a custom-built ray-tracing algorithm runs constantly. It projects a ray from the user's point of gaze and calculates the first collision point with any of the displayed images. The system records the precise location of this gaze point on the image (the local coordinates) and the duration of the fixation. This fixation duration data is the primary metric used to quantify the user's attention. The underlying hypothesis is that while a user may glance at many images, they will only stare at or fixate on images or regions within images that they subconsciously prefer or find compelling. By accumulating this fixation data, EUPHORIA bypasses the need for explicit selection (e.g., clicking a "like" button), thereby automating the selection process based on the user's unfiltered, subconscious attentional cues.

\subsection{Benefits of Moodspace Paradigm for designers}
\label{subsec:benefits_moodspace_paradigm}
The shift from 2D moodboards to 3D moodspaces is not merely aesthetic but epistemological. VR facilitates immersiveness, spatial memory reinforcement, and parallel visual access to hundreds of images—capabilities unattainable in traditional screen-based workflows. For experienced designers, this enables faster pattern recognition and broader visual synthesis. For novices, the moodspace provides scaffolding through attentional guidance and automatic structuring.
Moreover, the moodspace fosters a continuous interaction loop between the designer and system: gaze behaviour influences the visual hierarchy, the system adapts in real-time, and the designer's exploration deepens. Unlike conventional moodboards, which are static and passive, moodspaces are responsive and evolving, allowing design cognition to become externalized, navigable, and observable. This structured segmentation enables a unified framework where technology and cognition co-evolve to enhance creativity, reduce cognitive overload, and democratize access to form ideation. EUPHORIA, thus, does not replace the designer’s role but acts as an amplifier of design intent, capturing the subtleties of attention and emotion that traditionally remain implicit.


\greyline

\section{A Naive Approach to Automated Form Design}
\label{sec:naive_approach}
To achieve a more inclusive and efficient end-to-end workflow, the subsequent stages of form creation must also be considered. The conventional design process transitions from moodboarding into a series of highly skilled, action-dominant phases where the designer translates abstract inspirations into concrete product forms.

\subsection{Stages in Form Creation}
\label{subsec:conventional_inspiration_to_form}
The typical stages following moodboarding can be categorized as follows. In~\textit{Geometric Shape Abstraction} stage, the designer translates the implicit, mental impression derived from the moodboard in terms of characteristic lines, curves, and shapes. This is followed by \textit{Exploration through Scribbles}, a divergent phase where these geometric building blocks are exploited to rapidly ideate a wide array of potential product forms. Then, the designer converges on the most promising forms during the \textit{Creation of Concept Sketches} stage, refining the scribbles into a detailed, coherent representation of the final form. The process culminates in \textit{Photorealistic Form Rendering} stage, where colour, material, and texture are applied to the concept sketch. The detailed explanation of these stages, their process, and outcomes is given section~\ref{app:form_design_fundamentals}.

\subsection{Challenges in Manual Form Creation}
\label{subsec:challenges_manual_generation}
The process of form creation is dependent on the designer's individual capabilities. The key challenges in this regard, especially for novice designers, are as follows.
\begin{itemize}
    \item \textbf{Skill Dependency:} The workflow demands a high level of spatial reasoning, advanced sketching and drafting capabilities, and a refined sense of aesthetics.
    \item \textbf{The Cognitive Leap:} Translating abstract visual features from a moodboard into a coherent geometric and form language is a non-trivial cognitive leap that is difficult to teach and master.
    \item \textbf{Experience-Driven:} Seasoned designers can draw upon a vast mental library of forms and solutions, an advantage that novices lack.
    \item \textbf{Intuition-driven:} The rationale for specific form choices is often difficult to articulate, replicate, or collaboratively build upon.
\end{itemize}

\subsection{Requirements for an Automated System}
\label{subsec:requirements_automation}
To address the above challenges through an automated form design workflow, a system should satisfy the following requirements.
\begin{enumerate}
    \item \textbf{Understand Visual Features:} The ability to perceive and interpret the diverse visual features (style, shape, texture, and colour attributes) embedded in the stimuli in the moodboard.
    \item \textbf{Feature-to-Form Translation:} The capacity to translate these interpreted features into a coherent shape language for the specified product.
    \item \textbf{Generative Ideation:} The ability to generate multiple, diverse, and creative concept variations based on the same set of interpreted features.
    \item \textbf{High-Fidelity Synthesis:} The capability to synthesize high-quality renderings for each of the variations.
\end{enumerate}

\subsection{Experimenting with a Monolithic Generative Approach}
\label{subsec:monolithic_approach}
\begin{figure*}[htbp]
    \centering
    \includegraphics[width=0.9\linewidth]{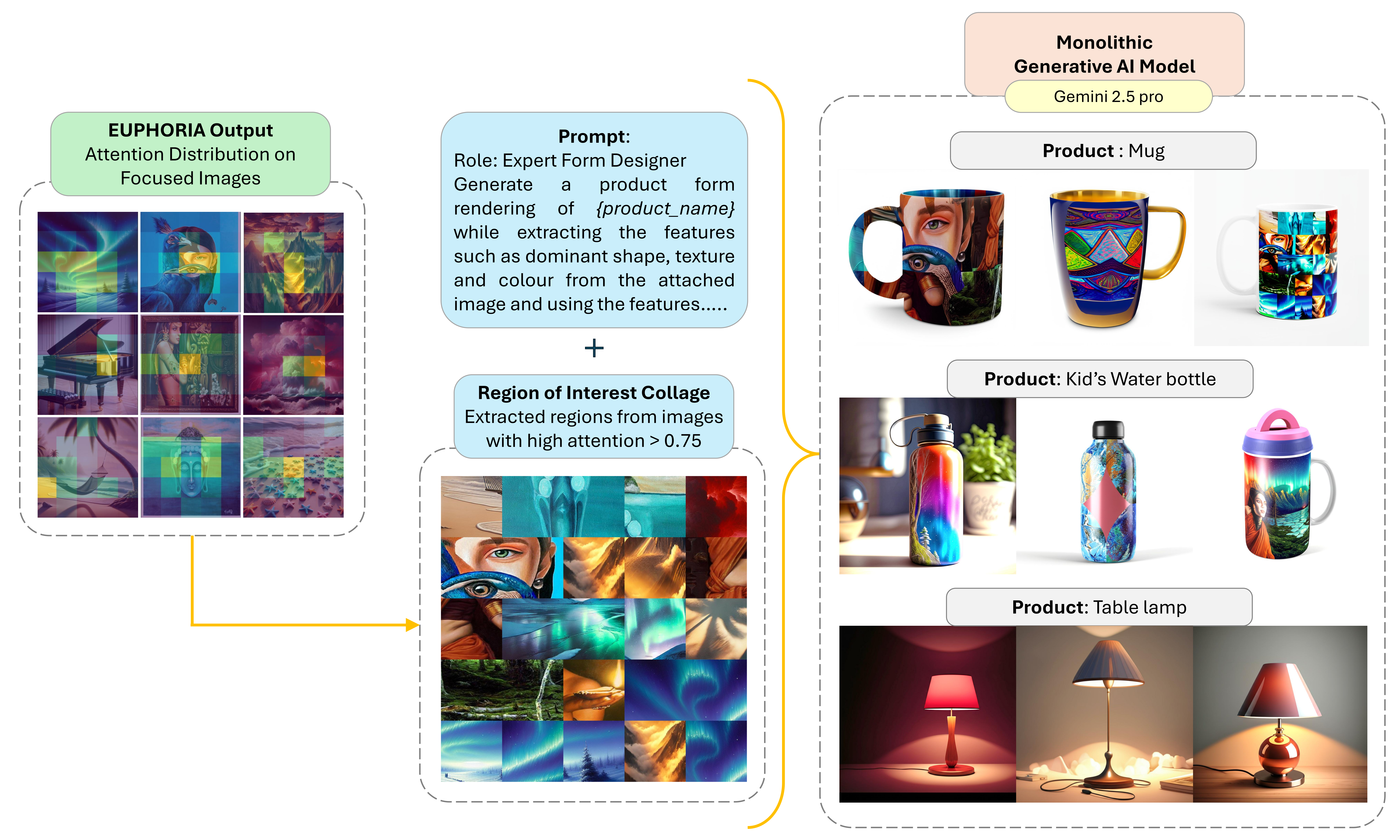}
    \caption{NAIVE Monolithic AI Workflow}
    \label{fig:naive_ai}
\end{figure*}

As a first attempt at automation, a naïve approach was tested. This involved using commercially available, off-the-shelf, monolithic Large Language and Vision Models (Gemini class of models) to try and generate final form renderings directly. 
The capabilities of those systems are limited to one image and one prompt. EUPHORIA gives us a set of images of interest of the designer with attention distribution heatmaps overlaid on the images. Thus, our output is not directly compatible with the available form rendering platforms. To bypass this limitation, the \textit{regions of interest (ROI) with high attention in each image are extracted, and an ROI collage is created out of these}. The ROI collage implicitly encodes the salient visual features (style, shape, texture, and colour) that attracted the designer's attention. This image collage, along with the prompt for the required response, is used for form generation. A representative set of results is shown in Figure~\ref{fig:naive_ai}.

It can be observed that despite the use of elaborate and detailed prompts, the models consistently struggled to provide desirable outcomes. The primary shortcomings were as follows.
\begin{itemize}
    \item \textbf{Failure of Feature Extraction:} The model could not reliably understand the important visual features in the attention heatmaps as indicators of importance that the designers would find appealing.
    \item \textbf{Lack of Understanding:} The models failed to interpret the textural and stylistic information embedded in the collage. The form generated was not influenced by the features in the collage. It merely wrapped the image around the product as a texture!
    \item \textbf{Non-creative Outputs:} Most often, it took the standard shape of a product as such. The system lacked the ability to abstract a coherent shape language from the ROI collage.
\end{itemize}

The failure of this naïve approach revealed that automated creative form generation is not a simple, single-input-to-output task. It requires a more sophisticated, fine-grained system: first to sequentially identify, extract, and interpret the different features (analyze), and then synthesize a coherent whole form (generate). An "analyze-then-generate" paradigm is, therefore, proposed for an \textit{agentic AI system}, RETINA, which uses specialized agents to handle distinct sub-tasks that mimic the human cognitive process in a coordinated manner.


\greyline

\section{Generating Form from Features}
\label{sec:retina_design}
The translation of visual features identified from EUPHORIA into tangible form is handled by a Responsive Embodiment Through Iterative visioN-guided Agentic AI (RETINA). It employs a compound Large Language Model (LLM) and Large Image Model (LIM) to deconstruct and then synthesize visual information. It is designed to systematically transform the implicit, subconscious preferences captured by eye-tracking into a rich, multi-modal set of explicit design features, which are then used to automatically generate novel and relevant form design solutions without human involvement.

\subsection{RETINA Architecture}
\label{subsec:retina_architecture}
\begin{figure*}[htbp]
    \centering
    \includegraphics[width=0.75\linewidth]{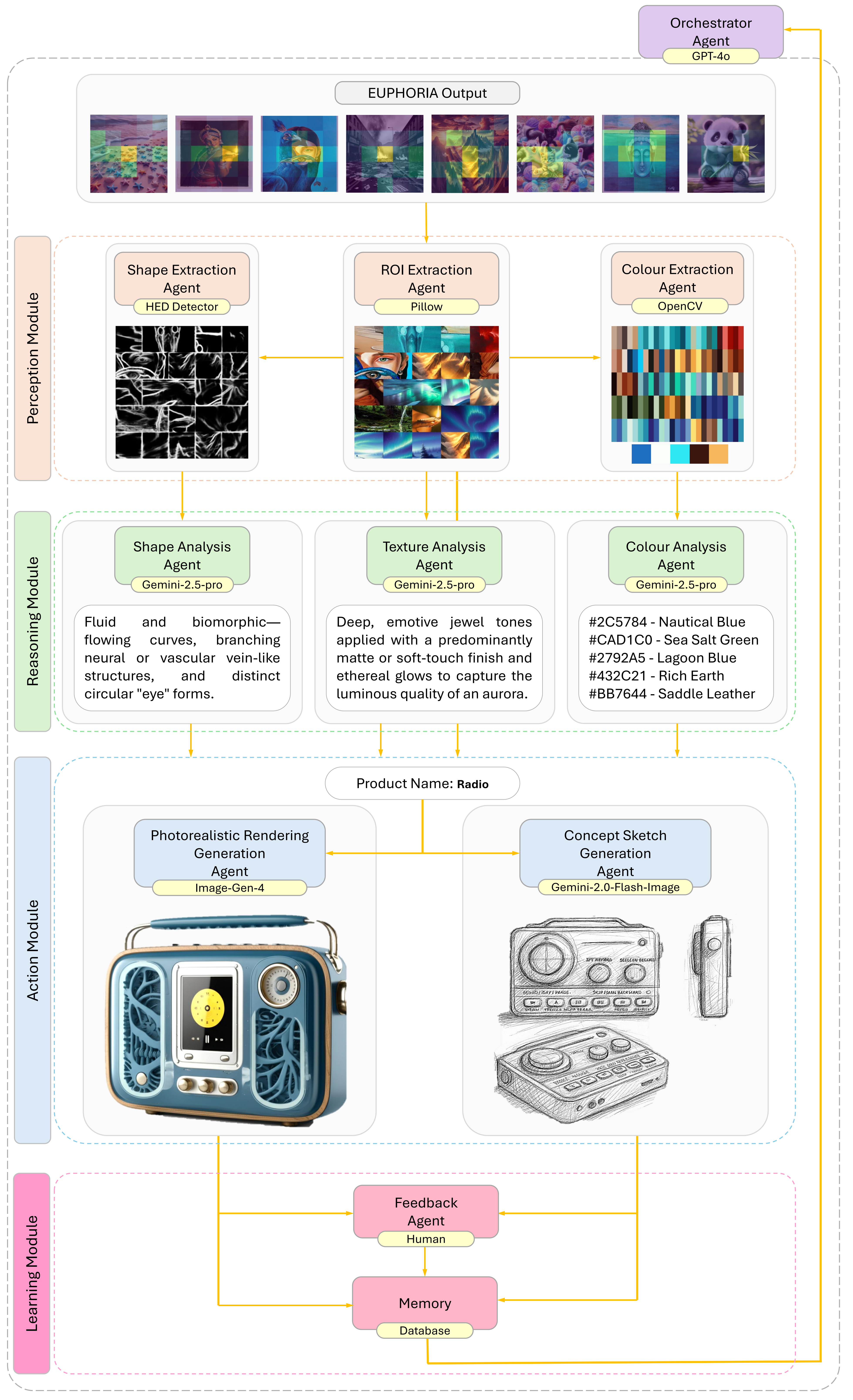}
    \caption{RETINA Agentic AI Workflow with a representative input and the corresponding auto-generated output at each level}
    \label{fig:retina_ai}
\end{figure*}

The architecture of RETINA is a multi-agent system (MAS), managed by a central orchestrator agent as shown in Figure~\ref{fig:retina_ai}. This hierarchical structure is designed to decompose the complex, overarching goal of "create a product form design" into a series of logical, executable sub-tasks, each assigned to a specialized agent.

The above approach aligns with the closed-loop \textit{operational cycle} of an agentic AI known as "Understand, Think, Act" \cite{Schneider2025-by, Sapkota2025-yt, Plaat2025-jj}. The specific meanings with which the terms are used here are as follows.

\begin{itemize}
    \item \textbf{Understand:} This process involves readying the system by stating its role, detailing the type of input it would receive, setting the context, and the nature of the output it is expected to generate.
    \item \textbf{Think:} This process involves engaging the orchestrator agent to create a multi-step plan, determining which specialized sub-agents to deploy and in what sequence to process the input data to generate a meaningful outcome.
    \item \textbf{Act:} This process involves executing the plan by invoking appropriate agents to extract and interpret the features, and generate the desired outcome.
\end{itemize}

\subsection{Essential Components of the RETINA Agentic AI system}
\label{subsec:essential_components_retina_agentic_ai_system}
RETINA is composed of several core modules that enable its autonomous and intelligent behaviour required for a robust AI agent.

\subsubsection{Perception Module}
\label{subsubsec:perception_module}
This module is RETINA's sensory input system. Unlike simpler generative AI that primarily takes a text prompt, RETINA's perception is multi-modal. It takes three primary data from the preceding stage, which are the form design problem statement as a textual description from the designer and the visual stimuli (images) that the designer has spent the most attention on. This module pre-processes this raw data into a distinct set of feature maps through a series of extraction agents as detailed below.

\paragraph{\textbf{Extraction Agents}}

\begin{enumerate}
\item \textbf{ROI Extraction Agent:} The agent receives a set of images along with the attention data. The system analyses the distribution of fixation durations across a grid overlaid on each highly attended image. Regions with a cumulative fixation time exceeding a predefined threshold are identified as ROIs and are automatically cropped. These ROIs, which capture the designer's subconscious preferences, are then compiled into a single visual summary as an "ROI Collage." This collage then serves as the input for shape and colour extraction agents.

\item \textbf{Shape Extraction Agent:} The ROI tiles are processed by this agent using computer vision techniques, specifically a Holistically-Nested Edge Detection (HED) algorithm, to extract dominant fuzzy edges that represent dominant lines and curves. This provides the system with a visual summary of the core shapes.

\item \textbf{Colour Extraction Agent:} A parallel agent uses computer vision techniques to analyze the ROI tiles to extract the dominant colour and create a colour palette.
\end{enumerate}

The ROI Collage, HED Shape Collage, and Dominant Colour Collage together are called \textit{feature maps}, which represent a direct, unfiltered reflection of the designer's implicit preferences.

\subsubsection{Reasoning Module}
\label{subsubsec:reasoning_module}
The "brain" of RETINA is its reasoning engine, which is responsible for evaluating the perceived feature maps and determining the sequence of actions. This engine leverages foundation models (gemini-2.5-pro for vision and text processing) for its core perceptual and reasoning. The primary function of this module is planning, as it analyses the \textit{feature maps into textual descriptors} through a series of analysis agents as detailed below.

\paragraph{\textbf{Analysis Agents}}

\begin{enumerate}
\item \textbf{Shape Analysis Agent:} The HED Collage, which contains the raw shape and edge information, is powered by a Vision AI agent, viz. gemini-2.5-pro model. This agent analyses the visual data to identify recurring patterns, contours, and geometric properties, and outputs a rich, textual description of the dominant shape language, e.g., "characterized by flowing organic curves and soft, rounded corners, defined by sharp, intersecting geometric angles".

\item \textbf{Colour Analysis Agent:} The Dominant Colour Palette image undergoes a two-step analysis. First, a computer vision (CV) agent extracts the primary HEX codes for each colour swatch. These codes are then fed into a Text Processing AI agent, viz., gemini-2.5-pro, which translates the HEX codes into descriptive names (e.g., "$\#EAE0D5$" becomes "eggshell white", "$\#3E5622$" becomes "forest green"). This provides a more human-readable and creatively useful colour specification.

\item \textbf{Texture Analysis Agent:} The ROI Collage is fed into a Vision AI agent, viz. gemini-2.5-pro. This agent focuses on identifying higher-level aesthetic properties, such as surface textures, material qualities, lighting, and overall style. Its output is a textual description of these elements (e.g., "features a matte, slightly weathered texture with brushed metal accents and a sense of minimalist design").

\end{enumerate}

This decomposition allows for a detailed and structured understanding of the designer's implicit preferences before any generative action is taken.

\subsubsection{Action Module}
\label{subsubsec:action_module}
This module enables RETINA to perform operations and interact with its "tools," which are the generative AI models. Once the analyses are complete, the pipeline converges the extracted textual and visual data into two different agents, which produce two distinct classes of outputs for the designer, namely, photorealistic rendering and concept sketch.

\paragraph{\textbf{Generation Agents}}

\begin{enumerate}
    \item \textbf{Concept Sketch Agent:} The first agent powered by image-gen-4 model. It receives a consolidated prompt containing the original problem statement text, the shape descriptor text, and the style and texture descriptor text. Based on this comprehensive input, it generates a concept sketch as output.
    
    \item \textbf{Photorealistic Rendering Agent:} The second agent, powered by the gemini-2.0-flash-image-generation model, creates a more refined, final-form visualization. Its prompt includes the problem statement, the shape descriptor, the HEX codes and colour names, the textural style descriptor, along with the ROI Collage image as a direct visual input. This allows the model to draw more directly from the original source of inspiration for texture and mood, resulting in the generation of a high-fidelity photorealistic product rendering.
\end{enumerate}

\subsubsection{Learning Module}
\label{subsubsec:learning_module}
RETINA incorporates a feedback loop with memory that serves as a mechanism for learning and improvement. The system generates multiple final renderings, and the designer's act of selecting one as the "best" solution, along with the associated comment. This feedback can be collected and leveraged over time using techniques like reinforcement learning from human feedback (RLHF), enabling favourable outcomes in subsequent generations.

\subsection{Prompt Engineering for RETINA}
\label{subsec:promptengg_retina}
The behaviour of the RETINA agentic pipeline is not random; it is meticulously guided by prompt and context engineering. Providing the system with a clear and comprehensive context is critical for ensuring that its autonomous actions are aligned with the designer's goals. This is primarily achieved through the system prompt of the central orchestrator agent. A short snippet of the system prompt is given below for the sake of brevity.

The system prompt for the RETINA orchestrator acts as its "job description," defining its identity, capabilities, and rules of engagement. It is structured with the following key components:
\begin{itemize}
    \item \textbf{Role, Purpose, Context:} "You are RETINA, an expert AI form design assistant. Your purpose is to translate a designer's implicit visual preferences, captured from EUPHORIA's gaze data, into novel product concepts that solve a given design problem."

    \item \textbf{Capabilities (Tools):} "You have access to a suite of specialized agents: three extraction agents (\verb|ROI-Extraction-Agent|, \verb|Shape-Extraction-Agent|, \verb|Colour-Extraction-Agent|), three descriptor agents (\verb|Shape-Descriptor-Agent|, \verb|Style-Texture-Descriptor-Agent|, \verb|Colour-Descriptor-Agent|), two generative agents (\verb|Sketch-Generator-Agent|, \verb|Rendering-Generator-Agent|), and a \verb|Feedback-Agent|. You also have access to a persistent \verb|Memory| module for storing feedback and learned preferences."

    \item \textbf{Instructions (SOP):} "(1) Upon receiving the initial inputs, first deploy the \verb|ROI-Extraction-Agent| to generate the ROI Collage from the gaze data. (2) Once the collage is created, deploy the \verb|Shape-Extraction-Agent| and \verb|Colour-Extraction-Agent| in parallel. (3) After the HED Collage and Colour Palette are generated, deploy the three Descriptor Agents in parallel to produce textual descriptors for shape, style/texture, and colour. (4) Once all textual descriptors are available, synthesize and execute prompts for the two generative agents. (5) Present all generated sketches and renderings clearly to the designer for their final selection. (6) Upon receiving the designer's final choice, deploy the \verb|Feedback-Agent| to process this selection and update the \verb|Memory| module."

    \item \textbf{Input/Output Definition:} "Your initial inputs are a text \verb|Problem Statement| and the raw \verb|EUPHORIA Gaze Data|. Your final outputs must be two sets of images: \verb|Concept-Sketches| and \verb|Photorealistic-Renderings|. You will also receive the \verb|Designer's Final Selection| as an input for the final feedback step."

    \item \textbf{Actions:} "Your only permitted actions are to call your available agents (\verb|ROI-Extraction-Agent|, \verb|Shape-Extraction-Agent|, \verb|Colour-Extraction-Agent|, \verb|Shape-Descriptor-Agent|, \verb|Style-Texture-Descriptor-Agent|, \verb|Colour-Descriptor-Agent|, \verb|Sketch-Generator-Agent|, \verb|Rendering-Generator-Agent|, \verb|Feedback-Agent|) with the correctly formatted inputs derived from the workflow."

    \item \textbf{Reminders:} "Ensure all generated concepts strictly adhere to the constraints and keywords mentioned in the original Problem Statement. Prioritize the features identified by the analysis and descriptor agents."
\end{itemize}
This multi-part prompt ensures that the system's complex, multi-step task execution is reliable, repeatable, and precisely aligned with the goals of the design process.

\subsection{Benefits of Agentic-AI Paradigm for designers}
\label{subsec:benefits_agenticAI_paradigm}
The shift from manual ideation to an Agentic-AI paradigm is not merely procedural but fundamentally creative. For seasoned designers, this agentic framework acts as a powerful generative partner, automating the laborious translation of abstract feature maps into a diverse array of tangible concept variants. This massively accelerates the divergent phase of ideation, enabling the exploration of a much wider design space in a fraction of the time. For novices, the pipeline provides crucial scaffolding by making the opaque "creative leap"—from inspiration to form—an explicit and transparent process. It offers concrete visual starting points, demystifying how abstract qualities captured by EUPHORIA can be embodied in a product form.

Moreover, the Agentic-AI paradigm fosters a new, continuous loop of human-AI co-creation. The designer's implicit preferences guide the AI agents, which in turn generate outputs for the designer to evaluate and curate. The designer's role evolves from a solitary generator of ideas to a strategic director of a generative process. Unlike the conventional ideation process, which is internal and untraceable, the agentic framework makes creative synthesis externalized, repeatable, and data-driven. RETINA, thus, does not replace the designer’s critical judgment or aesthetic sensibility but acts as a synthesis engine, amplifying their creative intent by translating the implicit language of attention into the explicit language of tangible design forms.

\begin{table*}[htbp]
\centering
\caption{Breakdown of the EUPHORIA-guided RETINA-Assisted Automated Form Design Process}
\label{tab:euphoria_retina_workflow_revised}
\begin{tabular}{|p{3.25cm}|p{1.0cm}|p{1.0cm}|p{5.0cm}|p{5.0cm}|}
\toprule
\textbf{Stage} & \textbf{Na\textsuperscript{a}} & \textbf{Pr\textsuperscript{b}} & \textbf{Inputs \& Outputs} & \textbf{Activity Involved} \\
\midrule
\textbf{Keyword Identification} & 
Ex & 
M & 
\textbf{Input:} User research data, product brief. \newline \textbf{Output:} A curated list of keywords defining style attributes. & 
The designer analyses research and defines the core stylistic attributes for the project. \\
\hline
\textbf{Moodboarding: \newline Collection and Selection} & 
Im & 
A & 
\textbf{Input:} Curated keywords to populate the initial image set via API. \newline \textbf{Output:} An attention-ranked set of images based on the designer's gaze data. & 
The designer immersively explores the VR moodspace; the EUPHORIA system performs the selection in the background based on their natural gaze patterns. \\
\hline
\textbf{Moodboarding: \newline Composition} & 
Im & 
A & 
\textbf{Input:} The attention-ranked set of images. \newline \textbf{Output:} An automatically generated virtual image space layout that organises and groups images based on attention rank. & 
The designer's role is passive; the EUPHORIA system computationally composes a structured layout of the most salient visual stimuli based on attention data. \\
\hline
\textbf{Moodboarding: \newline Reflection and Extraction} & 
Ex & 
A, G & 
\textbf{Input:} The organized layout of the virtual image space. \newline \textbf{Output:} Explicit feature maps, including an ROI Collage, an HED Edge Collage, a Dominant Colour Palette, and rich textual descriptors for texture, shape, colour, and style. & 
The designer actively reviews
and reflects on the set of images displayed on the composed layout. The EUPHORIA identifies the hotspots in each image and the RETINA system's agents extract the comprehensive feature maps. \\
\hline
\textbf{Concept Sketch Generation} & 
Ex & 
A & 
\textbf{Input:} Explicit feature maps, product name, and a high-level text prompt. \newline \textbf{Output:} A diverse set of AI-generated product concept sketches for review. & 
The RETINA system generates a diverse set of concept sketches; the designer then acts as a \textit{Curator}, reviewing and selecting the most promising options. \\
\hline
\textbf{Product Form Rendering} & 
Ex & 
A & 
\textbf{Input:} A selected concept sketch, the ROI Collage (for texture/style), and the extracted colour palette. \newline \textbf{Output:} A set of high-quality, photorealistic product renderings. & 
The RETINA system generates a set of high-quality renderings; the designer then curates these options, selecting the one that best embodies the final design intent. \\
\bottomrule
\end{tabular}
\flushleft
\footnotesize
\textsuperscript{a}\textbf{Nature:} Ex - Explicit; Im - Implicit. \\
\textsuperscript{b}\textbf{Process:} M - Manual; A - Automated; G - Guided.
\end{table*}


\greyline

\section{Research Questions and Methodology}
\label{sec:methodology}

The validation of the EUPHORIA system and its foundational premise—that an attention-driven workflow can systematize and enhance form design—necessitates a rigorous, multi-stage empirical investigation. A single study would be insufficient to untangle the complex interplay between a user's intrinsic preferences, the influence of external emotional cues, and the application of goal-directed creative strategies. Therefore, this research adopts a sequential, three-phase methodology. Each phase is designed to isolate and examine a specific component of the attention-based model, with the findings of each stage informing and providing the foundation for the next.

This phased approach allows for a systematic progression of inquiry:
\begin{itemize}
    \item \textbf{Phase 1} seeks to establish the baseline validity of using implicit attention as a proxy for personal preference in an unconstrained, natural exploratory context.
    \item \textbf{Phase 2} builds upon this baseline by introducing controlled variables (emotional stimuli) to investigate how pre-conditioning and priming a user's cognitive state influences their attentional patterns and leads to convergent behaviour.
    \item \textbf{Phase 3} integrates the findings from the first two phases into a complete, applied design task, comparing the EUPHORIA-guided RETINA-Assisted workflow against conventional methods to evaluate its practical efficacy, efficiency, and impact on creative output.
\end{itemize}

The central research question for each phase and a brief overview of the methodology employed to answer it are outlined below. The detailed experimental design, participant information, procedures, and specific metrics for each phase will be elaborated upon in the subsequent sections of this paper.

\subsection{Phase 1: Natural Exploration through Self-Attention}
\label{subsec:phase1_overview}
The central research question for this foundational phase is: \textbf{Does a user's implicit visual attention, as measured by gaze fixation duration, positively correlate with their explicit, self-reported preference for visual stimuli in the EUPHORIA environment?}

To address this question, an experimental study was designed where participants were invited to freely explore a large and diverse set of images within the immersive VR moodspace. The methodology involved capturing their implicit gaze data—specifically, the duration of their fixation on each image they chose to inspect—without any specific task or goal. This attentional data was then statistically compared with the explicit preference ratings that participants provided for the same images immediately after the VR session. This approach allows for a direct validation of the core relationship between implicit attention and explicit preference, which is the cornerstone of the EUPHORIA system.

\subsection{Phase 2: Conditioned Exploration through Stimuli-Attention}
\label{subsec:phase2_overview}
Building on the findings from the first phase, the second phase investigates the effect of goal-directed attention with the following research question: \textbf{How do external emotional and thematic stimuli, provided as pre-conditioning phrases, modulate a user's visual attention patterns and influence the convergence of image selection across different individuals?}

The methodology for this phase involved a controlled experiment where participants explored the same image set from Phase 1, but only after being primed with a specific emotional or thematic stimulus phrase (e.g., "Find something that feels calming"). By analyzing and comparing the visual selections and gaze patterns of different participants who were given the same prompt, this study aimed to measure the degree of convergence in their choices. This method was designed to demonstrate how the EUPHORIA system can be used to guide the creative process in a goal-oriented manner by framing the user's attentional focus.

\subsection{Phase 3: Goal-driven Exploration through Strategic-Attention}
\label{subsec:phase3_overview}
The final phase applies the validated principles to a real-world design scenario, asking the primary research question: \textbf{Can the EUPHORIA-guided RETINA-assisted product form design provide a more effective and efficient workflow compared to the conventional form design process?}

This question is addressed through a comparative study involving designers of varying experience levels who were tasked with solving a tangible form design problem. The methodology employed a Latin Square design to have participants execute the task using four distinct workflow paths: a fully conventional manual process, a hybrid EUPHORIA-assisted process, and two advanced RETINA AI-integrated pipelines. The effectiveness of each workflow was then systematically evaluated by comparing the time taken to complete the task and by subjecting the final design outputs to a formal assessment by a panel of independent design experts, who rated them on quality, creativity, and relevance to the design brief. The detailed workflow of EUPHORIA-guided RETINA-assisted form design is given in Figure~\ref{tab:euphoria_retina_workflow_revised}.

The central research question for each phase and a brief overview of the methodology employed to answer it are outlined below. The detailed experimental design, participant information, procedures, and specific metrics will be elaborated upon in the subsequent sections of this paper.


\greyline

\section{Phase 1 - Natural Exploration through Self-Attention}
\label{sec:phase1_study}

The primary objective of the first experimental phase was to empirically validate the foundational premise upon which the EUPHORIA system is built. Before the system could be deployed for more complex, goal-oriented design tasks, it was imperative to establish a baseline of evidence demonstrating that a user's implicit visual attention, as captured within the immersive moodspace, serves as a reliable and objective proxy for their explicit personal preference. This phase, therefore, focuses on examining the most fundamental mode of human-visual interaction: unconstrained, free-form exploration guided purely by the user's own internal curiosity, memories, and aesthetic inclinations. This process is termed \textit{Self-Attention}, as it reflects the attentional patterns that emerge when an individual's "self" is the sole director of their exploration, free from external tasks or prompts.

\subsection{Hypotheses}
\label{subsec:phase1_hypotheses}
To investigate the research question for this phase, the following specific hypotheses were formulated:

\begin{enumerate}
    \item \textbf{H1: Correlation Between Attention and Preference.}
    
    The amount of visual attention an individual dedicates to a particular image, quantified by the total fixation duration, will be directly and positively correlated with their subsequently reported preference rating for that image. It is posited that users will naturally gaze longer at images they find more appealing or interesting.
    \item \textbf{H2: Individual Variability in Preferences.} 
    
    In the absence of any external prompt or unifying goal, participants' preferences will be highly individualistic, driven by their unique personal experiences, memories, and tastes. This will manifest as minimal overlap in the specific images most preferred across different participants, highlighting the subjective nature of unguided aesthetic appreciation.
\end{enumerate}

\subsection{Study Design}
\label{subsec:phase1_design}
A within-subjects experimental design was employed to test the hypotheses. Each participant was exposed to the same set of visual stimuli within the EUPHORIA environment, and their implicit gaze behaviour and explicit preference ratings were recorded and analyzed.

\subsubsection{Participants and Ethical Considerations}
A total of 30 individuals (Mean Age = 27.4 years, SD = 3.2) participated in the study. To assess whether prior design training had a significant impact on natural visual exploration, the cohort was intentionally balanced, comprising 15 professional designers or design students and 15 non-designers from various professional backgrounds.

The study was conducted in accordance with the ethical guidelines stipulated by the institute's review board. All participants were provided with a detailed information sheet explaining the purpose of the study, the nature of the tasks involved, and the types of data being collected. Written informed consent was obtained from every participant prior to their involvement. Participants were assured of their anonymity, with all collected data being coded and handled securely to protect their privacy. They were also informed of their right to withdraw from the study at any point without penalty.

\subsubsection{Apparatus and Stimuli}
The experiment was conducted using the EUPHORIA application, a standalone software developed in the Unity 3D engine. The application was run on a Meta Quest Pro Virtual Reality headset, which was chosen for its integrated, high-fidelity eye-tracking capabilities.

The visual stimuli consisted of a carefully curated set of 150 diverse, real-world images. This set was compiled to cover a wide and eclectic range of subjects, including but not limited to natural landscapes, urban scenery, portraits, objects, abstract art, and fantasy scenes. The selection was intentionally broad to ensure the potential to evoke a wide spectrum of thoughts, emotions, and memories among the participants, thereby providing a rich dataset for analyzing personal preferences. All images were standardized in terms of resolution to ensure a consistent viewing experience. 

\subsubsection{Procedure}
Each participant was guided through the study individually in a controlled laboratory setting to minimize external distractions. The procedure was meticulously designed to foster a natural, unconstrained browsing experience, as follows:
\begin{enumerate}
    \item \textbf{Briefing and On-boarding:} Participants were first welcomed and briefed on the study's objectives. To make the task relatable and intuitive, the instructions were framed using the analogy of a visit to an art museum: \textit{"Imagine yourself walking through an art museum, where various paintings and artworks are displayed... as you glance at these artworks, you might pause and explore further if something catches your attention... Your task is to glance at the images and identify the ones that capture your attention."}
    \item \textbf{VR Immersion:} Participants were then comfortably seated and assisted in fitting the Meta Quest Pro headset. Upon launching the EUPHORIA application, they found themselves at the centre of the immersive moodspace, surrounded by the constellation of 150 floating images arranged in a large cylindrical layout. From their central vantage point, the images were clearly identifiable but lacked fine detail, encouraging active exploration. The 150 images were split into a 25x6 grid spanning over the visual field of 120 degrees of the cylindrical layout to minimize the head rotation by the participant, so as to minimize the variability in results due to additional factors.
    \item \textbf{Gaze-based Interaction:} A blue dot, corresponding to the participant's real-time point of gaze, served as the primary interaction cursor. To inspect an image more closely, the participant simply had to fixate their gaze on it. This action would trigger the image to smoothly animate towards them, enlarging to reveal its details. This interaction mechanism was designed to be effortless and intuitive, directly linking visual attention to interaction.
    \item \textbf{Free Exploration:} Participants were encouraged to explore any image that evoked thoughts or emotions, whether positive or negative, based entirely on their personal feelings and experiences. They were free to disengage from an image at any time by simply looking away, which would cause the image to return to its original position in the cylindrical layout.
    \item \textbf{Session Duration:} The exploration session for each participant was limited to a total of 10 minutes. It was explicitly communicated to them that they were not under any obligation to view all 150 images within this timeframe, reinforcing the unpressured, self-directed nature of the task.
\end{enumerate}

\subsubsection{Post-Study Data Collection}
Immediately following the conclusion of the 10-minute VR session, the headset was removed, and participants were guided to a computer. Here, they were presented with a grid displaying only the images they gave attention to. For each of these images, they were asked to provide an explicit preference rating on a 5-point Likert scale (1: Strongly Disliked, 2: Disliked, 3: Neutral, 4: Liked, 5: Strongly Liked). This two-step process ensured that the implicit gaze data was captured before being potentially influenced by the act of conscious rating.

\subsection{Outcome Metrics}
\label{subsec:phase1_metrics}
To evaluate the hypotheses, a combination of quantitative and qualitative metrics was employed:
\begin{itemize}
    \item \textbf{Fixation Duration:} The primary independent variable, measured as the total time in seconds that a participant's gaze remained fixated on each image they chose to interact with. This data was logged automatically by the EUPHORIA system.
    \item \textbf{Preference Rating:} The primary dependent variable, represented by the 5-point Likert scale rating provided by the participant for each interacted image.
    \item \textbf{Pearson Correlation Analysis:} A statistical test planned to measure the strength and direction of the linear relationship between the fixation duration and the corresponding preference rating for each image across all participants.
    \item \textbf{Disagreement Score (D):} A custom metric designed to quantify the variability in preference ratings for a given image across all participants who viewed it. This score serves as a quantitative measure of inter-participant agreement (or disagreement) on an image's appeal. It was calculated using the formula:
    \[ D = \frac{\sigma(R)}{2}*{(\frac{n}{T})} \]
    where $\sigma(R)$ is the standard deviation of the ratings for an image, $n$ is the number of participants who viewed that specific image, and $T$ is the total number of participants in the study (T=30). A higher score indicates greater disagreement.
    \item \textbf{Qualitative Eye-Tracking Data:} The system also generated visualizations of the raw eye-tracking data, including gaze heatmaps and fixation point plots for each image. This qualitative data was intended to provide deeper insights into which specific regions within an image drew the most visual attention.
\end{itemize}

\subsection{Results}
\label{subsec:phase1_results}
The data collected from the 30 participants in the Phase 1 study were analyzed to test the two primary hypotheses and to gain deeper insights into unguided, natural visual exploration behaviour. The findings are presented below, beginning with the results pertaining to the validation of each hypothesis.

\subsubsection{Validation of H1: Positive Correlation Between Attention and Preference}
The first hypothesis (H1) posited that a direct, positive correlation exists between the amount of visual attention paid to an image (fixation duration) and the user's explicit preference for it. The results from multiple analyses strongly support this hypothesis.

\begin{figure}[htbp]
    \centering
    \includegraphics[width=1\linewidth]{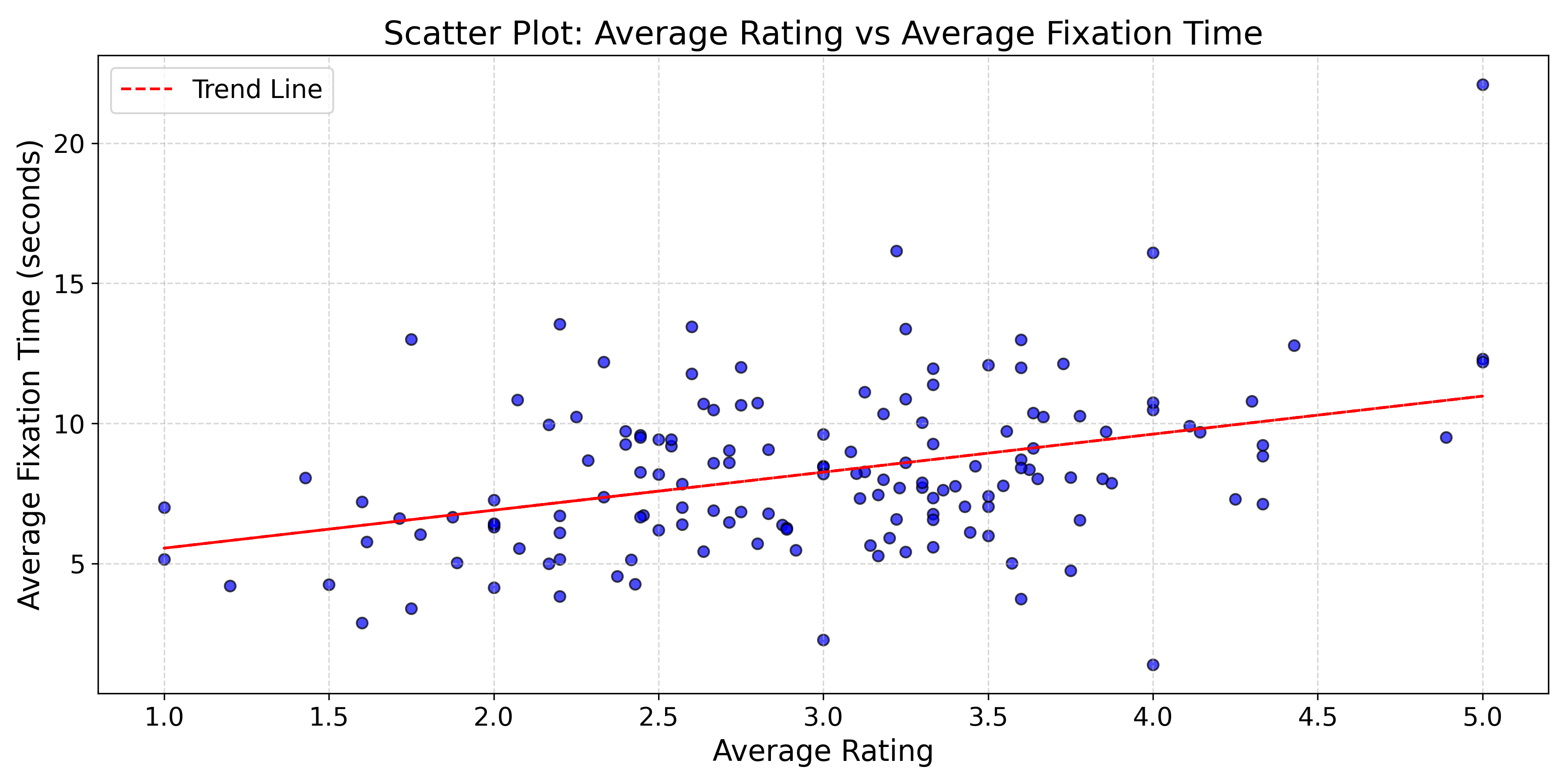}
    \caption{Scatter plot of average rating versus average fixation time for each viewed image, aggregated across all participants. Each blue dot represents an image. The red trend line illustrates a clear positive correlation, indicating that images with higher average preference ratings also tended to receive longer average fixation times, providing primary support for H1.}
    \label{fig:scatter_rating_time}
\end{figure}

An aggregate analysis of the data from all participants provides the clearest statistical evidence. As shown in the scatter plot in Figure~\ref{fig:scatter_rating_time}, there is a distinct positive trend between the average rating an image received and the average fixation time it garnered across all participants who viewed it. A Pearson correlation analysis performed on this aggregate data confirmed a statistically significant, moderate positive linear relationship, with a correlation coefficient of $r = 0.3819$ and a p-value $p < 0.0001$. This indicates that, on the whole, images that were rated more highly were also looked at for longer durations.

To further corroborate this finding, a group-based t-test was conducted by performing a median split on the average ratings. The median rating was found to be 3.0. Images rated above this median were placed in a "High Rating Group" (n=74), and those at or below were in a "Low Rating Group" (n=76). The analysis revealed that the mean average fixation time for the high rating group was 8.94 seconds, which was significantly higher than the 7.54 seconds for the low rating group. This difference was statistically significant (t-statistic = 3.0830, p-value = 0.0025), confirming that participants spent substantially more time on images they preferred.

\begin{figure}[htbp]
    \centering
    \includegraphics[width=1\linewidth]{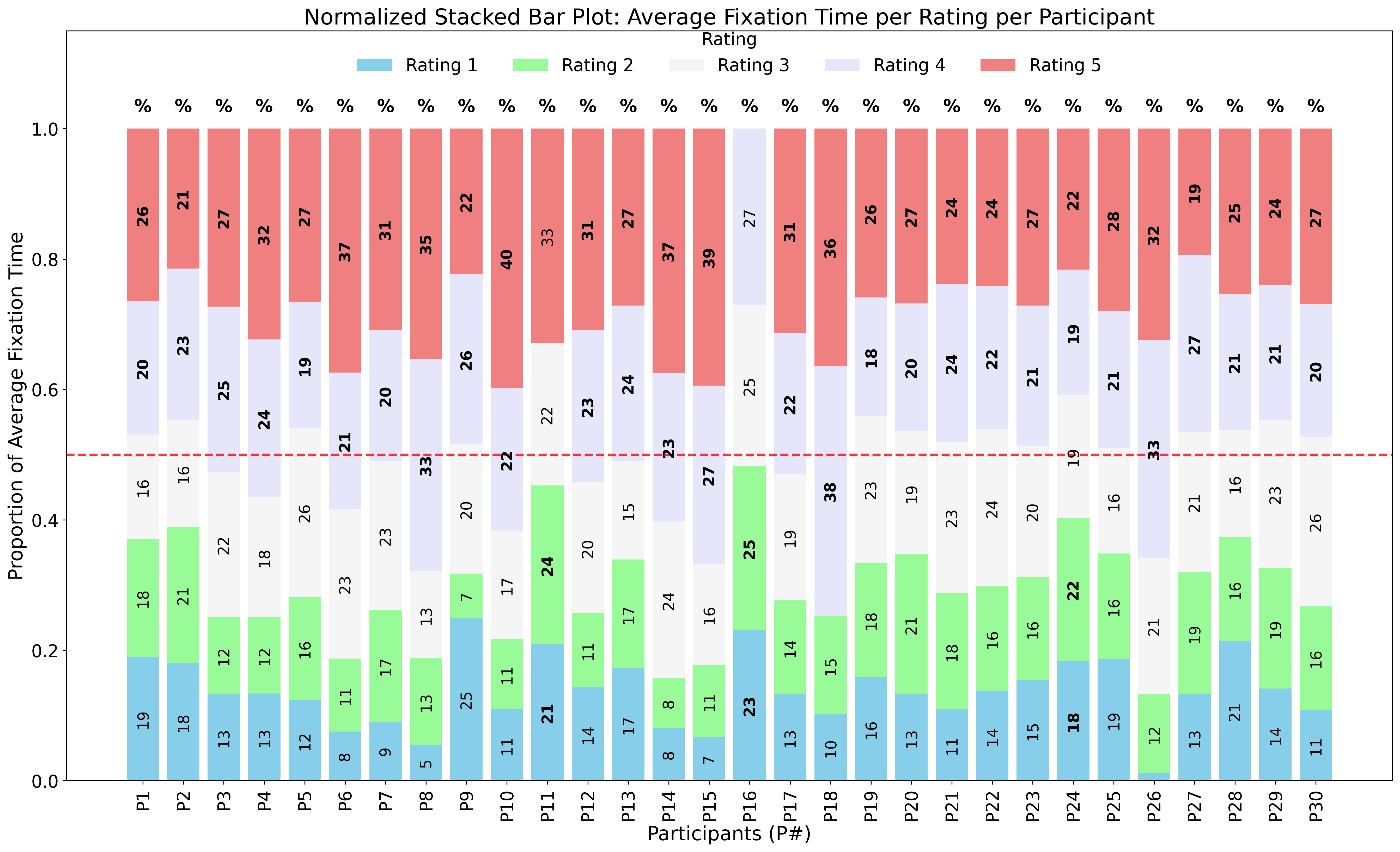}
    \caption{Normalized stacked bar plot of average fixation time per rating for each participant. Each bar represents a participant, and the coloured segments show the proportion of their total average fixation time spent on images they subsequently gave a specific rating to (1 to 5). The general trend of larger red (Rating 5) and purple (Rating 4) segments indicates that most participants dedicated more viewing time to images they preferred.}
    \label{fig:stacked_bar_time_rating}
\end{figure}

This trend is not merely an aggregate phenomenon but also holds true at the level of individual participants. The normalized stacked bar plot in Figure~\ref{fig:stacked_bar_time_rating} illustrates the proportion of each participant's average fixation time that was allocated to images of different rating categories. A consistent pattern is visible across the majority of participants: the segments of the bars representing higher ratings (Rating 4, shown in purple, and Rating 5, shown in red) are proportionally larger than those for lower ratings. This demonstrates that each individual, according to their own taste, dedicated more of their limited attentional resources to the images they personally found most appealing.

Based on the convergent evidence from the statistically significant aggregate correlation, the significant difference in means found via the t-test, and the clear visual trends at the individual participant level, \textbf{Hypothesis H1 is validated}.

\subsubsection{Validation of H2: High Individual Variability in Preferences}
The second hypothesis (H2) predicted that in an unguided exploratory context, preferences would be highly individualistic, resulting in minimal overlap in the images viewed and rated by different participants. The results from two different analyses—one of viewing behaviour and one of rating behaviour—confirm this high degree of subjectivity.

\begin{figure}[htbp]
    \centering
    \includegraphics[width=1\linewidth]{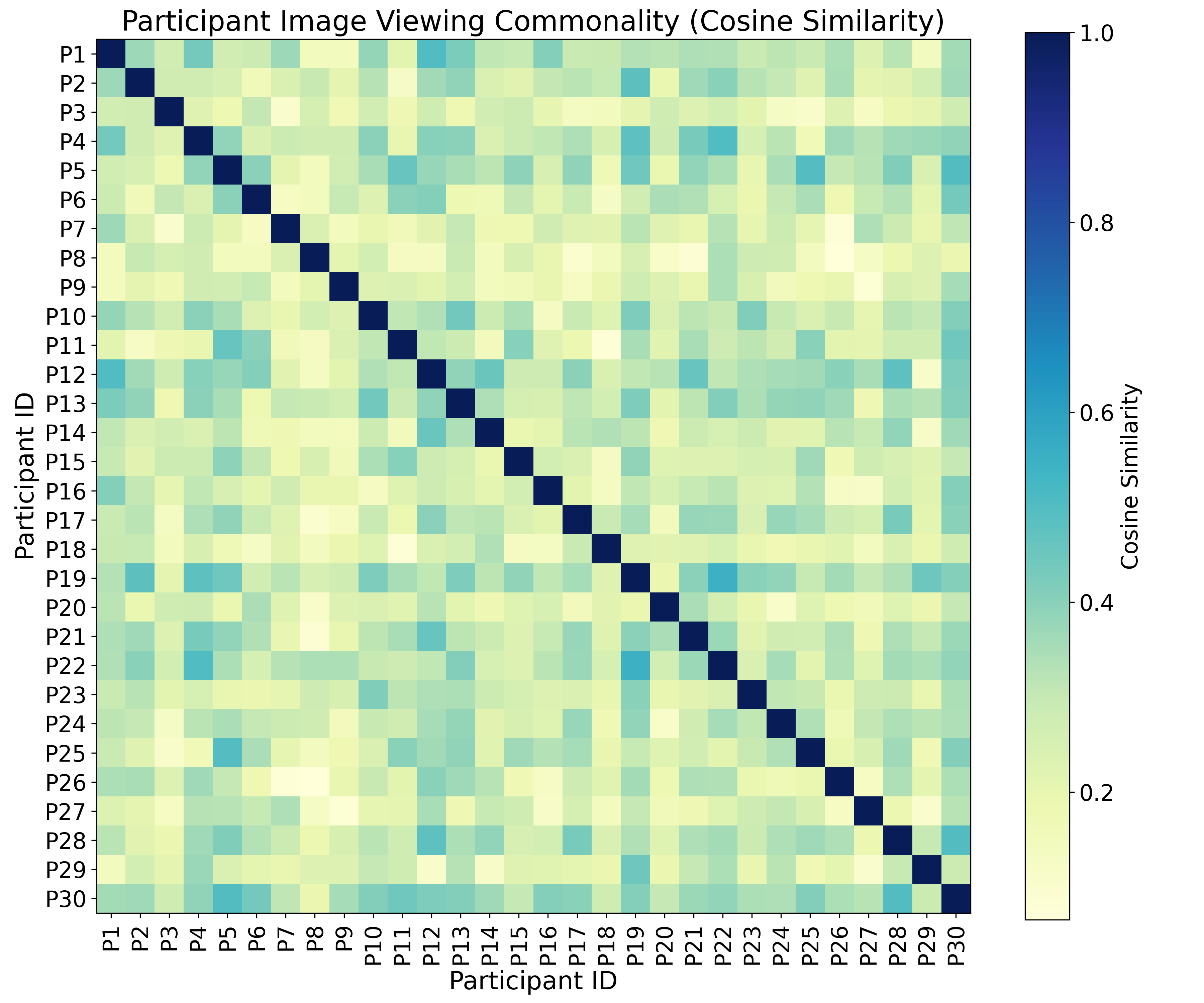}
    \caption{Heatmap of participant image viewing commonality, measured by cosine similarity. The dark diagonal represents perfect self-similarity, while the predominantly light-coloured off-diagonal cells indicate a low degree of overlap in the sets of images viewed by any two participants, providing strong evidence for H2.}
    \label{fig:cosine_similarity}
\end{figure}

The heatmap of the Participant Image Viewing Commonality, shown in Figure~\ref{fig:cosine_similarity}, provides a stark visualization of the diversity in exploration paths. This matrix displays the cosine similarity between the viewing vectors of every pair of participants. While the dark blue diagonal indicates perfect self-similarity (a score of 1.0), the off-diagonal cells are overwhelmingly dominated by light yellow and pale green hues. These colours correspond to very low cosine similarity scores, typically ranging from 0.1 to 0.4 and scattered all over the place without any visible grouping. This indicates that there was very little overlap in the specific set of images that any two participants chose to engage with from the 150 available options. Their "attentional journeys" through the moodspace were, therefore, highly unique.

\begin{figure}[htbp]
    \centering
    \includegraphics[width=\linewidth]{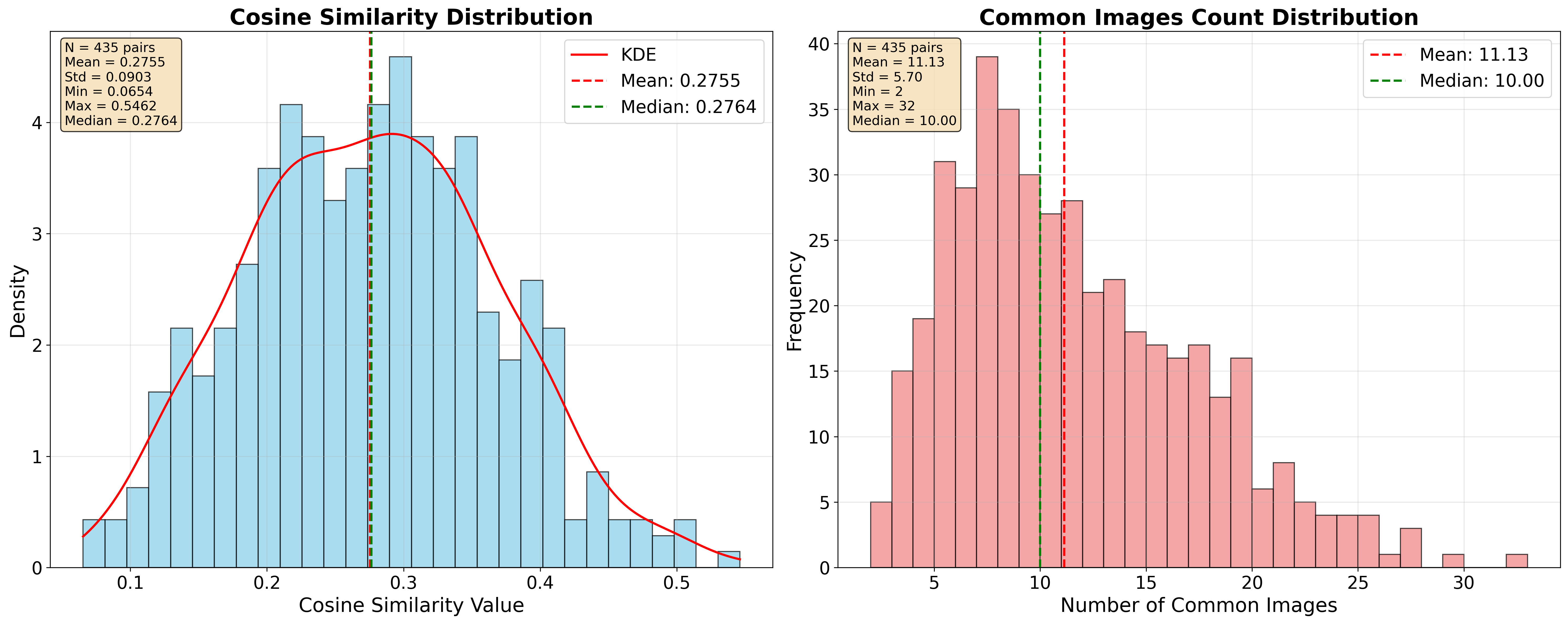}
    \caption{Statistical distributions of viewing commonality from Phase 1. The left panel shows the histogram of pairwise cosine similarity scores, which is heavily skewed towards low values (0.1-0.3). The right panel shows the histogram for the number of common images viewed between pairs of participants, which peaks at a low number (2-6 images). Both distributions quantitatively confirm the high degree of individual variability in exploration, supporting Hypothesis H2.}
    \label{fig:commonality_distribution_phase1}
\end{figure}

To further quantify this diversity in exploration, a statistical analysis of the viewing commonality was performed, as shown in the distributions in Figure~\ref{fig:commonality_distribution_phase1}. The histogram of pairwise cosine similarity scores (left panel) is heavily skewed towards the lower end of the scale, with the vast majority of scores falling between 0.1 and 0.3. This is corroborated by the distribution of the number of common images viewed between pairs of participants (right panel), which shows that most pairs had only between two and six images in common out of the 150 available options. Both distributions provide strong statistical evidence that the participants' exploration paths were largely independent and uncorrelated, quantitatively confirming the high degree of subjectivity and individualism predicted by Hypothesis H2.

\begin{figure}[htbp]
    \centering
    \includegraphics[width=1\linewidth]{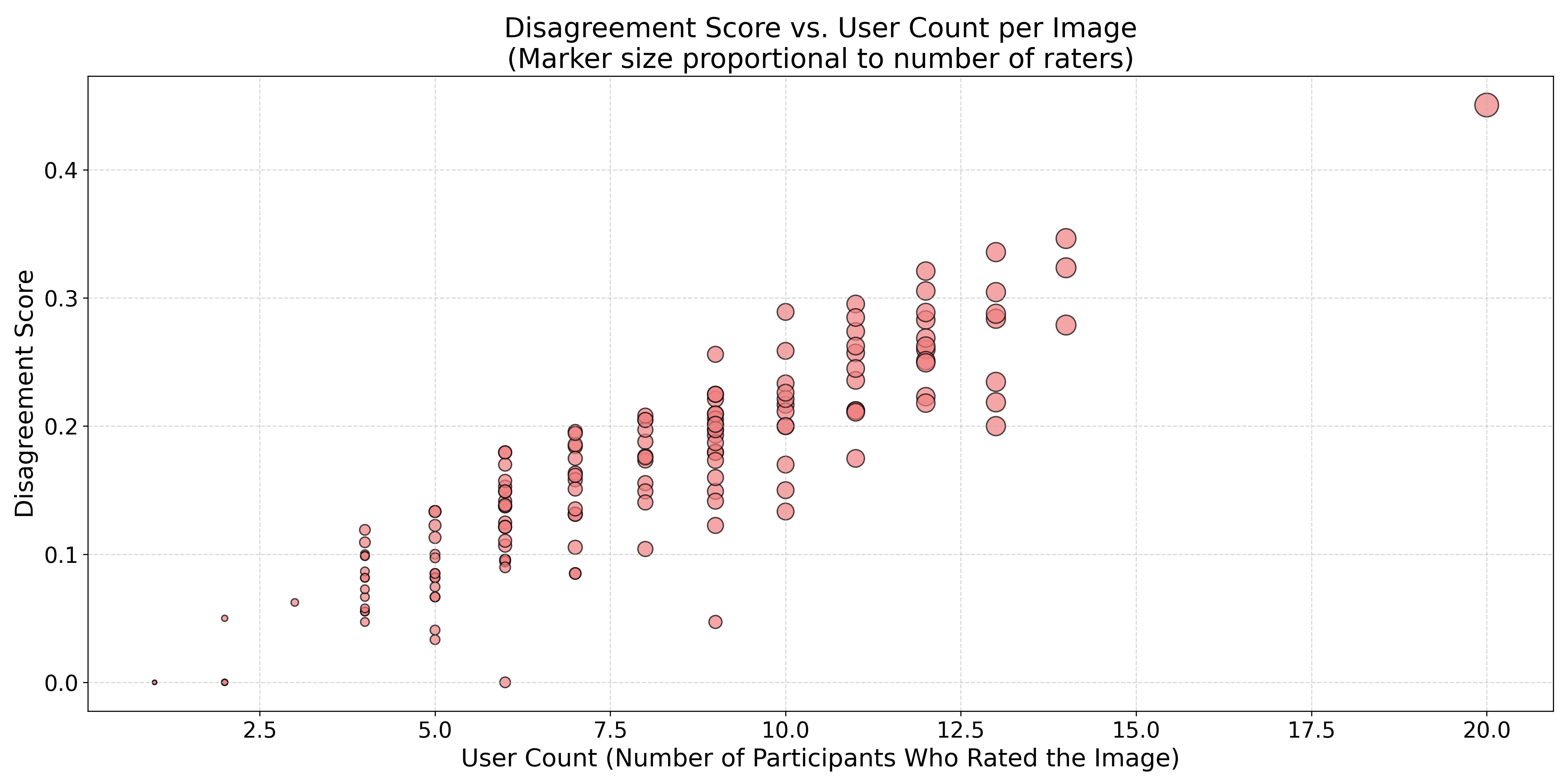}
    \caption{Scatter plot of the Disagreement Score versus the number of users who rated each image. The score, based on the standard deviation of ratings, quantifies the lack of consensus. The positive scores and their distribution demonstrate that preference was highly variable and individualistic, supporting H2.}
    \label{fig:disagreement_score}
\end{figure}

This variability was also evident in the explicit ratings. Figure~\ref{fig:disagreement_score} plots the Disagreement Score for each image against the number of participants who rated it. The score, which is based on the standard deviation of ratings, quantifies the lack of consensus; a score of 0 would indicate perfect agreement. The plot shows a wide distribution of positive scores, with an average score of 0.1355. This quantitatively demonstrates that even when different participants happened to view the same image, their opinions on its appeal varied significantly. This lack of consensus on ratings, combined with the low overlap in viewing patterns, strongly supports the hypothesis that unguided aesthetic preference is a highly individualistic phenomenon.

The low viewing commonality revealed by the cosine similarity matrix and the significant rating variance measured by the Disagreement Score both confirm the highly subjective nature of unguided exploration. Therefore, \textbf{Hypothesis H2 is validated}.

\subsubsection{Additional Insights into Participant Exploration Behaviour}
Beyond the primary hypotheses, the collected data revealed further insights into the diverse strategies participants employed during the free-form exploration task.

\begin{figure}[!ht]
    \centering
    
    \begin{subfigure}{\linewidth}
        \centering
        \includegraphics[width=1\linewidth]{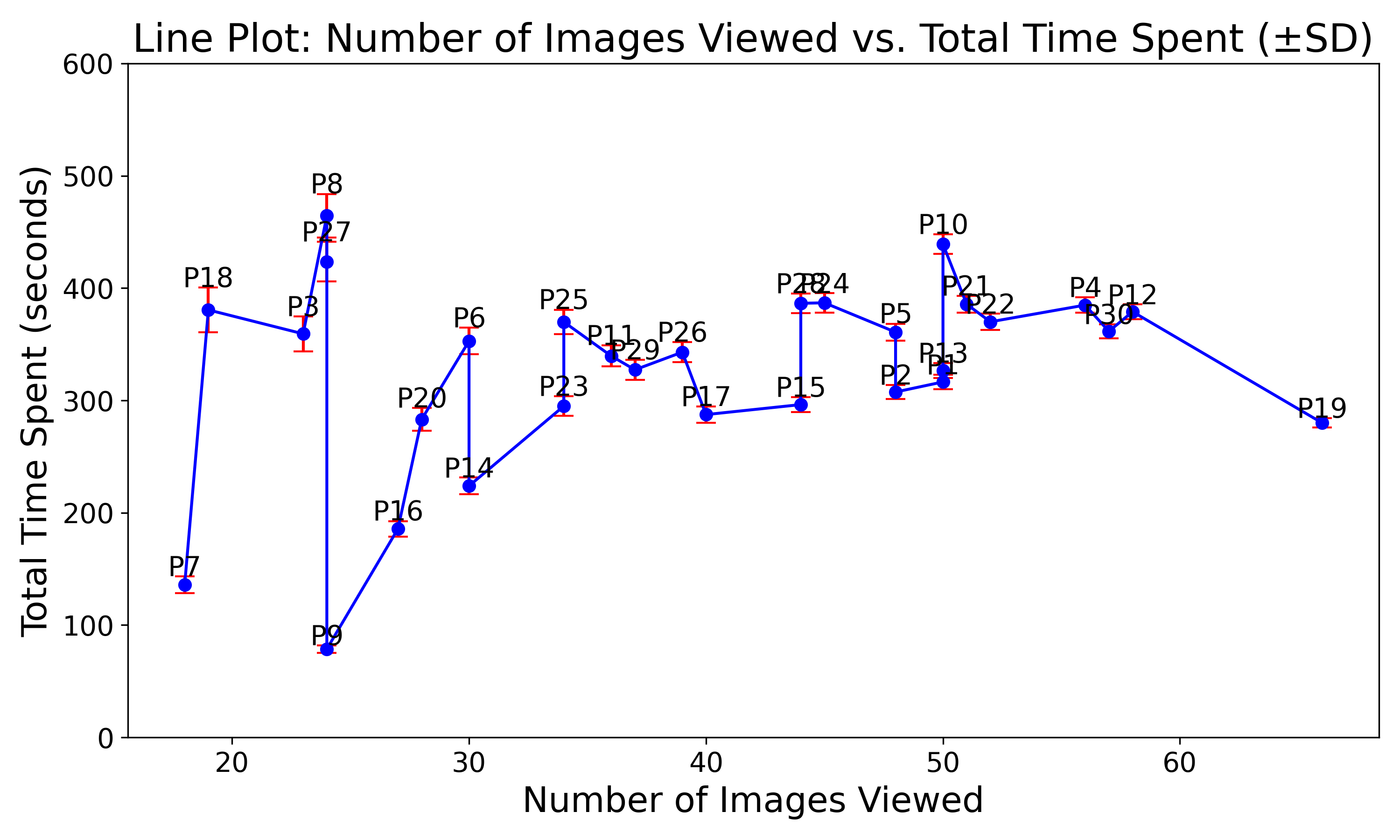}
        \caption{Line plot of the number of images viewed versus the total time spent per participant. Each data point represents a participant (P\#), with the error bars indicating the average time spent per image. The lack of a clear linear trend suggests the existence of different exploration strategies among participants.}
        \label{fig:viewcount_totaltime}
    \end{subfigure}
    
    \vfill
        
    \begin{subfigure}{\linewidth}
        \centering
        \includegraphics[width=1\linewidth]{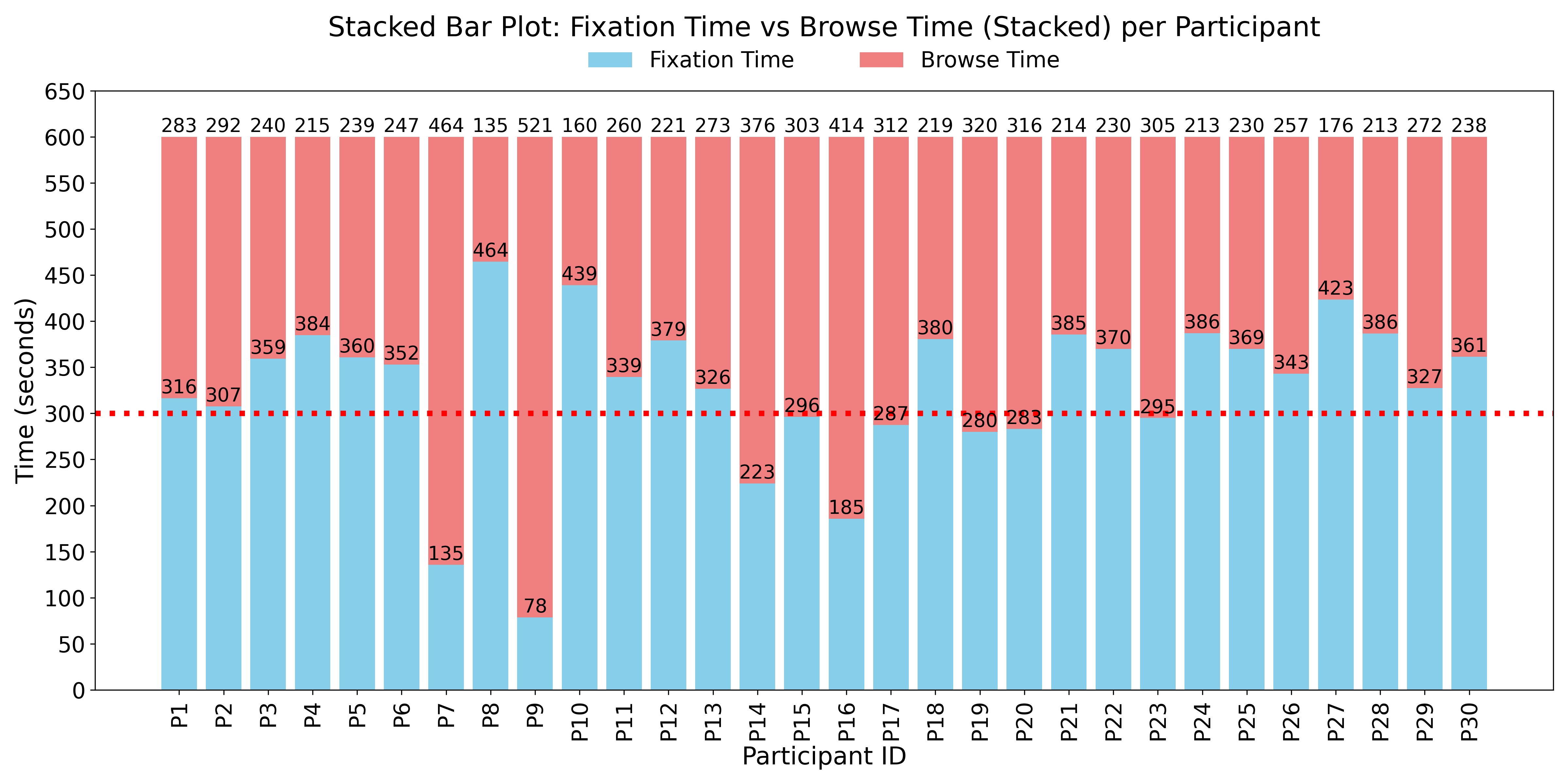}
        \caption{Breakdown of total session time into Fixation Time and Browse Time for each participant. This plot quantifies the different exploration strategies observed, from "deep divers" with high proportional fixation time to "broad scanners" with high proportional browse time.}
        \label{fig:stacked_browse_fixation}
    \end{subfigure}
    
    \caption{Comparison of participants’ exploration strategies. (a)~\ref{fig:viewcount_totaltime} shows the number of images viewed versus the total time per participant. (b)~\ref{fig:stacked_browse_fixation} shows the breakdown of Fixation vs Browse time for each participant.}
    \label{fig:combined_exploration_strategies}
\end{figure}

Analysis of the relationship between the number of images viewed and the total time spent in the session yielded a noteworthy result. As shown in Figure~\ref{fig:viewcount_totaltime}, there is no clear, consistent pattern across participants. A correlation analysis confirmed the absence of a statistically significant relationship (Pearson's r = 0.3064, p = 0.0996). This suggests that there was no single universal exploration strategy. The stacked bar plot in Figure~\ref{fig:stacked_browse_fixation} further quantifies this by breaking down each participant's session time into "Fixation Time" (time spent actively engaged with a selected image) and "Browse Time" (time spent scanning the overall canvas). It can be seen that some participants, such as P9, were "deep diverse", dedicating a large proportion of their session (~440 out of 521 seconds) to fixating on a curated set of images. In contrast, others, like P7, were "broad scanners," spending a much larger proportion of their time browsing and only briefly fixating on images.

\begin{figure*}[htbp]
    \centering
    \includegraphics[width=1\linewidth]{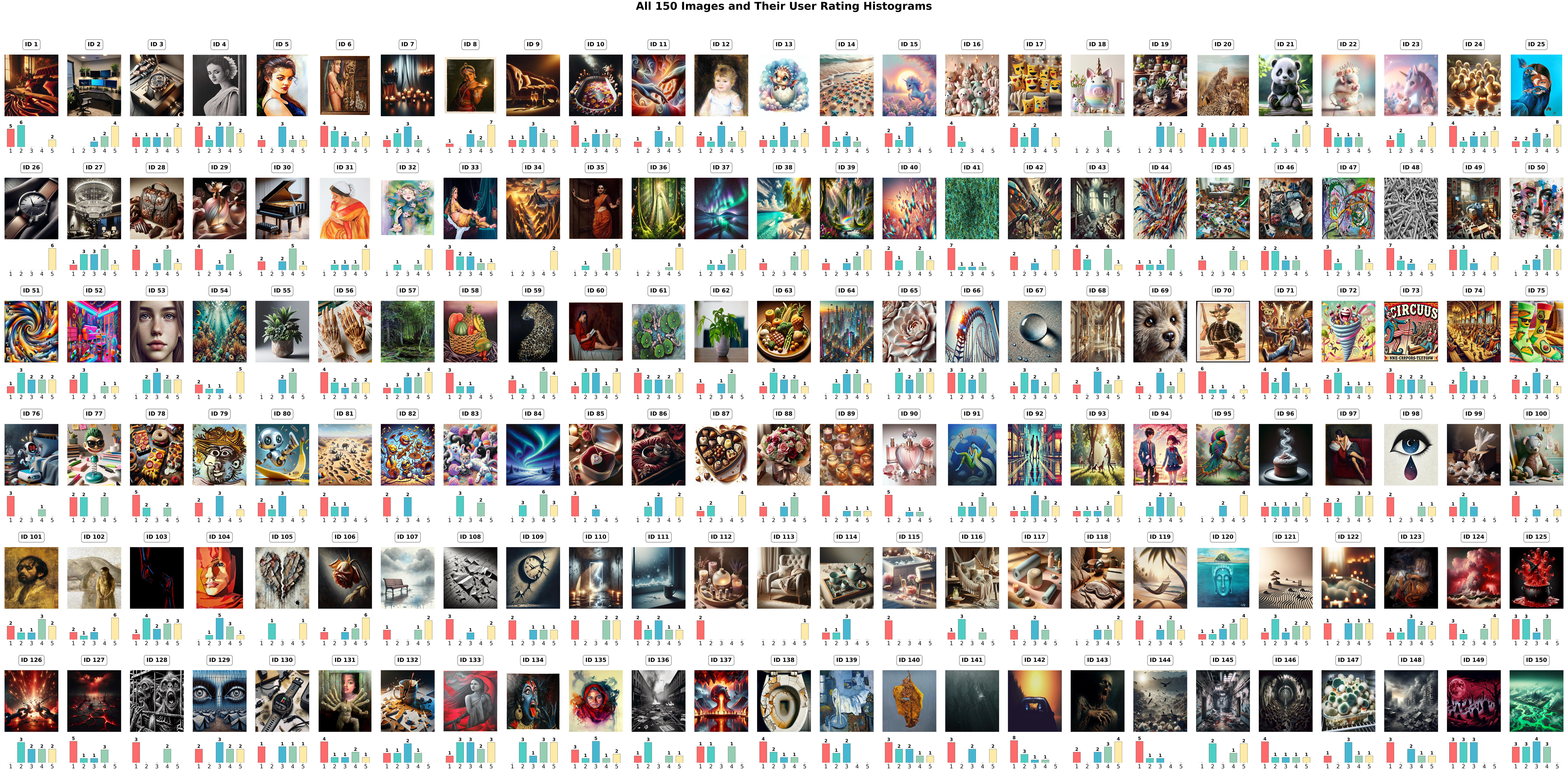}
    \caption{Grid overview of all 150 visual stimuli with their corresponding rating distributions. Below each image, a histogram displays the frequency of each rating (1-5) it received, providing a visual summary of its overall appeal and showing which images were universally liked, disliked, or divisive.}
    \label{fig:rating_histograms}
\end{figure*}

\begin{figure}[!ht]
    \centering
    
    \begin{subfigure}{\linewidth}
        \centering
        \includegraphics[width=1\linewidth]{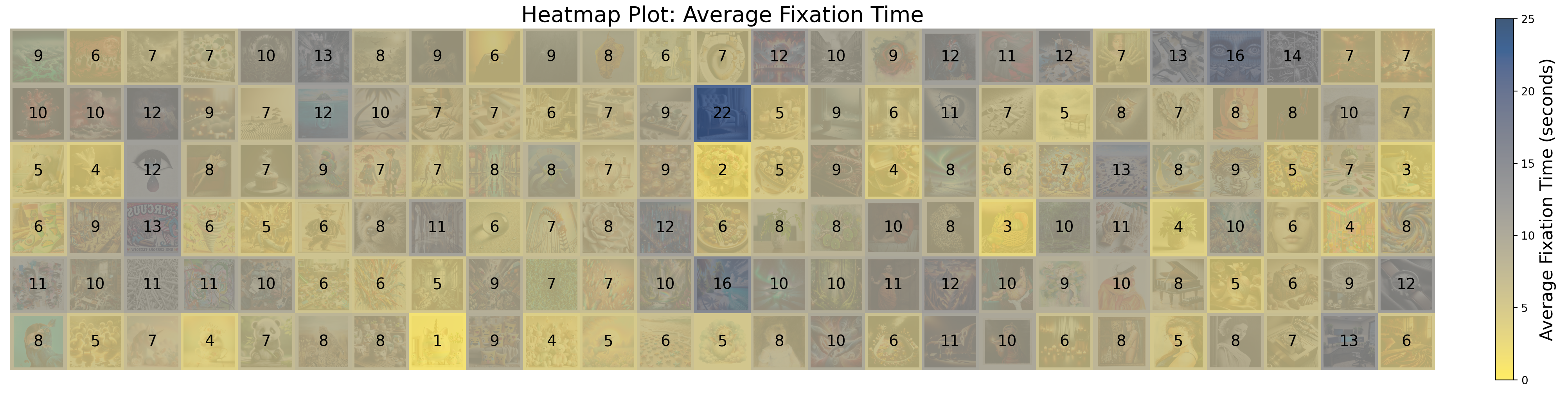}
        \caption{Heatmap of the average fixation time per viewer for each of the 150 images. This normalizes for the number of viewers and highlights images that were particularly engaging on a per-view basis.}
        \label{fig:heatmap_avg_fixation}
    \end{subfigure}
    
    \vfill  
    
    \begin{subfigure}{\linewidth}
        \centering
        \includegraphics[width=1\linewidth]{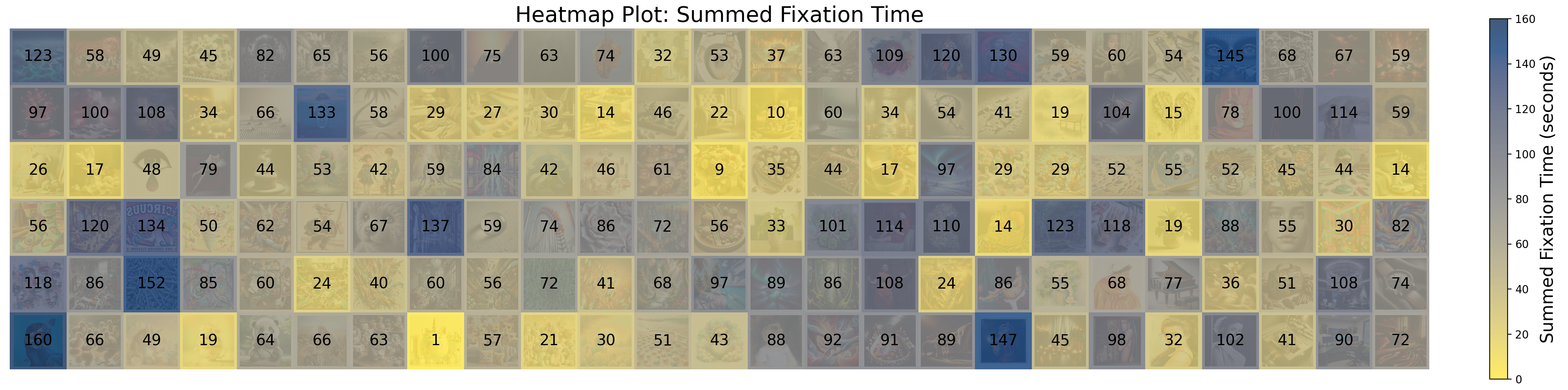}
        \caption{Heatmap of the total summed fixation time for each of the 150 images, arranged in their 25x6 grid layout. The color intensity indicates the cumulative amount of attention each image received from all 30 participants, highlighting the most collectively attention-grabbing images in the set.}
        \label{fig:heatmap_sum_fixation}
    \end{subfigure}
    
    \caption{Comparison of heatmaps for image engagement. (a)~\ref{fig:heatmap_avg_fixation} shows average fixation time per viewer. (b)~\ref{fig:heatmap_sum_fixation} shows summed fixation time across all participants.}
    \label{fig:combined_heatmaps}
\end{figure}

While preferences were largely individualistic, certain images did elicit more consistent responses. The grid plot in Figure~\ref{fig:rating_histograms} provides a complete overview of the rating distributions for all 150 images, revealing which images were more universally liked, disliked, or divisive. Similarly, the heatmaps in Figure~\ref{fig:heatmap_avg_fixation} and Figure~\ref{fig:heatmap_sum_fixation} show the distribution of average and summed fixation times across the 25x6 grid, indicating that certain images, regardless of individual taste, were more inherently "attention-grabbing" than others.

\begin{figure}[htbp]
        \centering
        \includegraphics[width=\linewidth]{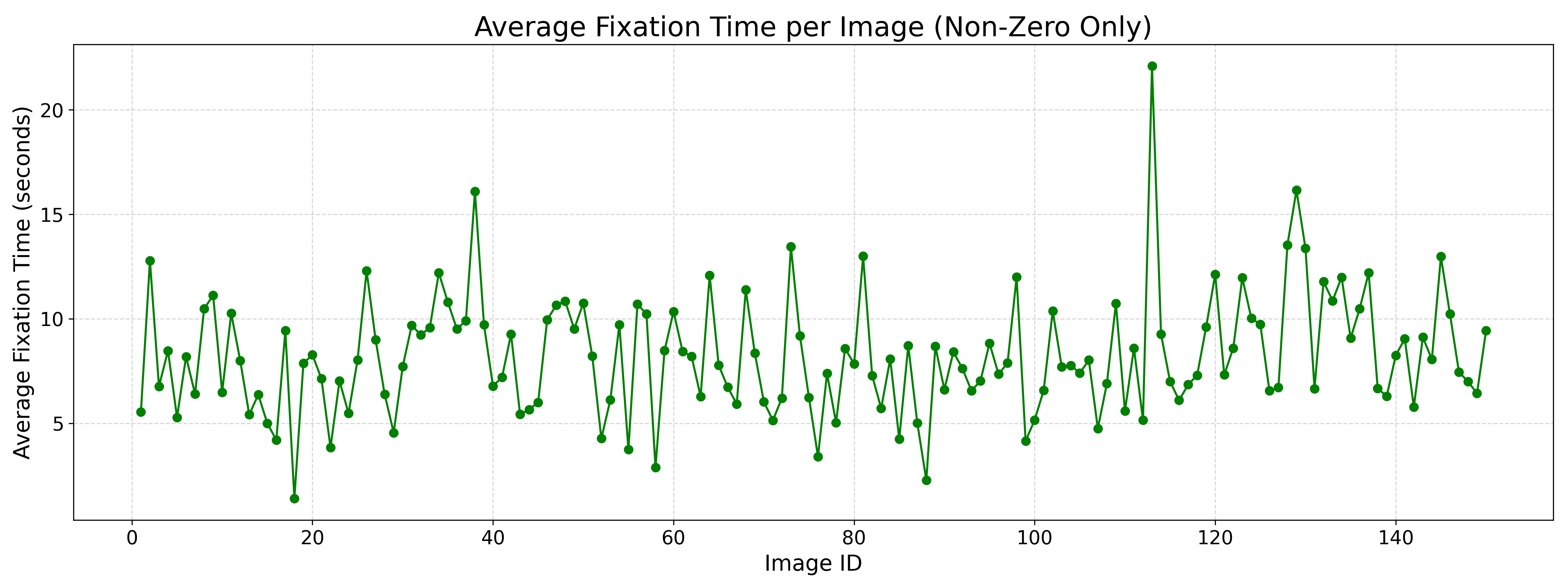}
        \caption{Line plot of the average fixation time in ascending order of image ID (non-zero only). This plot illustrates the variation in attention across images, ordered by image ID.}
        \label{fig:avg_time_per_image}
\end{figure}

\begin{figure}[htbp]
    \centering
    \includegraphics[width=1\linewidth]{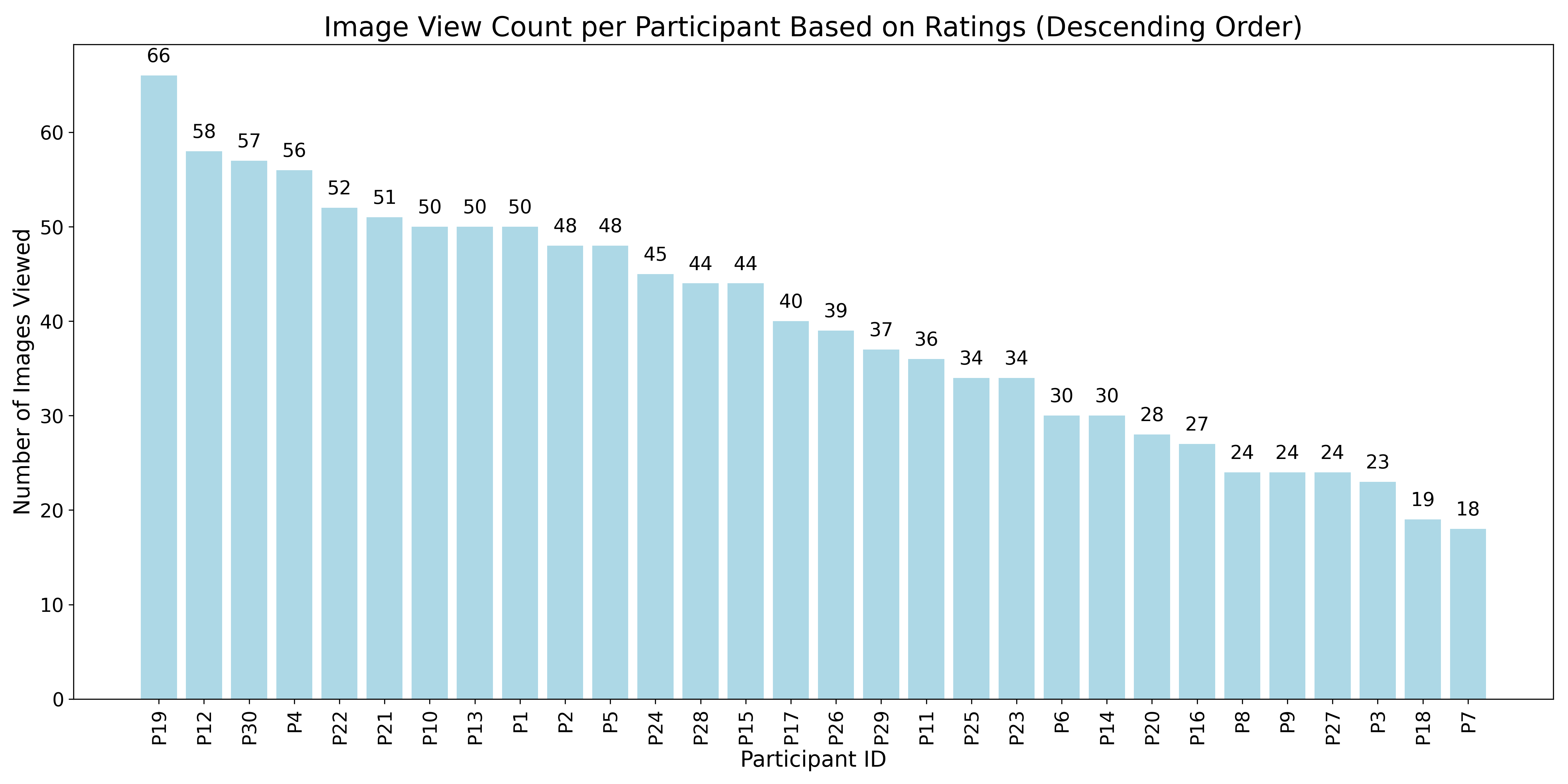}
    \caption{ Bar chart of the total number of unique images viewed by each participant within the 10-minute session. The plot highlights the significant variation in exploration breadth across the cohort, with some participants viewing as few as 18 images and others as many as 66.}
    \label{fig:image_count_per_participant}
\end{figure}

A more granular analysis of the dataset provides further detail on the stimulus set and individual behaviours. The line plots showing the average fixation time per image (Figure \ref{fig:avg_time_per_image} reveal the steady state, distinct performance. While overall preferences were highly subjective as per H2, this identify specific images that were clear outliers, either by being consistently highly-rated or by being exceptionally "attention-grabbing." Furthermore, the bar chart in Figure \ref{fig:image_count_per_participant} details the exact number of images each participant chose to view, quantifying the breadth of their exploration. This reinforces the finding of diverse user strategies, with the number of images viewed ranging from a narrow focus (18 images by P7) to a broad survey (66 images by P19) within a time span of 10 minutes, which is harder to do in conventional digital interfaces.

\begin{figure}[htbp]
\centering
\includegraphics[width=1\linewidth]{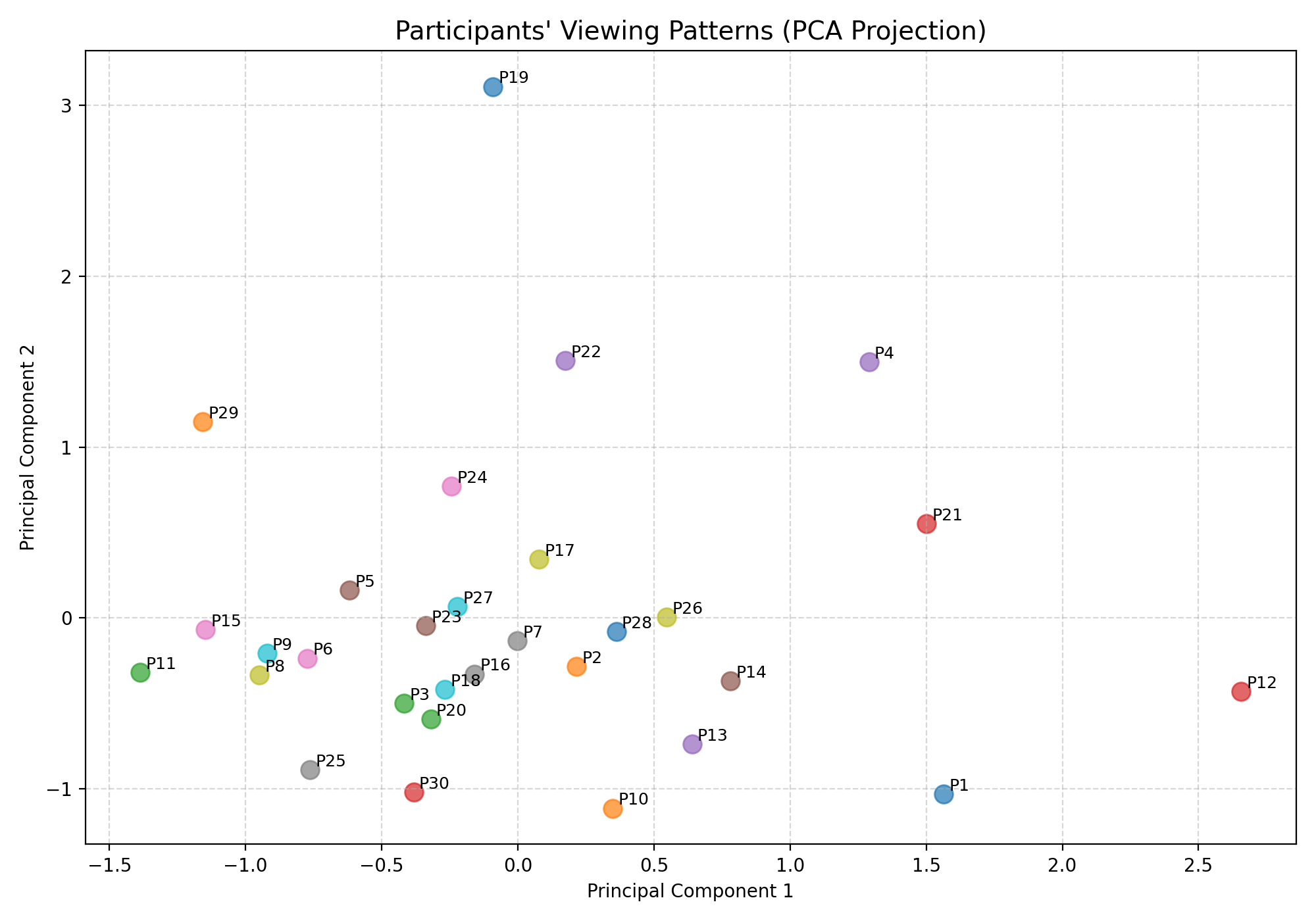}
\caption{Phase 2: Participants' Viewing Patterns (PCA Projection).}
\label{fig:pca_phase2_no_cluster}
\end{figure}

The scatter plot shown in Figure~\ref{fig:pca_phase2_no_cluster} illustrates the projection of 30 participants' image viewing sequences onto the first two principal components obtained through PCA. Each point represents a participant, labeled with their identification number. The distribution of the points reveals a broad dispersion, suggesting a considerable variability in the order in which participants viewed the images. The absence of distinct clusters indicates a lack of a common or preferred viewing order across the majority of the participants. Participant 26 is positioned somewhat distantly from the main group, indicating a notably different viewing pattern.

\subsection{Discussion}
\label{subsec:phase1_discussion}
The results of the Phase 1 study provide an empirical foundation for the theory upon which the EUPHORIA system is based. The validation of both primary hypotheses carries significant implications for the development of an attention-driven design methodology.

The confirmation of H1 is the cornerstone of this research. The statistically significant positive correlation between fixation duration and preference rating validates the use of eye-tracking as a legitimate and objective tool for implicitly capturing a user's aesthetic inclinations in an immersive virtual environment. This finding aligns with and extends prior research in psychology and neuroscience into the practical domain of design ideation. It confirms that the core mechanic of the EUPHORIA system—translating the simple, subconscious act of "looking longer" into a meaningful signal of preference—is sound. This validation provides the justification for automating the selection and composition stages of moodboarding based on this implicit data.

Simultaneously, the validation of H2 highlights the very problem that EUPHORIA aims to solve. The finding that unguided preferences are highly individualistic, with low overlap in both viewing and rating behaviour, underscores the inherent subjectivity and potential bias of the conventional design process, which often relies on the singular vision of one designer. It demonstrates a clear need for a tool that can capture and externalize these unique, personal preferences to create a truly user-centred or designer-specific moodboard. The EUPHORIA system is designed to do exactly that, by treating each user's attentional journey as a unique data signature from which to build a personalized inspirational palette.

Furthermore, the additional behavioural insights, particularly the identification of "deep diver" versus "broad scanner" exploration strategies, are of great interest. This suggests that users do not interact with large visual datasets in a uniform manner. This diversity in behaviour opens up avenues for future iterations of the system, which could potentially adapt its interface or feedback mechanisms in real-time based on a user's detected exploration style. For instance, the system might offer more diverse suggestions to a deep diver to encourage breadth, or highlight thematic clusters to a broad scanner to encourage depth.

In conclusion, the Phase 1 study successfully demonstrated that the foundational principles of the EUPHORIA system are valid. It proved that implicit attention is a reliable measure of preference and that unguided preferences are highly subjective. These findings provide the necessary confidence and empirical grounding to proceed to the subsequent phases of the research: to investigate whether this captured attention can be purposefully guided (Phase 2) and strategically applied to solve a concrete design problem (Phase 3).


\greyline

\section{Phase 2 - Conditioned Exploration through Stimuli-Attention}
\label{sec:phase2_study}

Having established in Phase 1 that a user's implicit visual attention is a reliable proxy for their intrinsic preference, the second phase of this research was designed to investigate a more directed form of attentional behaviour. The objective of Phase 2 was to explore how the natural, self-guided exploration process could be modulated and guided by pre-conditioning the user's mind with a specific external cue. This phase delves into the concept of \textit{Stimuli-Attention}, a goal-driven process where a participant's attentional focus is deliberately framed by an emotional or thematic prompt. The core aim was to determine if such priming could steer users towards a convergent set of visual choices, thereby demonstrating a method for systematically guiding the creative exploration process within the EUPHORIA environment.

\subsection{Hypotheses}
\label{subsec:phase2_hypotheses}
To investigate the research question for this phase, the following hypotheses were were formulated:

\begin{enumerate}
    \item \textbf{H3: Stimulus-Driven Convergence.} 
    
    When conditioned with a specific stimulus phrase, participants are likely to direct their attention towards and select a similar cluster of images that they perceive as being thematically or emotionally aligned with that stimulus. This would result in a high degree of inter-participant agreement for a given stimulus.
    
    \item \textbf{H4: Shift in Preference and Behaviour.} 
    
    A participant's image selections under a conditioned state will differ significantly from their unconstrained, personal preferences. Furthermore, the pre-conditioning will make the selection process more efficient, as the cognitive filter provided by the stimulus will reduce the time needed to identify relevant images.
\end{enumerate}

\subsection{Study Design}
\label{subsec:phase2_design}
A between-subjects experimental design was employed, where different groups of participants were exposed to different emotional stimuli. This approach allowed for a clean assessment of each stimulus's effect on visual selection without cross-contamination.

\subsubsection{Participants and Ethical Considerations}
A new cohort of 30 participants (Mean Age = 28.2 years, S.D. = 3.5) was recruited for this phase to ensure that their responses were not influenced by prior exposure to the experimental setup in Phase 1. The participants were divided into six groups of five, with each group being assigned one of the six stimulus phrases.

As with the previous phase, the study adhered strictly to the ethical guidelines of the institution. All participants were fully briefed on the nature of the task and provided written informed consent before the study commenced. Anonymity and data confidentiality were guaranteed, and participants were reminded of their right to withdraw at any time.

\subsubsection{Apparatus and Stimuli}
The experimental apparatus remained consistent with Phase 1, utilizing the EUPHORIA application running on the Meta Quest Pro VR headset. The visual stimuli also remained the same—the curated set of 150 diverse images—to ensure that any observed differences in behaviour between Phase 1 and Phase 2 could be attributed to the experimental manipulation rather than the image set itself.

The key experimental variable introduced in this phase was the set of six distinct stimulus phrases based on four (anger, happiness, surprise, disgust) of the six basic emotions \cite{ekman2013emotion}. Two remaining emotions, namely sadness and fear, were excluded to avoid the lingering unpleasant mood of the participants. These phrases were carefully crafted to represent a spectrum of emotional states and were categorized as follows:
\begin{itemize}
    \item \textbf{Positive Emotion (Ph 1, 2):} "Ph1 - Wow, this is so beautiful", "Ph2 - Love at first sight"
    \item \textbf{Neutral Emotion (Ph 3, 4):} "Ph3 - This feels so calming and relaxing", "Ph4 - Ah so cuteee"
    \item \textbf{Negative Emotion (Ph 5, 6):} "Ph5 - Ugh what a horrible sight", "Ph6 - Boiling over with rage"
\end{itemize}

\subsubsection{Procedure}
The core interaction mechanics within the EUPHORIA moodspace were identical to those in Phase 1 to maintain procedural consistency. The critical difference lay in the pre-session briefing:
\begin{enumerate}
    \item \textbf{Priming with Stimulus:} Before entering the VR environment, each participant was presented with one of the six stimulus phrases assigned to their group. They were instructed to read the phrase, reflect on its meaning, think about a situation in their life where they could relate the phrase to, and keep it at the forefront of their mind throughout the exploration session. This task lasted for 10 to 15 minutes.
    \item \textbf{Conditioned Search:} Their task was explicitly defined as a conditioned search. They were asked to explore the moodspace as before, except that now their mind is conditioned with phrases.
    \item \textbf{VR Exploration:} Participants then entered the EUPHORIA moodspace and proceeded to explore the 150 images using the same gaze-based interaction as in Phase 1. However, we believe their exploration was now guided by the internal cognitive filter established by the stimulus prompt.
    \item \textbf{Session Duration:} The session duration remained at 10 minutes, providing an equivalent timeframe for exploration as in the previous phase.
\end{enumerate}

\subsubsection{Post-Study Data Collection}
To gain deeper insights into the cognitive processes behind the conditioned selections, the post-study data collection was made more comprehensive. For each image they interacted with, participants were asked a series of questions designed to probe their reasoning:
\begin{enumerate}
    \item \textit{"How much do you like the image?"} (A 5-point Likert scale preference rating).
    \item \textit{"What did you like in the image - Colour, Shape, Size, Orientation, Texture?"} (A multiple-choice question to identify salient visual features).
    \item \textit{"What was your thought when you saw the image?"} (An open-ended question to capture their qualitative cognitive and affective response).
\end{enumerate}

\subsection{Outcome Metrics}
\label{subsec:phase2_metrics}
The analysis in Phase 2 was designed to measure both individual behaviour and group convergence. The following metrics were used:
\begin{itemize}
    \item \textbf{Implicit and Explicit Data:} Fixation duration and preference ratings were collected as in Phase 1 to analyze attentional behaviour under conditioned settings.
    \item \textbf{Pearson Correlation and Disagreement Score:} These statistical measures were again employed to analyze the relationship between attention and preference, and the level of agreement within each stimulus group.
    \item \textbf{Image Overlap Analysis:} A primary metric for this phase, this involved calculating the degree of overlap (i.e., the number of commonly selected images) among the five participants within each stimulus group. This directly measured the convergence predicted by H1.
    \item \textbf{Stimulus-Specific Heatmaps:} Gaze heatmaps were generated and aggregated for each stimulus group. This allowed for a visual analysis of whether participants exposed to the same stimulus also exhibited similar gaze patterns on the images they selected.
    \item \textbf{Qualitative Thematic Analysis:} The open-ended responses from participants were subjected to a multi-stage computational thematic analysis to quantitatively analyze their semantic content. First, each participant's textual 'thought' was converted into a high-dimensional vector using a sentence-transformer embedding model (Qwen/Qwen3-Embedding-8B). Subsequently, the Uniform Manifold Approximation and Projection (UMAP) algorithm was employed to reduce the dimensionality of these embeddings, projecting them into a 2D space while preserving their semantic relationships. This enabled the generation of scatter plots to visualize the thematic clustering of thoughts. Concurrently, responses were mapped to a predefined lexicon of emotion-related keywords to create distribution plots showing the frequency of specific emotional themes for each stimulus phrase. This combined approach provided both a spatial map of semantic meaning (UMAP) and a quantitative breakdown of emotional themes.
\end{itemize}

\subsection{Results}
\label{subsec:phase2_results}
The analysis of data from the Phase 2 study focused on evaluating the two primary hypotheses: whether pre-conditioning participants with stimuli would lead to convergent selections (H3) and whether it would induce a significant shift in exploratory behaviour compared to the unguided exploration in Phase 1 (H4).

\subsubsection{Validation of H3: Stimulus-Driven Convergence}
The first hypothesis (H3) predicted that participants, when primed with the same stimulus phrase, would select a similar cluster of images, resulting in high inter-participant agreement. The results provide compelling evidence supporting this hypothesis, demonstrating a strong convergence in both viewing patterns and explicit ratings.

\begin{figure}[htbp]
    \centering
    \includegraphics[width=1\linewidth]{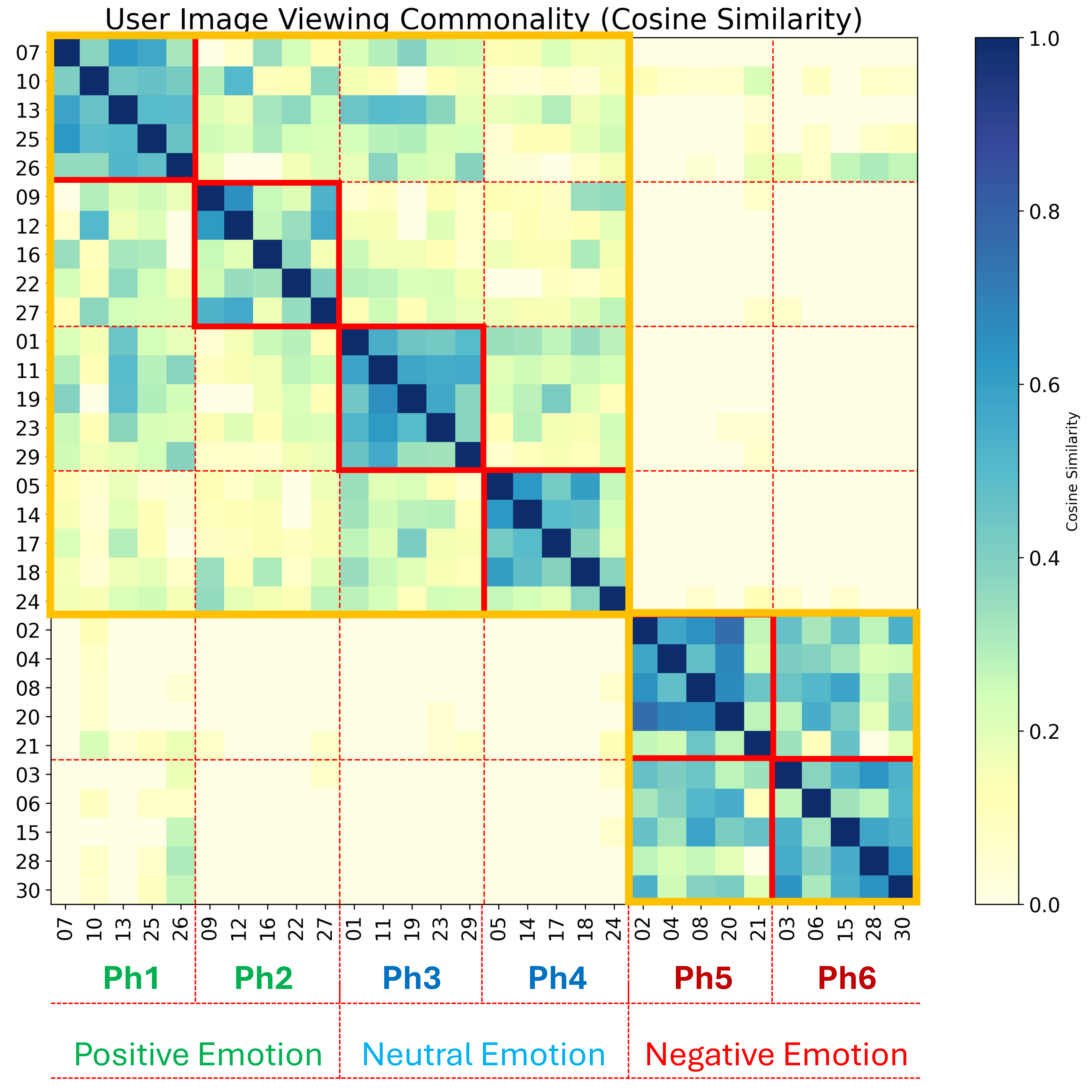}
    \caption{Heatmap of participant image viewing commonality for Phase 2, measured by cosine similarity. Participants are grouped by their assigned stimulus phrase (Ph1-Ph6). The distinct squares of high similarity (dark blue/green) along the block diagonal indicate that participants within the same stimulus group selected a highly similar set of images, providing strong visual evidence for H1 (Stimulus-Driven Convergence).}
    \label{fig:cosine_similarity_phase2}
\end{figure}

The most direct visual evidence of this convergence is presented in the User Image Viewing Commonality heatmap in Figure \ref{fig:cosine_similarity_phase2}. This 30x30 matrix, which plots the cosine similarity of viewing patterns between participants, is organized by stimulus group (Ph1-Ph6). A stark contrast to the highly dispersed heatmap from Phase 1 is immediately apparent. Distinct squares of high similarity (dark blue and green hues) are visible along the block diagonal, corresponding to the 5x5 blocks for each stimulus group. Conversely, the regions between different stimulus groups remain light yellow, indicating low similarity. This block-diagonal pattern visually confirms that participants within the same stimulus group explored and selected a remarkably similar set of images, while their selections differed significantly from those in other groups. The external prompt successfully channelled their attentional journeys, causing them to converge.

\begin{figure}[htbp]
    \centering
    \includegraphics[width=\linewidth]{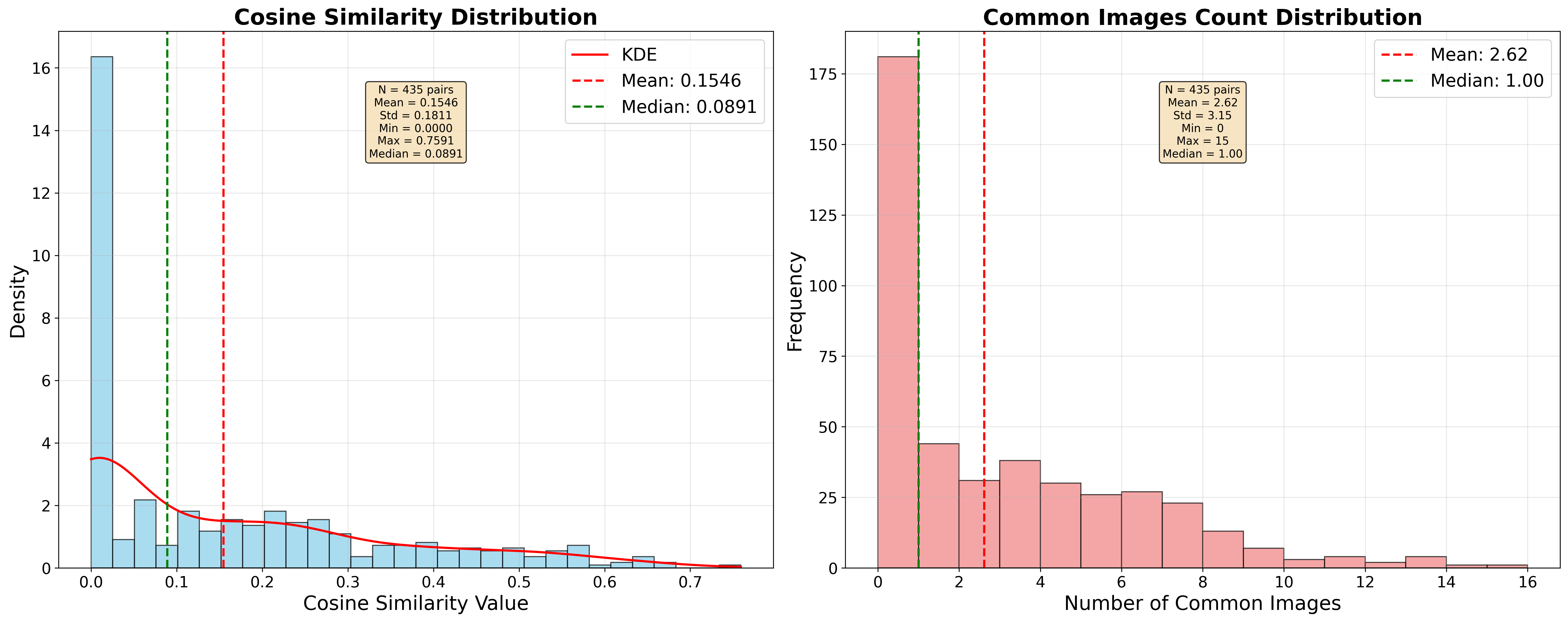}    
    \caption{Statistical distributions of viewing commonality from Phase 2. The left panel shows the Cosine Similarity Distribution, indicating that most participant pairs had low similarity in their viewing patterns (median = 0.0891), with the distribution being heavily right-skewed. The right panel displays the Common Images Count Distribution, which is also strongly right-skewed, showing that the majority of pairs viewed very few images in common (median = 1.00). Together, these plots suggest a low overall viewing commonality among participants during this phase.}
    \label{fig:commonality_distribution_phase2}
\end{figure}

This figure~\ref{fig:commonality_distribution_phase2} presents two statistical distributions that measure the viewing commonality among the participant pairs during Phase 2 of a study, assessing the effect of preconditioning on attention. The left panel, "Cosine Similarity Distribution," shows that the similarity of viewing patterns between pairs is generally low. The distribution is heavily skewed to the right, with a median similarity value of just 0.0891, indicating that over half of the pairs had very little overlap in their visual attention. Similarly, the right panel, "Common Images Count Distribution," is also strongly right-skewed. It reveals that most pairs viewed few images in common, with a median of only 1.00 common image. The mean for both metrics is pulled higher than the median by a long tail of a few participant pairs who did exhibit higher commonality.

\begin{figure}[htbp]
    \centering
    \includegraphics[width=1\linewidth]{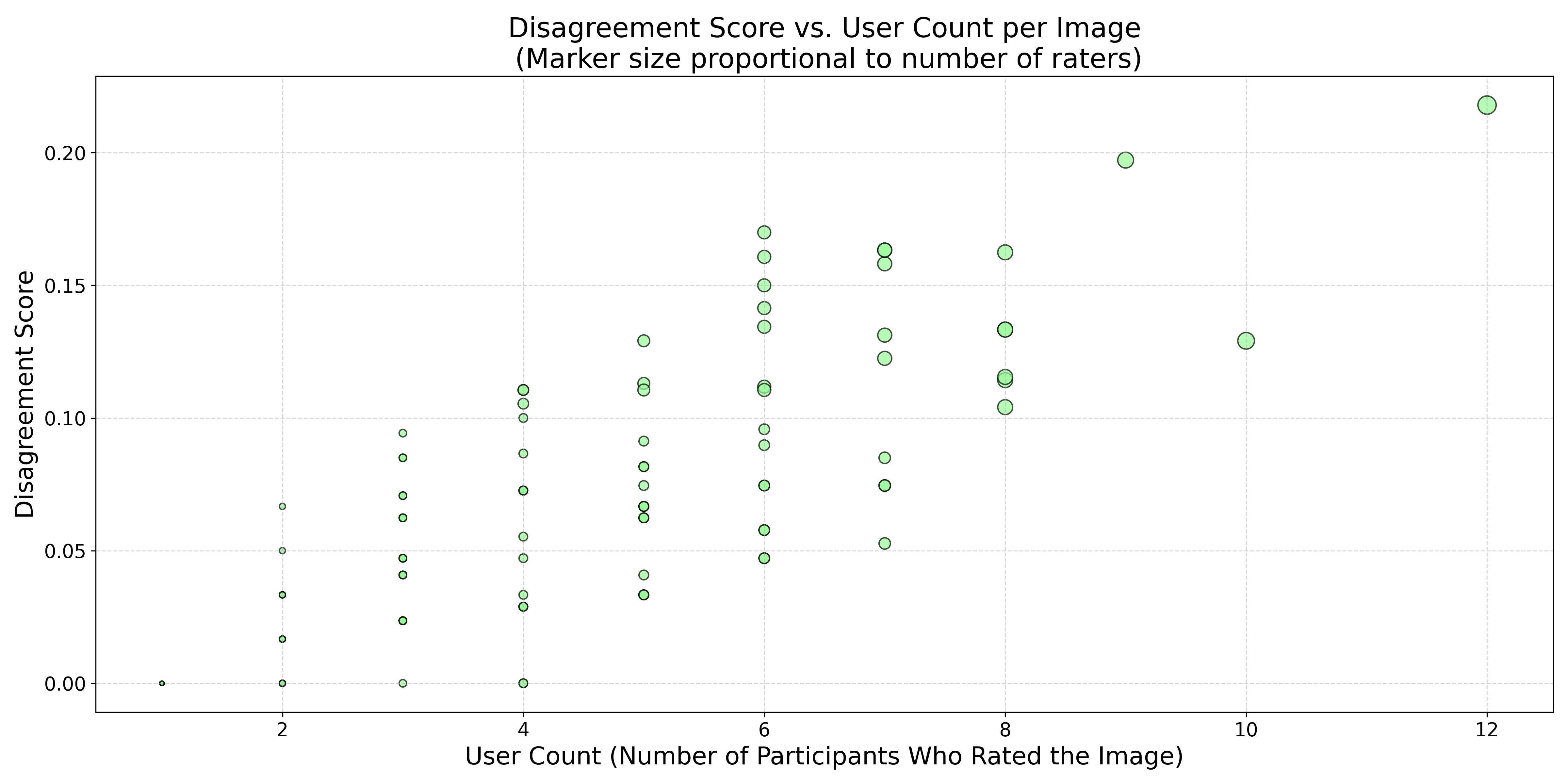}
    \caption{Scatter plot of the Disagreement Score versus the number of users who rated each image in Phase 2. The consistently low scores (average of 0.0490) signify a high degree of consensus in ratings, quantitatively confirming that the stimulus prompts led to convergent emotional and thematic interpretations among participants.}
    \label{fig:disagreement_score_phase2}
\end{figure}

This convergence is also quantitatively reflected in the participants' ratings. The Disagreement Score, which measures the variance in ratings for a given image, was substantially lower than in Phase 1. As shown in Figure \ref{fig:disagreement_score_phase2}, the scores are tightly clustered near the bottom of the plot, with a weighted average disagreement score of just 0.0490. This indicates a very high level of consensus; when primed with a specific emotion, participants not only viewed the same images but also tended to agree on their relevance and rating.

The evidence from the high viewing commonality within stimulus groups and the low disagreement in ratings across the cohort demonstrates a powerful convergence effect. Therefore, \textbf{Hypothesis H3 is validated.}

\subsubsection{Validation of H4: Shift in Preference and Behaviour}
The second hypothesis (H4) posited that a conditioned, goal-directed task would induce a significant shift in participant behaviour and make the selection process more efficient. The analysis reveals a fundamental change in the role of attention and the emergence of a more systematic exploration strategy.

\begin{figure}[htbp]
    \centering
    \includegraphics[width=1\linewidth]{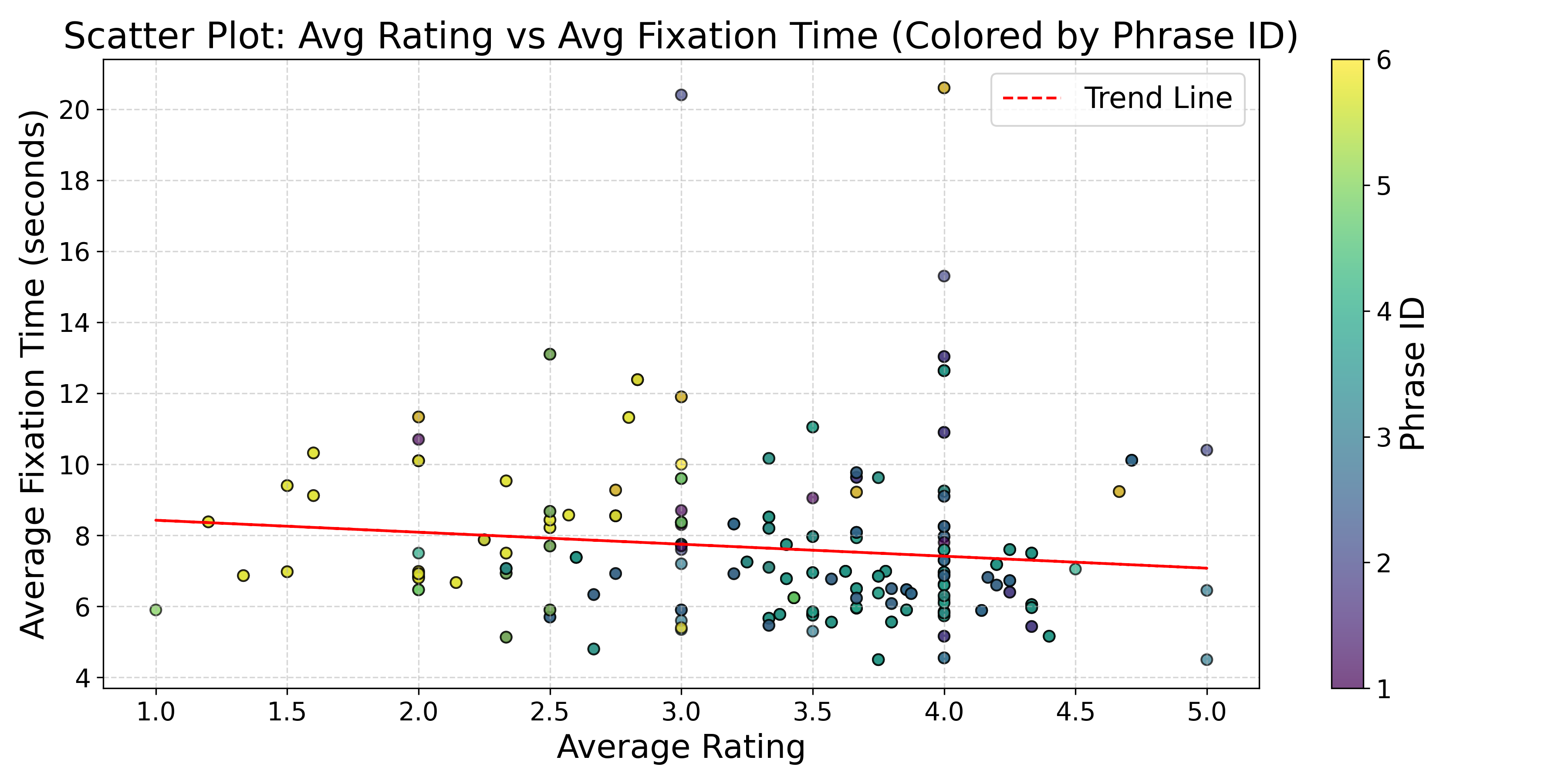}
    \caption{Scatter plot of average rating versus average fixation time for each viewed image in Phase 2. The nearly flat trend line illustrates the disappearance of the strong positive correlation observed in Phase 1, indicating a fundamental shift in attentional behaviour during a goal-directed task and supporting H2.}
    \label{fig:scatter_rating_time_phase2}
\end{figure}

A critical finding is the decoupling of the attention-preference link that was established in Phase 1. The scatter plot in Figure \ref{fig:scatter_rating_time_phase2} shows a nearly flat, slightly negative trend line between the average rating of an image and the average fixation time it received. A Pearson correlation analysis confirmed that there was no statistically significant relationship (r = -0.0915, p = 0.2786). This is a stark reversal from Phase 1 and indicates a significant behavioural shift. In this conditioned search, fixation time no longer serves as a proxy for "liking" but rather for "cognitive effort in assessing relevance." Participants spent only as much time as was necessary to determine if an image matched the given stimulus, meaning an obvious, high-rated match could be processed very quickly.

\begin{figure}[htbp]
    \centering
    \includegraphics[width=1\linewidth]{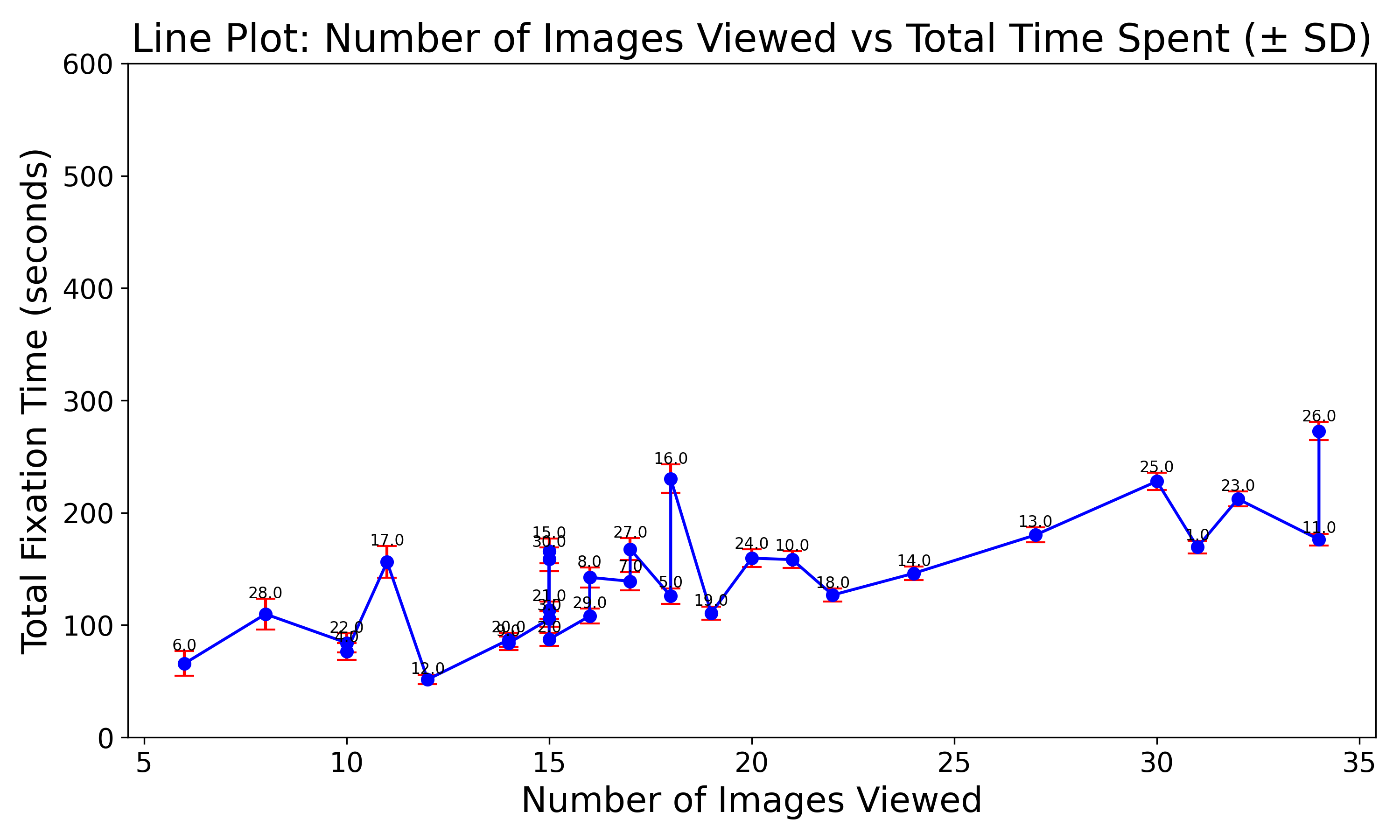}
    \caption{Line plot of the number of images viewed versus the total fixation time per participant in Phase 2. The strong, statistically significant positive correlation (r=0.7559, p<0.0001) reveals the emergence of a systematic exploration strategy, a key behavioural shift from Phase 1 that supports H2.}
    \label{fig:viewcount_totaltime_phase2}
\end{figure}

Furthermore, the exploration process became more systematic and efficient. Unlike in Phase 1, a strong, statistically significant positive correlation was found between the number of images a participant viewed and the total time they spent in the session, as shown in Figure \ref{fig:viewcount_totaltime_phase2} (Pearson's r = 0.7559, p < 0.0001). This emergence of a predictable pattern suggests that a more structured search strategy replaced the random exploration of Phase 1. The efficiency of this process is also reflected in the average fixation times. For instance, the mean fixation time for highly-rated images in this phase was 7.50 seconds, which is noticeably lower than the 8.94 seconds observed for the equivalent group in Phase 1, indicating faster decision-making.

The disappearance of the attention-preference correlation, the emergence of a predictable exploration strategy, and the increased efficiency in decision-making all confirm a significant behavioural change. Therefore, \textbf{Hypothesis H4 is validated.}

\subsubsection{Additional Insights into Conditioned Exploration Behaviour}
\begin{figure}[htbp]
    \centering
    \includegraphics[width=1\linewidth]{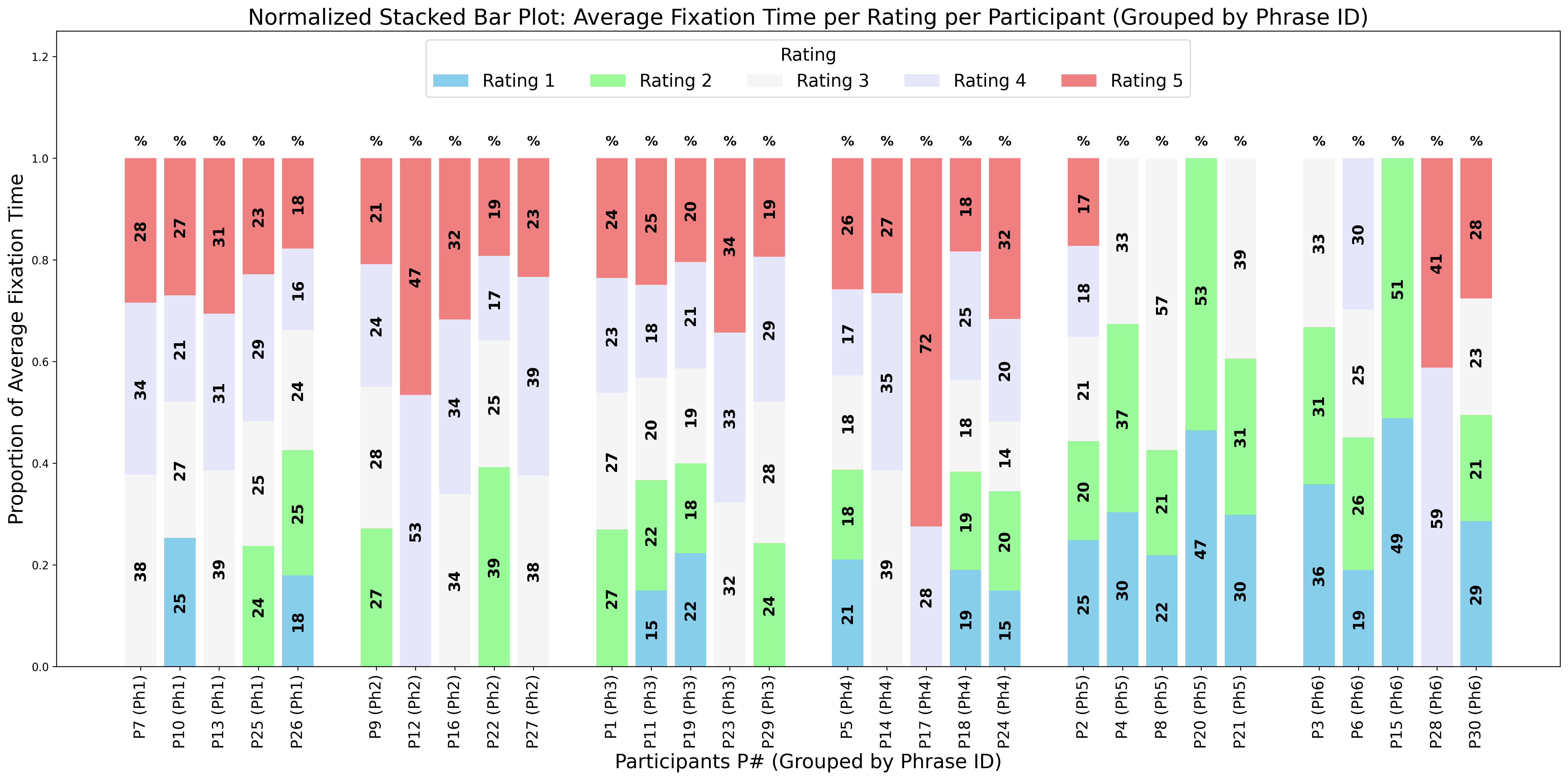}
    \caption{Normalized stacked bar plot of average fixation time per rating for each participant, grouped by stimulus phrase. The plot clearly shows how rating distributions were framed by the stimuli, with positive prompts (Ph1-Ph4) yielding high ratings (purple/red) and negative prompts (Ph5-Ph6) yielding low ratings (blue/green).}
    \label{fig:stacked_bar_time_rating_phase2}
\end{figure}

The data also offers deeper insights into how the different emotional stimuli framed the participants' perceptions and actions. The normalized stacked bar plot in Figure \ref{fig:stacked_bar_time_rating_phase2} reveals how effectively the prompts influenced ratings. For positive and neutral stimuli (Ph1-Ph4), the bars are dominated by high ratings (purple and red). In stark contrast, for negative stimuli (Ph5-Ph6), the bars are almost entirely composed of low ratings (blue and green), confirming that the stimuli successfully guided participants' affective judgments.

\begin{figure}[htbp]
    \centering
    \includegraphics[width=1\linewidth]{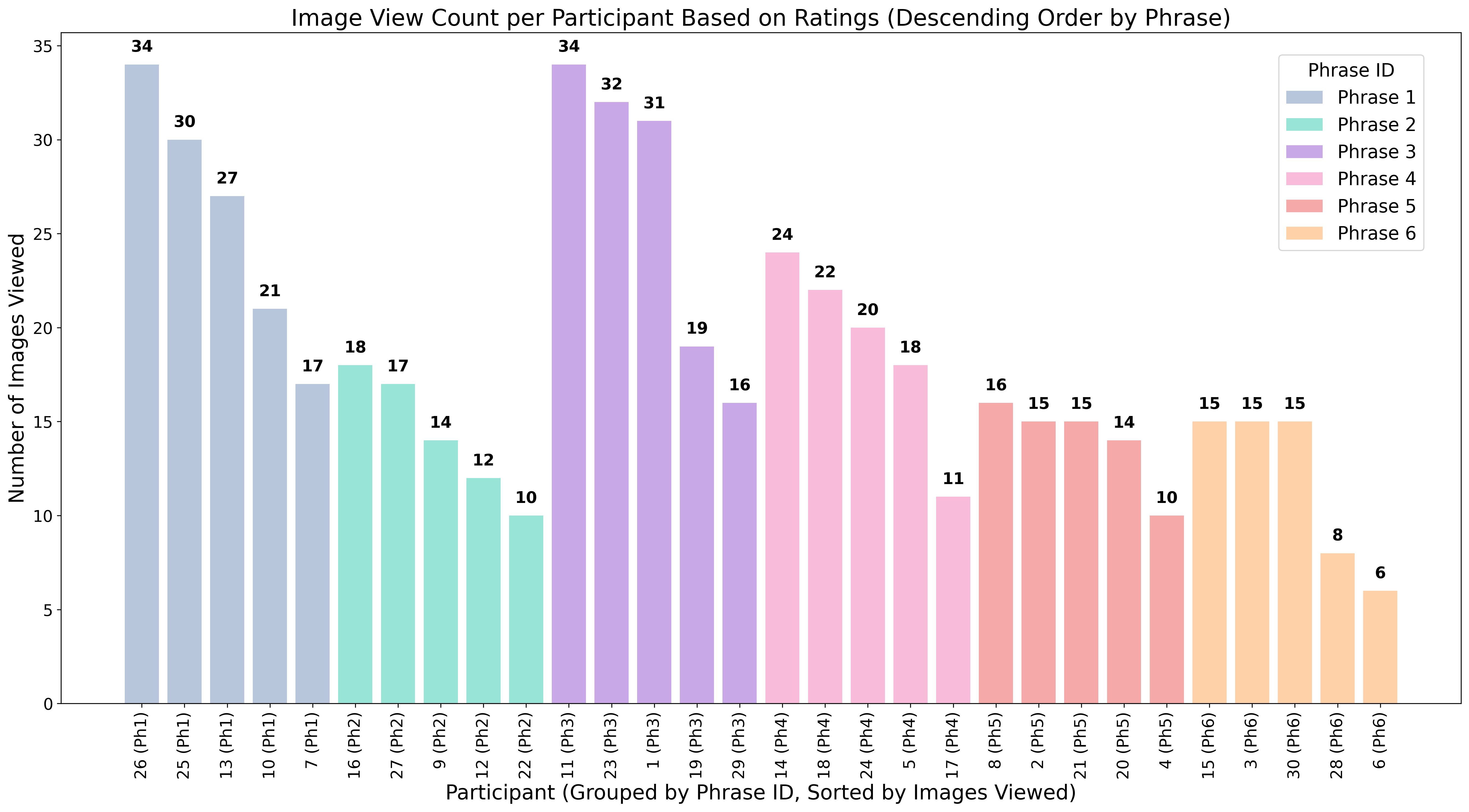}
    \caption{Bar chart showing the total number of unique images viewed by each participant, grouped by their assigned stimulus. This highlights how different stimuli influenced the breadth of exploration, with participants assigned negative stimuli (Ph5, Ph6) generally selecting fewer images.}
    \label{fig:image_count_per_participant_phase2}
\end{figure}

The bar chart in Figure \ref{fig:image_count_per_participant_phase2} shows the number of images viewed per participant, grouped by stimulus. It is interesting to note that participants tasked with finding images for negative emotions (Ph5 and Ph6) tended to select far fewer images than those looking for positive or neutral ones. This may suggest that images corresponding to negative emotions were either less common in the dataset or that the cognitive task of identifying them was more taxing, leading to a narrower selection.

\begin{figure}[htbp]
    \centering
    \includegraphics[width=1\linewidth]{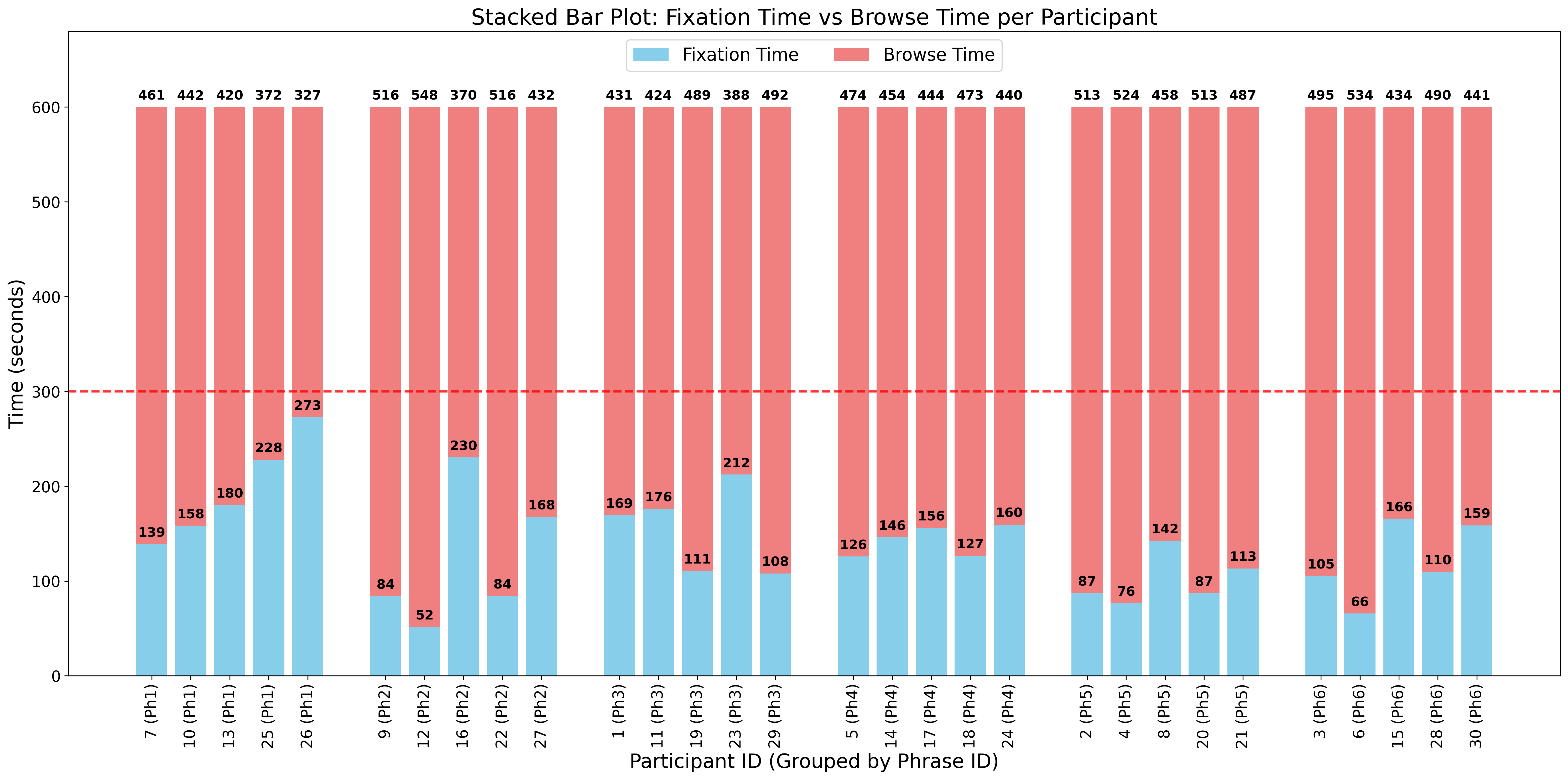}
    \caption{Breakdown of total session time into Fixation Time and Browse Time for each participant, grouped by stimulus. The plot quantifies the search strategies employed under different emotional conditions.}
    \label{fig:stacked_browse_fixation_phase2}
\end{figure}

A more detailed characterization of the conditioned exploration behaviour is revealed by a closer look at the attentional and rating distributions. The breakdown of session time into fixation and Browse periods, shown in Figure \ref{fig:stacked_browse_fixation_phase2}, quantifies how search strategies were influenced by the emotional framing; for instance, participants conditioned with negative stimuli (Ph5, Ph6) often exhibit a lower proportion of fixation time, suggesting a more hesitant or critical search.

\begin{figure*}[htbp]
    \centering
    \includegraphics[width=1\linewidth]{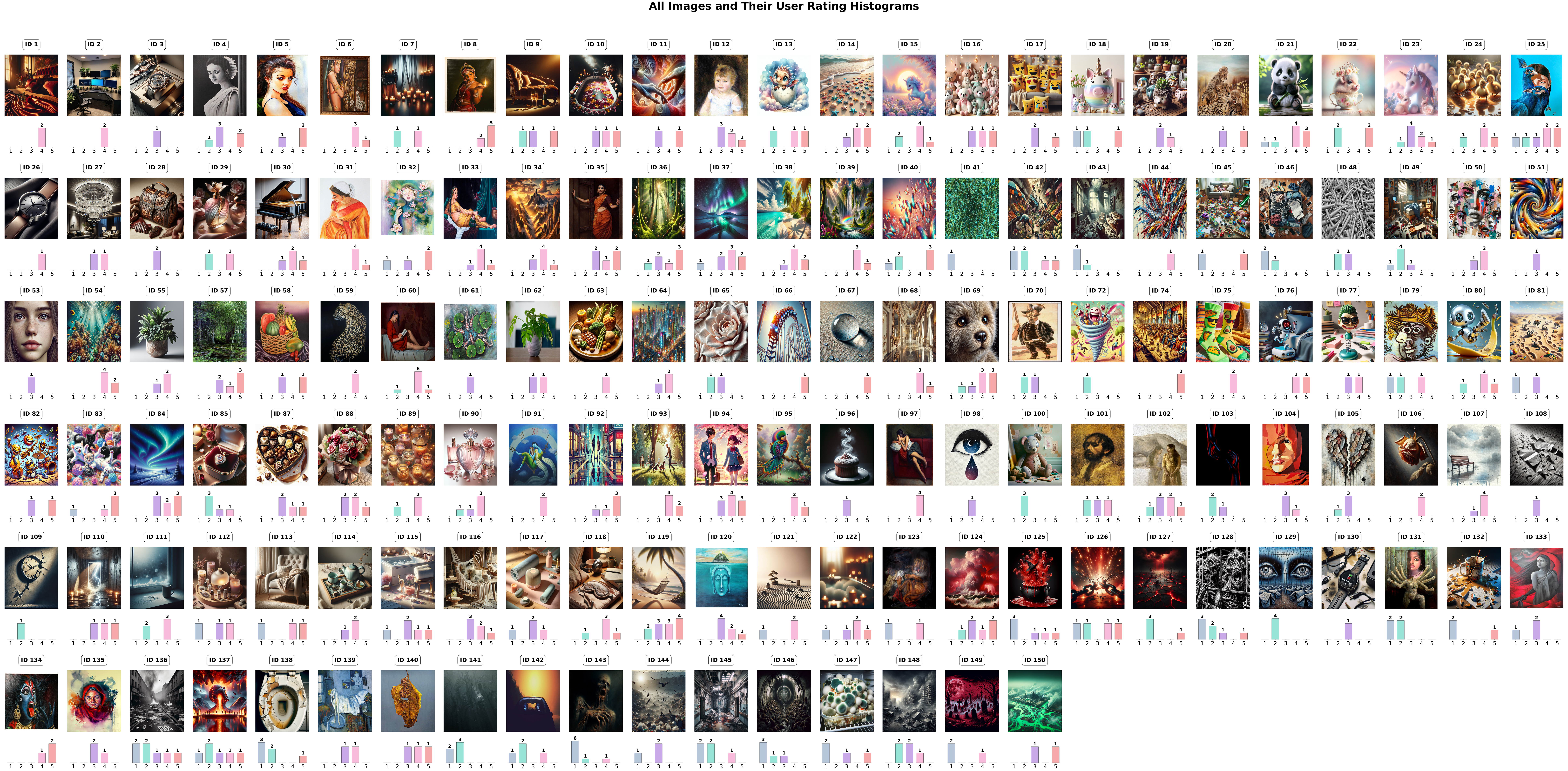}
    \caption{Grid overview of all 150 visual stimuli with their corresponding rating distributions. Below each image, a histogram displays the frequency of each rating (1-5) it received, providing a visual summary of its overall appeal and showing which images were universally liked, disliked, or divisive.}
    \label{fig:rating_histograms_phase2}
\end{figure*}

\begin{figure}[!ht]
    \centering
    
    \begin{subfigure}{\linewidth}
        \centering
        \includegraphics[width=1\linewidth]{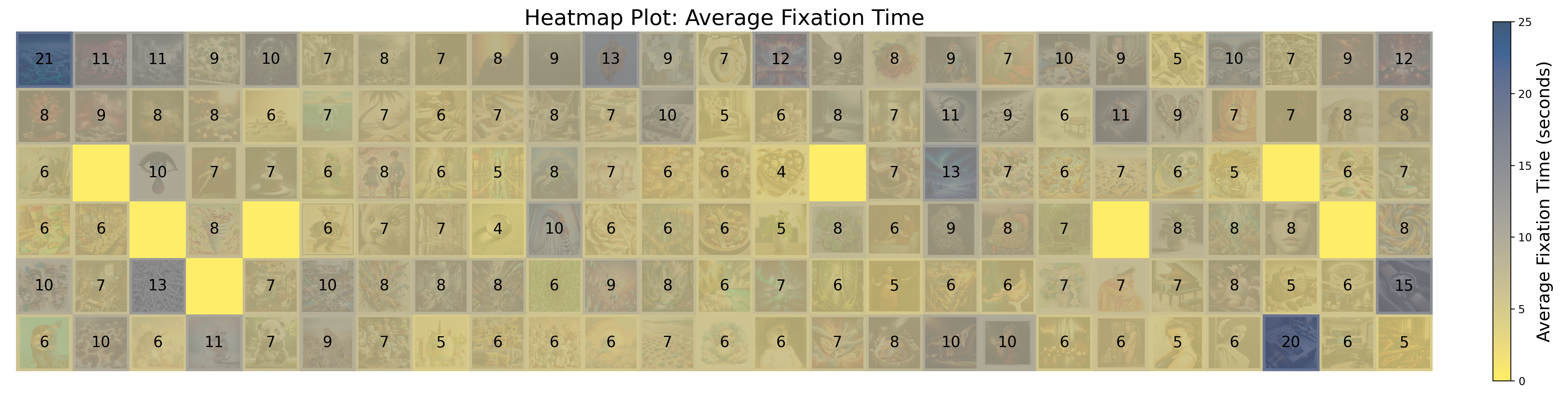}
        \caption{Heatmap of the average fixation time per viewer for each of the 150 images in Phase 2. By normalizing for the number of viewers, this heatmap highlights images that were inherently more engaging or complex to evaluate on a per-view basis during the conditioned search task.}
        \label{fig:heatmap_avg_fixation_phase2}
    \end{subfigure}
    
    \vfill
    
    \begin{subfigure}{\linewidth}
        \centering
        \includegraphics[width=1\linewidth]{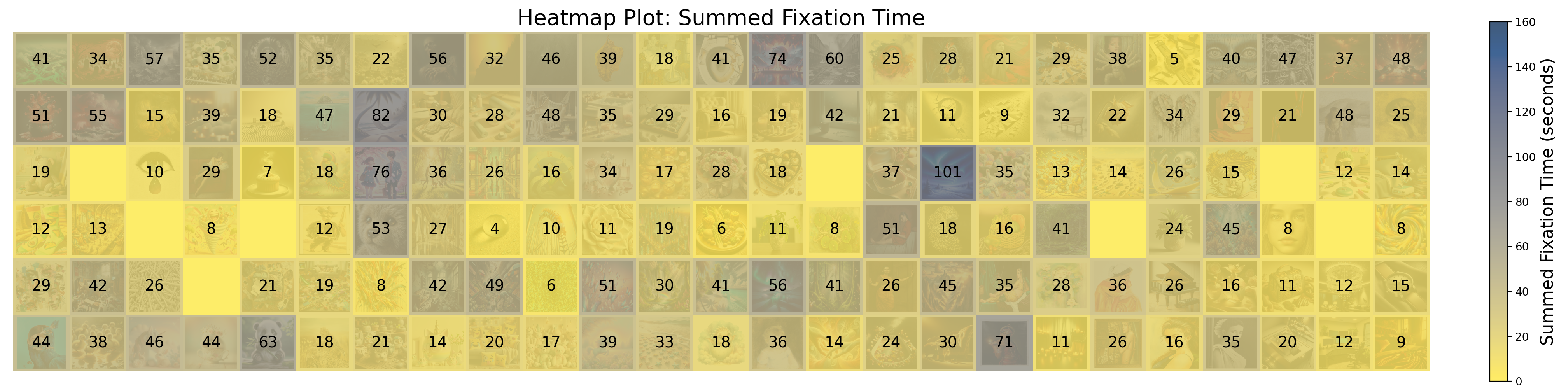}
        \caption{Heatmap of the total summed fixation time for each of the 150 images in Phase 2. This visualization shows the collective attentional hotspots across all participants and stimuli, revealing which images were most frequently selected for evaluation in the goal-directed tasks.}
        \label{fig:heatmap_sum_fixation_phase2}
    \end{subfigure}
    
    \caption{Phase 2 heatmaps of image engagement. (a)~\ref{fig:heatmap_avg_fixation_phase2} shows average fixation time per viewer. (b)~\ref{fig:heatmap_sum_fixation_phase2} shows summed fixation time across all participants.}
    \label{fig:combined_heatmaps_phase2}
\end{figure}

The grid plot in Figure~\ref{fig:rating_histograms_phase2} provides a complete overview of the rating distributions for 142 images out of 150 images, revealing that around 8 images were completely ignored by all the participants, showing that none of the participants attention went to those images as the stimuli from those images might not have been related to the emotions with which the participants were primed. Furthermore, the heatmaps of summed and average fixation times (Figure \ref{fig:heatmap_sum_fixation_phase2} and Figure \ref{fig:heatmap_avg_fixation_phase2}) offer a spatial perspective, revealing the emergence of new "attentional hotspots" on the image canvas that were different from those in the unguided Phase 1 study. This directly illustrates how the stimuli re-prioritized the visual importance of the image set.

\begin{figure}[htbp]
    \centering
        \includegraphics[width=1\linewidth]{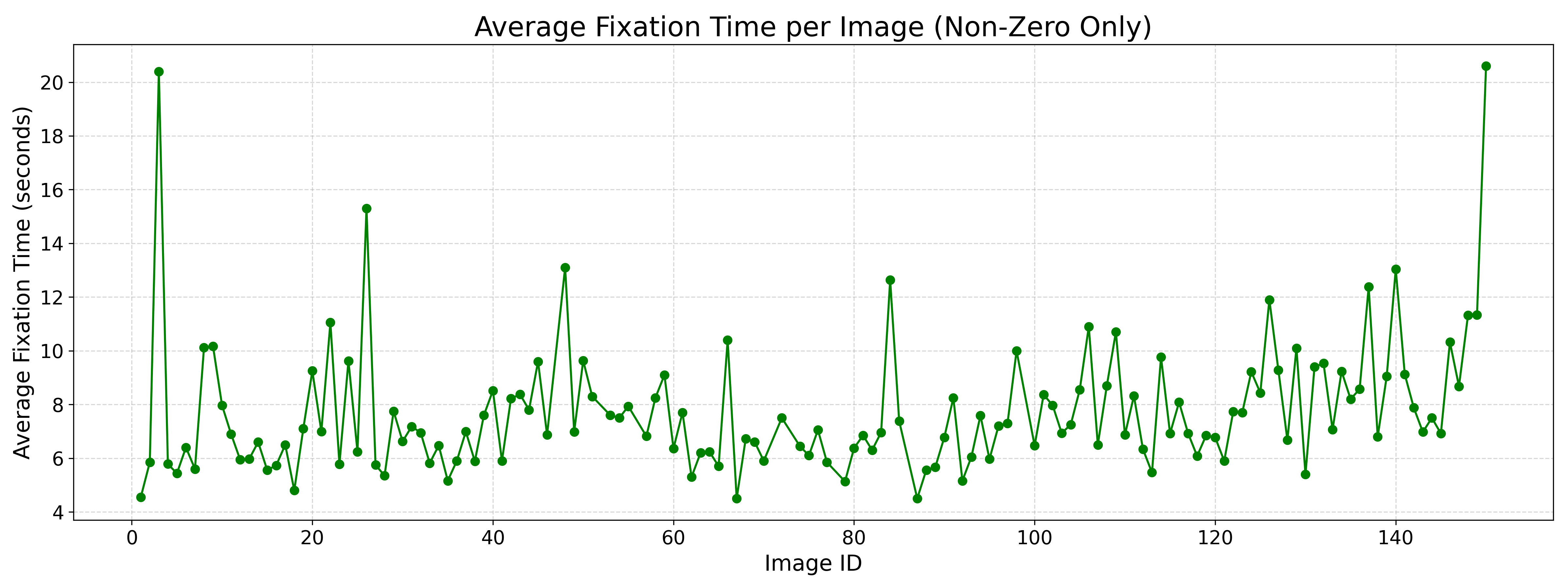}
        \caption{Line plot of the average fixation time per image in Phase 2. The peaks in this plot identify images that required the most cognitive effort for participants to evaluate for relevance against their assigned stimulus.}
        \label{fig:avg_time_per_image_phase2}
\end{figure}

Finally, the line plot for fixation times per image as shown in Figure \ref{fig:avg_time_per_image_phase2}) identify specific images that became exemplars for a particular emotional category, serving as strong visual anchors that consistently captured high relevance scores or required significant cognitive effort to evaluate. This visualization illustrate not just that convergence occurred, but how it manifested through tangible shifts in search strategy and the collective re-evaluation of specific visual stimuli.

\begin{figure}[htbp]
\centering
\includegraphics[width=1\linewidth]{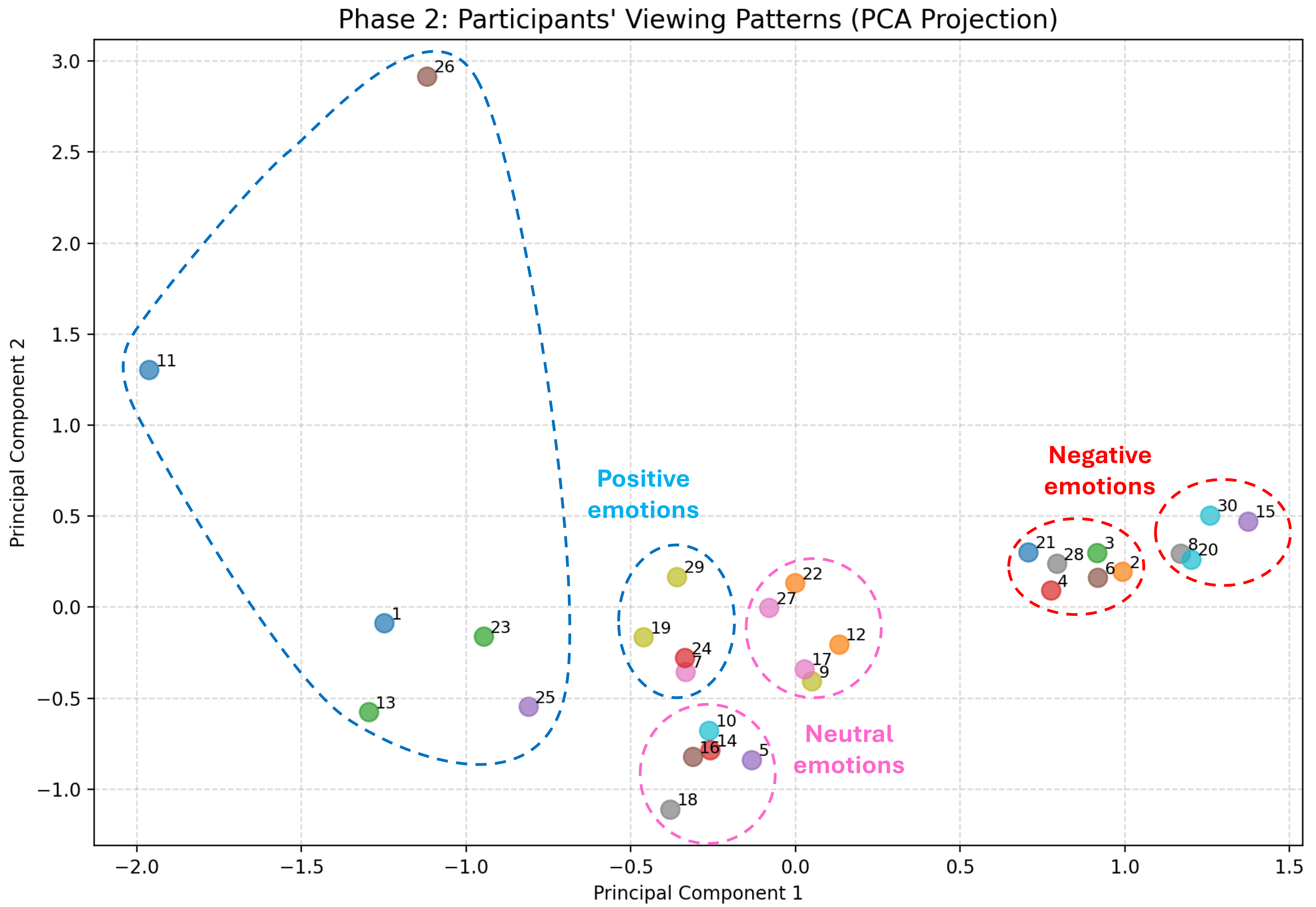}
\caption{Participants' Viewing Patterns (PCA Projection). }
\label{fig:pca_emotion_clusters}
\end{figure}

The scatter plot shown in Figure~\ref{fig:pca_emotion_clusters} displays the PCA projection of 30 participants' image viewing sequences onto the first two principal components. Each point represents a participant, labeled as P1 through P30. In this projection, a degree of clustering is evident. Participants appear to group into different regions, suggesting similarities in their viewing patterns. A cluster in the upper right might correspond to viewing sequences influenced by positive emotions, with some overlap with a cluster near the origin potentially related to neutral emotions. A more distinct cluster is observed in the lower portion, possibly indicating viewing patterns associated with negative emotional content. The presence of these clusters implies that, for this group of participants, the emotional categories of the images may have played a role in structuring their viewing order.

\begin{figure}[!ht]
    \centering
    \begin{subfigure}{1\linewidth}
        \centering
        \includegraphics[width=1\linewidth]{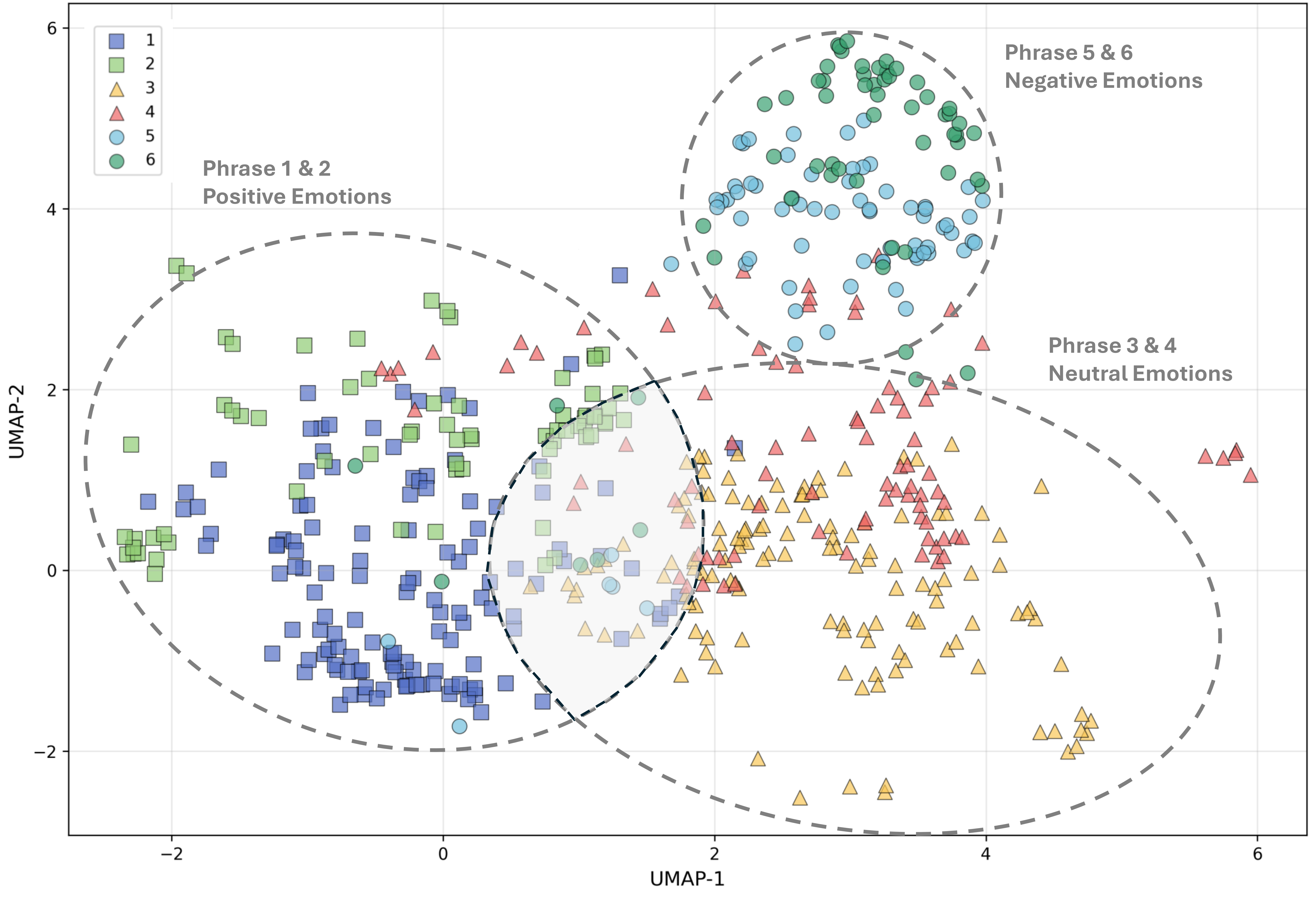}
        \caption{UMAP visualization of semantic clusters from participant thoughts, colored by stimulus phrase ID. Each point represents a participant's thought, projected into 2D space. The plot shows a clear semantic separation for thoughts related to negative stimuli (Phrases 5 \& 6, circles). Thoughts from positive (Phrases 1 \& 2, squares) and neutral (Phrases 3 \& 4, triangles) stimuli form distinct but partially overlapping clusters, indicating a finer-grained emotional spectrum in non-negative responses.}
        \label{fig:umap_thematic_analysis}
    \end{subfigure}
    \vfill
    \begin{subfigure}{1\linewidth}
        \centering
        \includegraphics[width=1\linewidth]{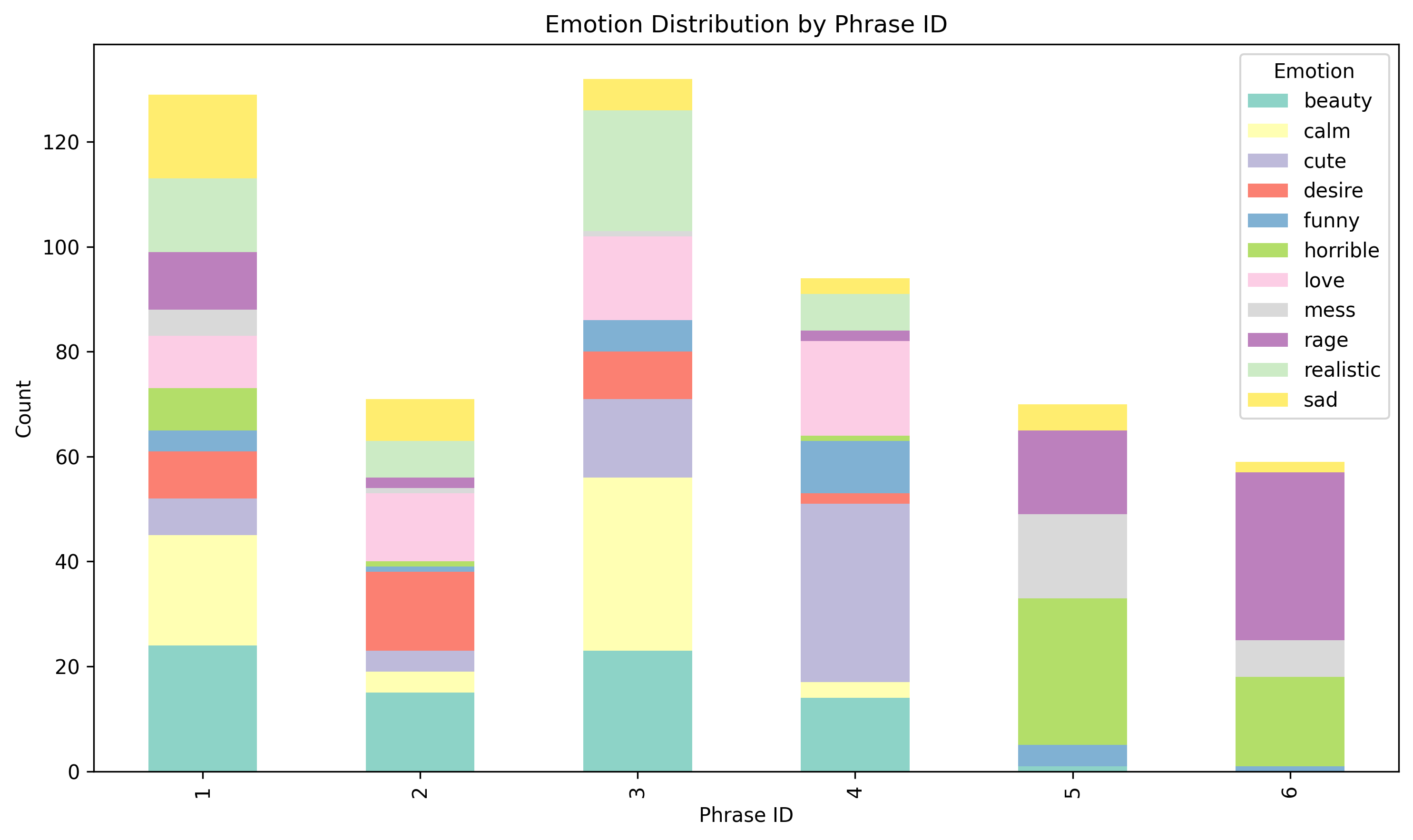}
        \caption{Distribution of thematic emotion keywords for each stimulus phrase. The stacked bars show the frequency of specific emotional keywords identified in participants thoughts for each of the six stimulus phrases. The chart highlights the distinct emotional palettes associated with each stimulus, such as 'rage' for Phrase 6, while also showing the semantic overlap (e.g., 'love', 'desire') between positive and neutral categories.}
        \label{fig:emotion_bar_chart}
    \end{subfigure}
    \caption{(Top) UMAP-based semantic clustering of participant thoughts, and (Bottom) distribution of extracted emotion keywords by stimulus phrase.}
    \label{fig:phase2_thematic_analysis_combined}
\end{figure}

A qualitative thematic analysis was performed on the participants' open-ended thoughts to provide a deeper understanding of their cognitive responses to the emotional stimuli. The thoughts were converted into semantic embeddings and visualized using UMAP, as shown in Figure \ref{fig:umap_thematic_analysis}. The analysis reveals a clear spatial separation of the semantic clusters based on emotional valence. Thoughts generated in response to the negative stimuli (Phrases 5 \& 6) form a tight, isolated cluster, demonstrating their distinct and unambiguous emotional content. The positive (Phrases 1 \& 2) and neutral (Phrases 3 \& 4) stimuli also produce distinct clusters; however, they exhibit little overlap, suggesting a more nuanced gradient and shared emotional language between these states. This is further clarified by the breakdown of specific emotion keywords in Figure \ref{fig:emotion_bar_chart}. The chart shows that negative phrases are dominated by terms like 'horrible' and 'rage', whereas the positive and neutral phrases share a more complex palette that includes overlapping keywords like 'love', 'desire', and 'beauty'. Together, these plots confirm that the stimuli created predictable "thought-worlds," further validating H3 (Stimulus-Driven Convergence) on a qualitative and semantic level.

\subsection{Discussion}
\label{subsec:phase2_discussion}
The findings from the Phase 2 study have significant implications for the development of guided creative tools. The validation of both hypotheses demonstrates that the EUPHORIA system can be used not only for passive preference capture but also for actively directing a user's creative exploration in a predictable and efficient manner.

The validation of H3, demonstrating stimulus-driven convergence, is a powerful result. It proves that simple, high-level textual prompts can successfully align the visual selections of multiple individuals. This capability is crucial for practical design scenarios, where a moodboard must adhere to a specific client brief or a set of brand keywords. It confirms that the system can be used to generate thematically coherent inspirational material that is not based on the subjective whims of a single designer but on a shared, guided interpretation of a design goal. This moves the moodboarding process from pure subjectivity towards a more objective, repeatable method.

The validation of H4, which revealed a fundamental shift in user behaviour, provides a clearer understanding of the role of visual attention in different contexts. The decoupling of the attention-preference link in a goal-directed task is a key insight. It clarifies that the meaning of a long fixation depends on the user's goal: in free exploration (Phase 1), it signals \textit{liking or interest}; in a conditioned search (Phase 2), it signals \textit{cognitive effort to evaluate relevance}. This distinction is critical for the design of any gaze-based system, as it dictates how the collected data should be interpreted. The increased efficiency and emergence of a systematic search strategy also highlight the practical benefits of providing a clear cognitive filter, which helps users navigate large datasets more effectively.

In summary, Phase 2 successfully demonstrated that the EUPHORIA system can effectively guide a user's attention to produce convergent, thematically consistent results in an efficient manner. This provides the crucial evidence that the system can be used as an active tool for focused inspiration gathering, setting the stage for the final phase of the research: applying this guided, attention-driven methodology to the generation and evaluation of tangible product form concepts.


\greyline

\section{Phase 3: Goal-driven Exploration through Strategic-Attention}
\label{subsec:phase3_study}

The final and most critical phase of this research transitions from foundational validation to practical application. Building upon the principles established in the preceding phases—that attention correlates with preference (Phase 1) and can be directed by external cues (Phase 2)—this phase investigates the efficacy of the complete EUPHORIA system within a real-world design context. This stage explores \textit{Strategic-Attention}, a higher-order cognitive process where the designer's attentional focus is purposefully directed towards solving a specific, multi-faceted design problem. The primary objective was to conduct a comparative analysis, evaluating the EUPHORIA-driven workflow against conventional design methodologies to measure its impact on efficiency, process structure, and the quality of the final creative output.

\subsection{Aim}
\label{subsec:phase3_hypotheses}

To investigate the research question for this phase, the following hypotheses were formulated:

\begin{enumerate}
    \item \textbf{H5: EUPHORIA-RETINA-driven design workflow Enhances the Quality of the Design Outcomes and increases Process Efficiency.} 
    
    When designers utilize the EUPHORIA-RETINA system to solve a form design problem, the structured, attention-driven workflow will lead to a more efficient process (reduced time) and result in final concept sketches that are rated higher by experts in terms of creativity, aesthetic coherence, and alignment with the specified design keywords, compared to concepts developed using conventional workflow.
\end{enumerate}

\subsection{Study Design}
\label{subsec:phase3_design}
To test this hypothesis, a study was designed to compare four different design workflows in a controlled yet realistic setting. The design involved multiple participants, a series of challenging design problems, and a set of distinct methodological paths, ranging from fully manual to highly automated.

\subsubsection{Participants}
Four designers with a diverse range of professional and academic experience were recruited to participate. This variation in expertise was intentional, allowing for an evaluation of the system's utility for both seasoned professionals and designers in training. The participants were:
\begin{itemize}
    \item \textbf{D1:} A senior designer with 7 years of professional experience.
    \item \textbf{D2:} A practicing designer with 3 years of experience and the founder of his own design firm.
    \item \textbf{D3:} A 3rd-year undergraduate student pursuing a Bachelor of Design (B.Des) in Industrial Design.
    \item \textbf{D4:} A 2nd-year postgraduate student pursuing a Master of Design (M.Des) in Product Design.
\end{itemize}
All participants provided informed consent, and all ethical protocols regarding data privacy and anonymity were strictly followed.

\subsubsection{Stimuli: The Form Design Problem Statements}
The stimuli for this phase were four distinct form design problem statements. These were not generic tasks; they were carefully authored to present a significant creative challenge. Each problem focused on a familiar product category (e.g., a speaker, a lamp) to ensure the designers had a baseline understanding of its function. However, the required stylistic attributes, defined by a set of keywords, were intentionally chosen to be contradictory or unconventional (e.g., "organic" and "brutalist"). This was a deliberate strategy to mitigate the designers' pre-existing stylistic biases and to prevent them from relying on established solutions. The challenge forced them to engage deeply with the ideation process to synthesize these conflicting styles into a novel and coherent product form. The four problem statements were:

\begin{itemize}
    \item \textbf{S1: Portable Music Player}
    \begin{itemize}
        \item \textbf{Contradicting Styles:} Retro (1960s-80s) and Futuristic.
        \item \textbf{Keywords:} \textit{Nostalgic, Tactile, Analog Controls, Vintage Knobs, Woodgrain Textures, Sleek, Contemporary, Innovative}.
    \end{itemize}

    \item \textbf{S2: Study Lamp}
    \begin{itemize}
        \item \textbf{Contradicting Styles:} Japanese Zen Philosophy and Cyberpunk Futurism.
        \item \textbf{Keywords:} \textit{Serene, Minimal, Natural Materials, Muted Tones, Bold, High-Contrast, Sharp Edges, Neon Accents, Layered Complexity}.
    \end{itemize}

    \item \textbf{S3: Handheld Flashlight}
    \begin{itemize}
        \item \textbf{Contradicting Styles:} Post-Apocalyptic Utilitarianism and High-End Luxury.
        \item \textbf{Keywords:} \textit{Rugged, Modular, Weathered Surfaces, Worn Metals, Refined Elegance, Polished Accents, Seamless, Sculptural Silhouette}.
    \end{itemize}

    \item \textbf{S4: Candlestick Holder}
    \begin{itemize}
        \item \textbf{Contradicting Styles:} Prehistoric/Primordial and Contemporary/Modern.
        \item \textbf{Keywords:} \textit{Raw, Weathered Stone, Hand-carved Textures, Irregular Silhouettes, Modern Functionality, Minimal, Compact Geometry, Timeless}.
    \end{itemize}
\end{itemize}

\subsubsection{Workflow Paths}
\begin{figure*}[htbp]
    \centering
    \includegraphics[width=0.95\linewidth]{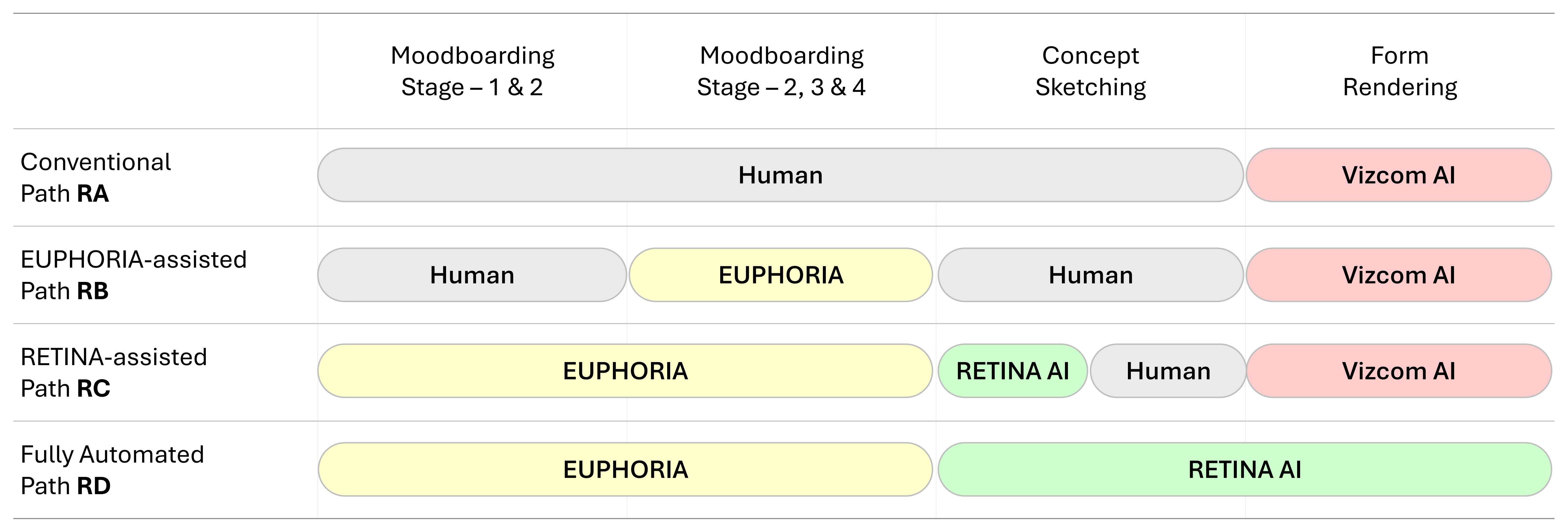}
    \caption{A high-level comparison of the four experimental workflow paths, illustrating the distribution of tasks between the human designer and the automated systems across the major stages of the form design process. Each row represents a different path, from the fully conventional (Path RA) to the fully automated (Path RD). The color-coded blocks indicate the primary agent responsible for each stage: grey for tasks performed manually by the human designer, yellow for tasks handled by the EUPHORIA system, green for tasks executed by the RETINA AI pipeline, and red for the final rendering step using an off-the-shelf commercial software known as Vizcom AI\textcopyright. The diagram visually demonstrates the progressive shift of responsibility from manual to automated.}
    \label{fig:workflow_paths_comparison}
\end{figure*}

To facilitate a comprehensive comparison, four distinct workflow paths were defined, each representing a different level of collaboration between man and machine, as shown in Figure~\ref{fig:workflow_paths_comparison}. A summary of the stages and the level of automation in each path is presented in Table~\ref{tab:workflow_process_medium} and Table~\ref{tab:workflow_nature_outcome_revised} with the process type, medium used, outcome and their nature.

\begin{table}[htbp]
\centering
\caption{Latin Square Design for Phase 3 Study Task Allocation}
\label{tab:phase3_latin_square}
\begin{tabular}{|c|c|c|c|c|}
\hline
\textbf{Problem ID (PID)} & \textbf{RA} & \textbf{RB} & \textbf{RC} & \textbf{RD} \\ \hline
\textbf{S1}      & D1                   & D4                   & D3                   & D2                   \\ \hline
\textbf{S2}      & D2                   & D1                   & D4                   & D3                   \\ \hline
\textbf{S3}      & D3                   & D2                   & D1                   & D4                   \\ \hline
\textbf{S4}      & D4                   & D3                   & D2                   & D1                   \\ \hline
\end{tabular}
\end{table}

\begin{table*}[htbp]
\centering
\caption{Comparative Matrix of Process Type and Medium Used Across Four Form Design Workflow Paths}
\label{tab:workflow_process_medium}
\renewcommand{\arraystretch}{1.8} 
\begin{tabular}{
    >{\bfseries}l 
    >{\centering\arraybackslash}p{2.8cm}
    >{\centering\arraybackslash}p{2.8cm}
    >{\centering\arraybackslash}p{2.8cm}
    >{\centering\arraybackslash}p{2.8cm}
}
\toprule
\textbf{Stage} & \textbf{Path RA \small{(Conventional)}} & \textbf{Path RB \small{(EUPHORIA-Assisted)}} & \textbf{Path RC \small{(RETINA-Assisted)}} & \textbf{Path RD \small{(Fully Automated)}} \\
\midrule

Keyword Identification & 
\cellcolor{processManual} Manual Analysis & 
\cellcolor{processManual} Manual Analysis & 
\cellcolor{processManual} Manual Analysis & 
\cellcolor{processManual} Manual Analysis \\

Collection & 
\cellcolor{processManual} Web Search & 
\cellcolor{processManual} Web Search & 
\cellcolor{processAutomated} EUPHORIA API Search & 
\cellcolor{processAutomated} EUPHORIA API Search \\

Selection & 
\cellcolor{processManual} Manual Curation & 
\cellcolor{processManual} Manual Curation & 
\cellcolor{processAutomated} EUPHORIA (Gaze) & 
\cellcolor{processAutomated} EUPHORIA (Gaze) \\

Composition & 
\cellcolor{processManual} Figma / Miro & 
\cellcolor{processAutomated} EUPHORIA MS & 
\cellcolor{processAutomated} EUPHORIA MS & 
\cellcolor{processAutomated} EUPHORIA MS \\

Reflection & 
\cellcolor{processManual} Designer's Mind & 
\cellcolor{processGuided} Designer's Mind & 
\cellcolor{processGuided} Designer's Mind & 
\cellcolor{processGuided} Designer's Mind \\

Extraction & 
\cellcolor{processManual} Mental Mapping & 
\cellcolor{processAutomated} CV/AI Pipeline & 
\cellcolor{processAutomated} CV/AI Pipeline & 
\cellcolor{processAutomated} CV/AI Pipeline \\

Shape Abstraction & 
\cellcolor{processManual} Pen \& Paper & 
\cellcolor{processAutomated} HED Processing & 
\cellcolor{processAutomated} HED Processing & 
\cellcolor{processAutomated} HED Processing \\

Rough Sketch & 
\cellcolor{processManual} Pen \& Paper & 
\cellcolor{processManual} Pen \& Paper & 
\cellcolor{processAutomated} RETINA AI & 
\cellcolor{gray!20} --- \\

Concept Sketching & 
\cellcolor{processManual} Pen \& Paper & 
\cellcolor{processManual} Pen \& Paper & 
\cellcolor{processManual} Pen \& Paper & 
\cellcolor{gray!20} --- \\

Rendering & 
\cellcolor{processAutomated} Vizcom AI & 
\cellcolor{processAutomated} Vizcom AI & 
\cellcolor{processAutomated} Vizcom AI & 
\cellcolor{processAutomated} RETINA AI \\
\bottomrule
\end{tabular}
\begin{flushleft}
\vspace{1ex}
\small
\textbf{Legend:} \quad
\colorbox{processManual}{\strut\hspace{1em}} Manual (M) \quad
\colorbox{processAutomated}{\strut\hspace{1em}} Automated (A) \quad
\colorbox{processGuided}{\strut\hspace{1em}} Guided (G)
\end{flushleft}
\end{table*}

\begin{table*}[!ht]
\centering
\caption{Comparative Matrix of Process Nature and Outcome Across Four Workflow Paths}
\label{tab:workflow_nature_outcome_revised}
\renewcommand{\arraystretch}{1.8} 
\begin{tabular}{
    >{\bfseries}l 
    >{\centering\arraybackslash}p{2.8cm}
    >{\centering\arraybackslash}p{2.8cm}
    >{\centering\arraybackslash}p{2.8cm}
    >{\centering\arraybackslash}p{2.8cm}
}
\toprule
\textbf{Stage} & \textbf{Path RA \small{(Conventional)}} & \textbf{Path RB \small{(EUPHORIA-Assisted)}} & \textbf{Path RC \small{(RETINA-Assisted)}} & \textbf{Path RD \small{(Fully Automated)}} \\
\midrule

Keyword Identification & 
Keywords & 
Keywords & 
Keywords & 
Keywords \\

Collection & 
\cellcolor{natureExplicit} Downloaded Images Folder & 
\cellcolor{natureExplicit} Downloaded Images Folder & 
\cellcolor{natureImplicit} Real-time Populated Images Grid& 
\cellcolor{natureImplicit} Real-time Populated Images Grid\\

Selection & 
\cellcolor{natureExplicit} Curated Images & 
\cellcolor{natureExplicit} Curated Images & 
\cellcolor{natureImplicit} Attention-selected Images & 
\cellcolor{natureImplicit} Attention-selected Images \\

Composition & 
\cellcolor{natureExplicit} Digital Moodboard & 
\cellcolor{natureImplicit} Attention-Weighted Moodspace & 
\cellcolor{natureImplicit} Attention-Weighted Moodspace  & 
\cellcolor{natureImplicit} Attention-Weighted Moodspace  \\

Reflection & 
\cellcolor{natureImplicit} Design Insight & 
\cellcolor{natureExplicit} Gaze Tracked Designer Insight & 
\cellcolor{natureExplicit} Gaze Tracked Designer Insight & 
\cellcolor{natureExplicit} Gaze Tracked Designer Insight \\

Extraction & 
\cellcolor{natureImplicit} Mental Feature Map & 
\cellcolor{natureExplicit} Feature Maps (ROI Collage, Colour Palette) & 
\cellcolor{natureExplicit} Feature Maps (ROI Collage, Colour Palette) & 
\cellcolor{natureExplicit} Feature Maps (ROI Collage, Colour Palette) \\

Shape Abstraction & 
Geometric Sketches & 
HED Edge Collage & 
HED Edge Collage & 
HED Edge Collage \\

Rough Sketch & 
Rough Sketches & 
Rough Sketches & 
AI-Generated Sketches & 
\cellcolor{gray!20} --- \\

Concept Sketching & 
Concept Sketch & 
Concept Sketch & 
Concept Sketch & 
\cellcolor{gray!20} --- \\

Rendering & 
AI-Generated Rendering & 
AI-Generated Rendering & 
AI-Generated Rendering & 
AI-Generated Rendering \\
\bottomrule
\end{tabular}
\begin{flushleft}
\vspace{1ex}
\small
\textbf{Legend (for Moodboarding Stages):} \quad
\colorbox{natureImplicit}{\strut\hspace{1em}} Implicit \quad
\colorbox{natureExplicit}{\strut\hspace{1em}} Explicit
\end{flushleft}
\end{table*}

\subsubsection{Experimental Structure: Latin Square Design}
To ensure the integrity of the comparison and to control for confounding variables such as designer skill and problem difficulty, a Latin Square design was implemented. This structure ensured that each of the four designers (D1-D4) attempted each of the four problem statements (S1-S4), using each of the four workflow paths (RA-RD) exactly once. This balanced design allows for a robust analysis by isolating the effect of the workflow path from the individual abilities of the designers and the specific challenges of each problem. The precise allocation of tasks is shown in Table~\ref{tab:phase3_latin_square}.

\subsection{Procedure}
\label{subsec:phase3_procedure}
Each designer completed their four assigned tasks in separate sessions. They were given detailed, step-by-step instructions specific to the workflow path they were required to follow for that session, as detailed below. The complete details of the process type, medium used, and the outcome generated in each stage for all different paths are tabulated in Table~\ref{tab:workflow_process_medium} and Table~\ref{tab:workflow_nature_outcome_revised}.

\textit{\textbf{Path RA (Conventional)}}

For Path RA, representing the conventional modern style workflow, participants followed a sequence of manual, explicit steps. Participants began by carefully reading the assigned problem statement and manually identifying a list of keywords that best described the required aesthetic and stylistic attributes. This was followed by a comprehensive moodboarding stage. Using web browsers, participants searched for and collected a repository of images from image hosting websites. They then manually curated this collection and used a digital whiteboard application to compose the moodboard by arranging and grouping the selected images. The subsequent reflection on this board and the extraction of common features was a purely cognitive, implicit activity performed by the designer. Based on the mental feature maps they had formed, participants proceeded to manually sketch simple geometric abstractions on paper. These shapes were then used as building blocks for a divergent phase of rough sketch exploration on A3 sheets. Finally, they converged their ideas from the rough sketches into a single, detailed final concept sketch. The process concluded when participants used the AI-assisted tool, Vizcom AI, to render their manual sketch.

\textit{\textbf{Path RB (EUPHORIA-Assisted)}}

Path RB introduced the EUPHORIA system to automate the latter half of the moodboarding process. The procedure began identically to Path RA with manual keyword identification from the problem statement. Participants then performed a manual image collection step, searching the web and downloading a folder of images relevant to the keywords, but they were instructed not to manually arrange them. Instead of using a digital whiteboard, they loaded this image collection into the EUPHORIA system. Participants then spent 15 minutes immersively exploring their collected images in the VR Moodspace while the system implicitly captured their gaze data. Based on this data, the system automatically generated explicit feature maps: an ROI Collage, an HED Edge Collage for shape abstraction, and a Dominant Colour Palette. Using these tangible, system-generated feature maps as direct inspiration, participants then proceeded with the manual stages of rough sketch exploration and final concept sketch creation, followed by rendering with Vizcom AI, as in Path RA.

\textit{\textbf{Path RC (RETINA-Assisted)}}

Path RC integrated EUPHORIA for the entire moodboarding process and introduced an AI agent for initial ideation. After manually identifying keywords as in Path RA, participants engaged with a fully automated moodboarding process within EUPHORIA. They provided the keywords to the system, which automatically populated the VR moodspace with relevant images via an API search in real-time. A 15-minute immersive exploration session followed, during which the system performed implicit collection, selection, and composition based on the participant's gaze, culminating in the automatic generation of the explicit feature maps (ROI Collage, HED Edge Collage, and Colour Palette). The extracted feature maps were then fed into the `RETINA AI` pipeline, an agentic AI framework that uses a compound LLM, which generated a set of diverse concept sketches to be used as a reference. Participants used these AI-generated sketches as inspiration to manually sketch their own final, detailed concept sketch. The final rendering of their manual sketch was then completed using Vizcom AI.

\textit{\textbf{Path RD (Fully Automated)}}

Path RD represented the most automated workflow, where the designer's role shifted significantly from creator to curator. The procedure was identical to Path RC through the initial keyword identification and the fully automated EUPHORIA moodboarding and feature extraction stages. However, in the final step, the extracted feature maps were provided as input to the `RETINA AI` pipeline, which was tasked with generating a collection of varied, photorealistic product renderings directly. The participant's role was transformed into that of a curator. They did not perform any manual sketching or rendering. Their final task was to review the set of AI-generated renderings and select the single image they judged to be the most successful and creative solution to the design problem statement.

\subsection{Expert Evaluation}
\label{subsec:phase3_datacollection}
The final concept sketches and the corresponding product renderings were used for expert evaluation. The complete set of 16 final design outputs (one from each designer for each path) was presented to a panel of 50 design experts for assessment. The expert panel was composed of alumni from the Department of Design and Manufacturing at the Indian Institute of Science (IISc), representing a wide range of professional experience. Out of the 50 experts, 35 possessed more than five years of design experience, while the remaining 15 had between two to five years of experience. The outputs were presented to the experts in a random order without disclosing their genesis. The experts provided the following two inputs.
\begin{itemize}
    \item \textbf{Ranking:} They ranked the four outputs for each of the four problem statements from rank (1) to rank (4) based on overall design appeal, coherence, and their potential as a real product.
    \item \textbf{Rating:} Using a 5-point Likert scale, they assessed each of the 16 renderings against eight criteria, namely, adherence to the brief, novelty, visual appeal, emotional resonance, clarity of purpose, distinctiveness of silhouette, implied materiality, and proportional balance.
\end{itemize}

\begin{figure*}[htbp]
    \centering
    \begin{subfigure}[b]{0.24\textwidth}
        \centering
        \includegraphics[width=\linewidth]{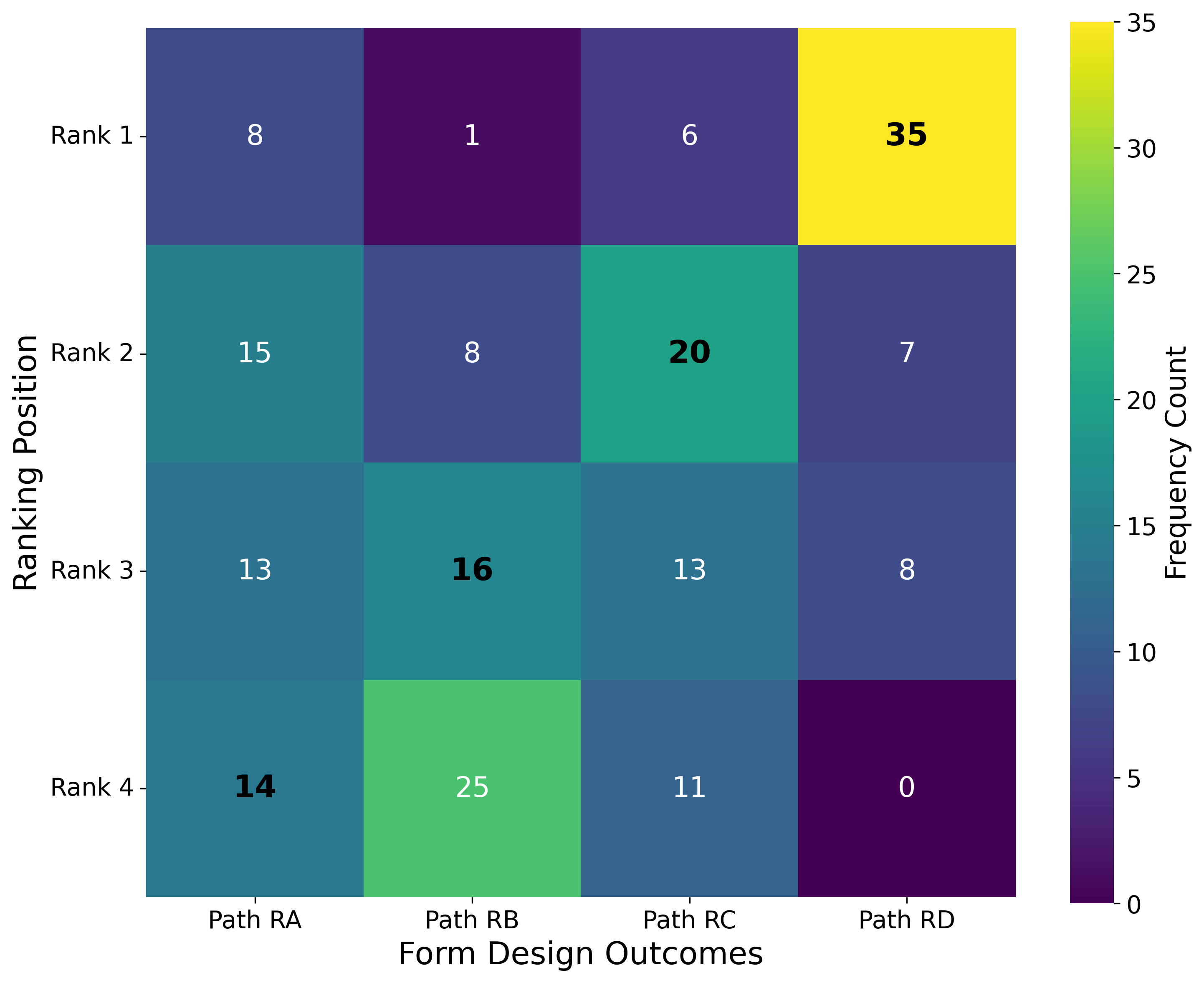}
        \caption{Music Player (S1)}
        \label{fig:heatmap_rankings_s1}
    \end{subfigure}
    \hfill 
    \begin{subfigure}[b]{0.24\textwidth}
        \centering
        \includegraphics[width=\linewidth]{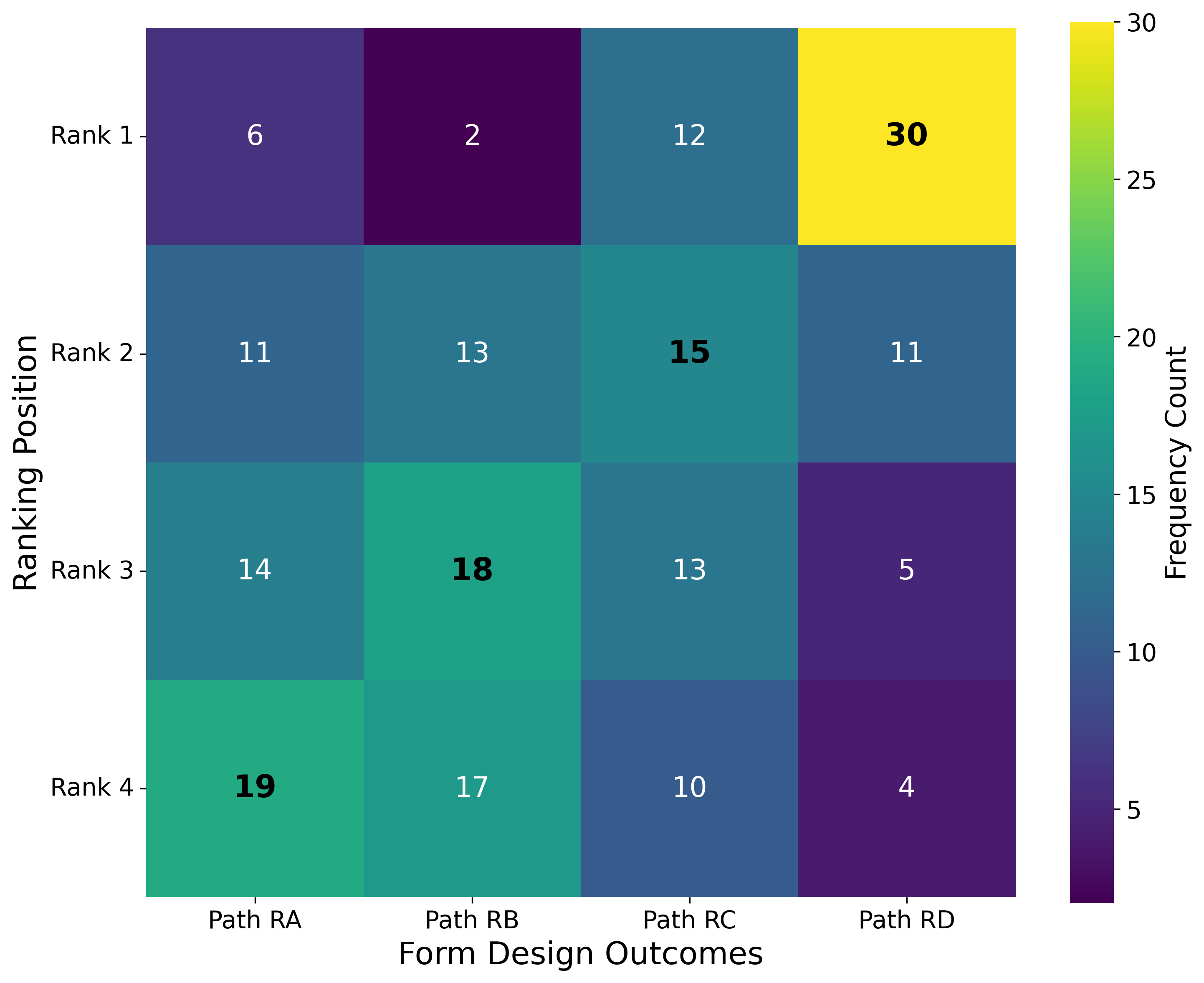}
        \caption{Study Lamp (S2)}
        \label{fig:heatmap_rankings_s2}
    \end{subfigure}  
    \hfill
    \begin{subfigure}[b]{0.24\textwidth}
        \centering
        \includegraphics[width=\linewidth]{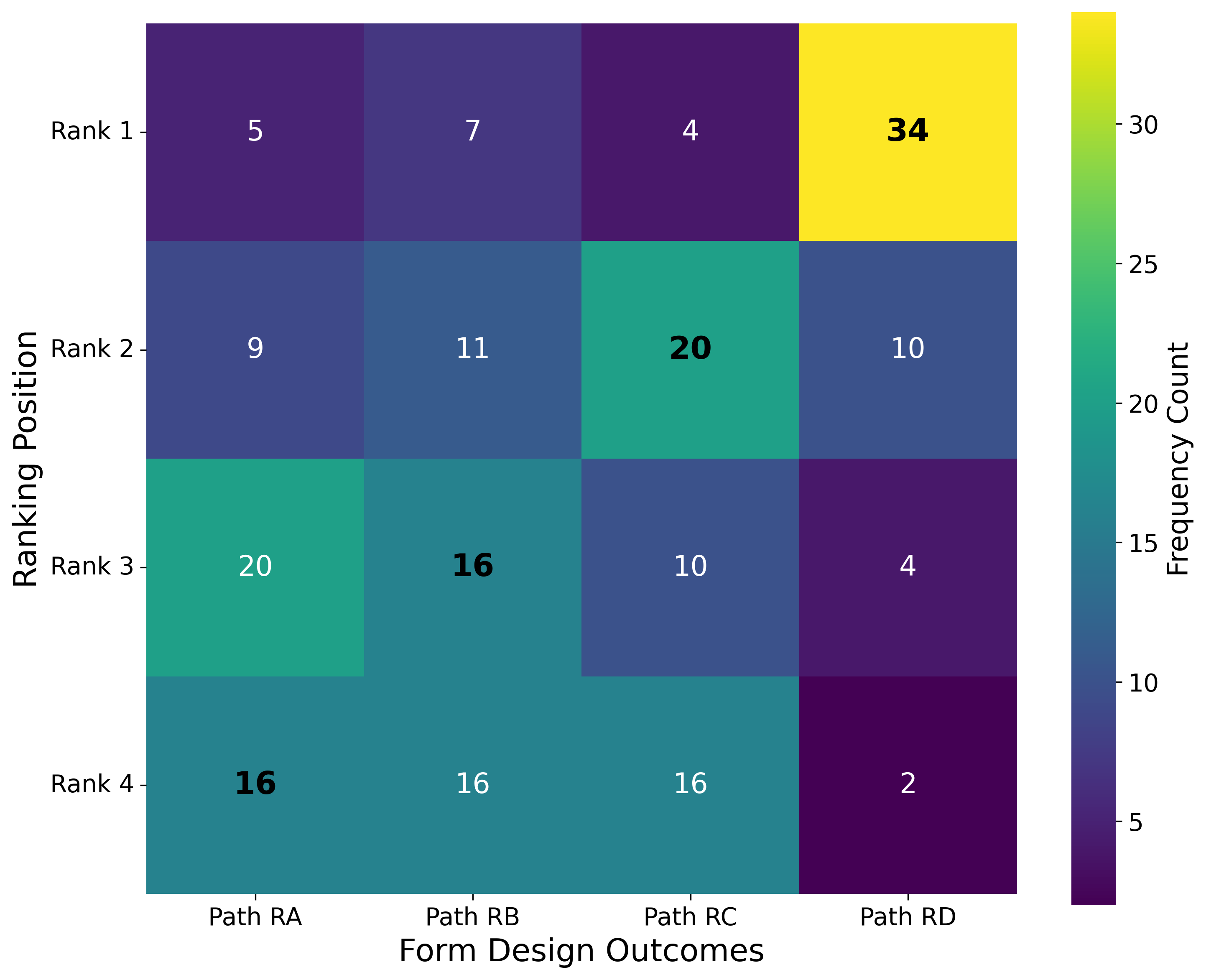}
        \caption{Flashlight (S3)}
        \label{fig:heatmap_rankings_s3}
    \end{subfigure}
    \hfill 
    \begin{subfigure}[b]{0.24\textwidth}
        \centering
        \includegraphics[width=\linewidth]{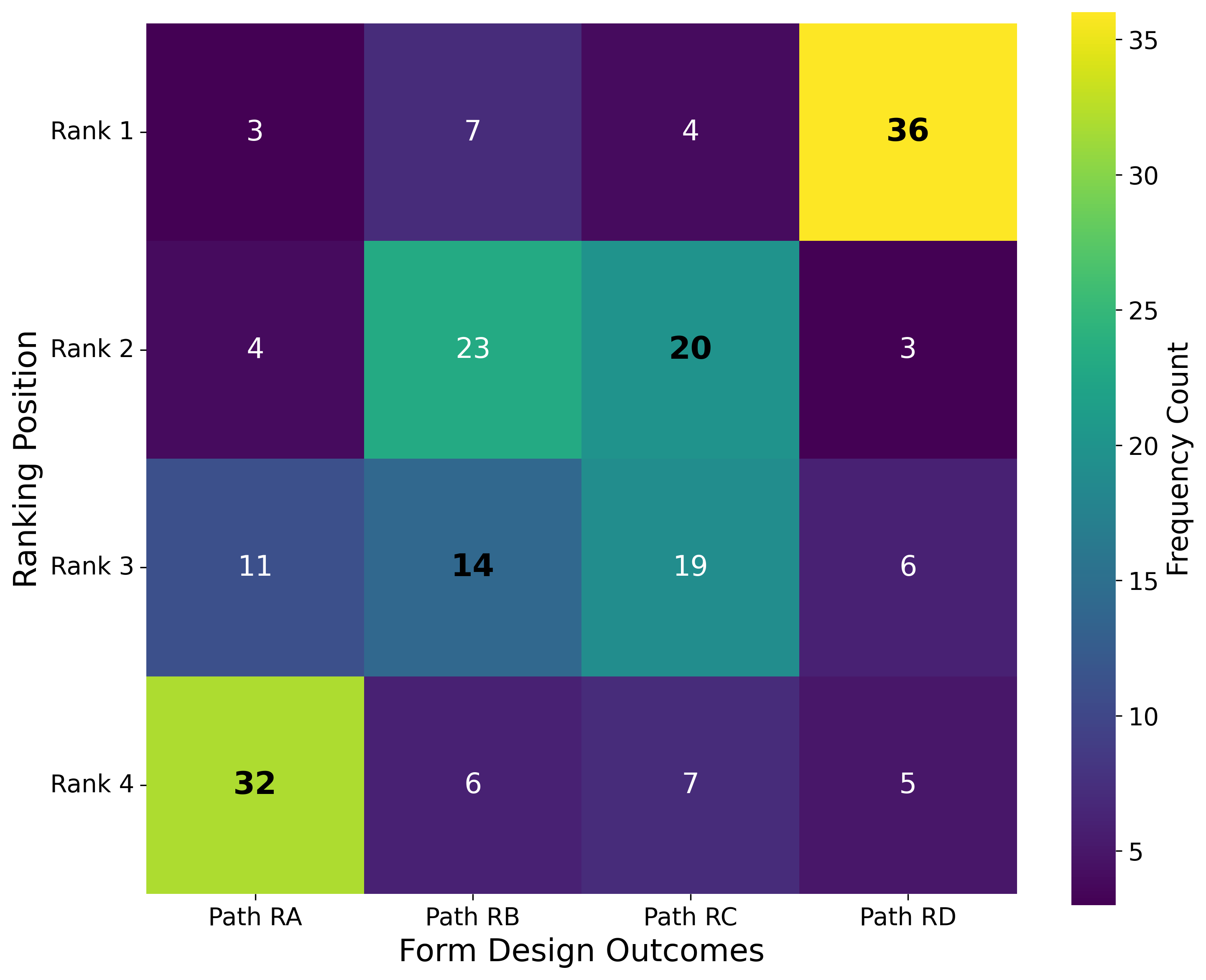}
        \caption{Candle Holder (S4)}
        \label{fig:heatmap_rankings_s4}
    \end{subfigure}
    \caption{Heatmaps of the cumulative rank distributions for each of the four problem statements. Each heatmap shows the frequency (colour intensity) with which each workflow path (RA, RB, RC, RD) was assigned a given rank (1-4) by the 50 expert evaluators.}
    \label{fig:ranking_heatmaps}
\end{figure*}

\subsection{Quantifying Design Quality: An Inverse Plackett-Luce Model for Expert Rankings}
\label{subsec:plackett_luce_model}
The raw ranking data from the 50 experts was first aggregated to create a cumulative rank distribution for each of the four problem statements. These distributions, visualized as heatmaps in Figure~\ref{fig:ranking_heatmaps}, show the frequency with which each workflow path (RA, RB, RC, RD) was assigned a given rank (1-4). A general observation from the heatmaps is the consistent dominance of the yellow colour along Rank 1 for the fully-automated Path RD, suggesting it was most frequently chosen as the best design. However, beyond this top choice, the distribution of ranks for the other paths does not provide a clear, quantifiable differentiation of their perceived quality. A simple frequency count of ranks is insufficient to capture the nuanced preferences of the expert panel and does not yield a single, cardinal score of overall quality.

To move beyond this ordinal data and derive a more robust, quantitative measure of design quality, a more sophisticated statistical technique was required. For this purpose, we turned to the Plackett-Luce model, a well-established model for probabilistic ranking data. The model is grounded in a strong psychological principle: when a person ranks a set of choices, their decision is guided by an inherent, latent 'worthiness' or 'attractiveness' they assign to each choice. The probability of any given ranking can thus be modelled as a function of these underlying worth scores. The standard 'forward' Plackett-Luce model predicts ranking probabilities given a known set of worth scores. However, our research faced the 'inverse' problem: we have the observed rankings from the experts and needed to estimate the unknown, latent worthiness scores for each of the four design paths.

\begin{figure}[htbp]
    \centering
    \includegraphics[width=0.95\linewidth]{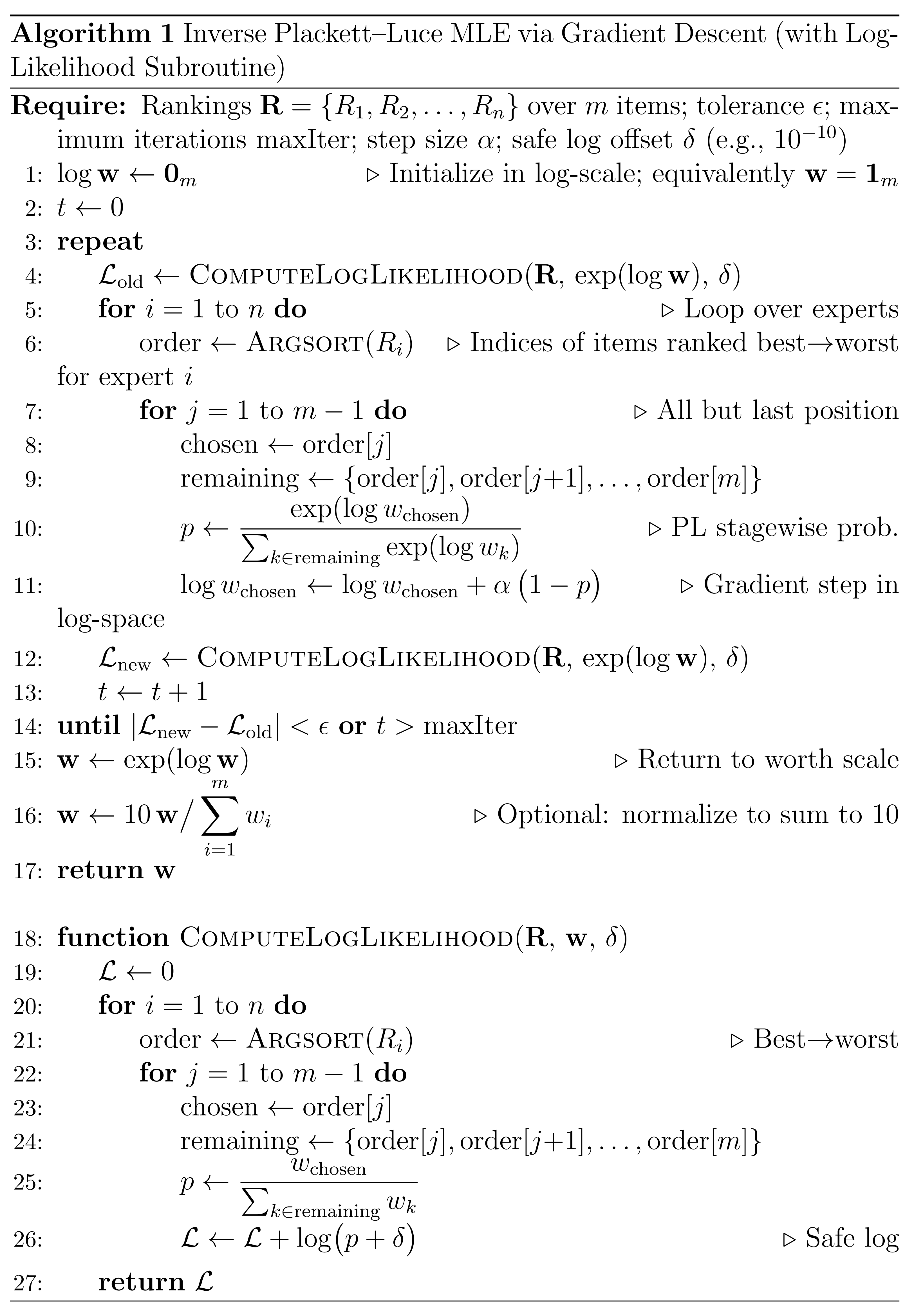}
    \caption{{The Inverse Plackett-Luce Maximum Likelihood Estimation (MLE) algorithm used to infer worthiness scores from a collection of rankings. This iterative process calculates worth scores (\textbf{w}) from a set of rankings (\textbf{R}) by maximizing the log-likelihood of observing those rankings.}}
    \label{fig:plackett_luce_algo}
\end{figure}

To solve this, we developed an optimization-based method to solve the inverse Plackett-Luce problem by calculating the Maximum Likelihood Estimation (MLE) of the worthiness scores. The objective is to find the set of worth scores that maximizes the log-likelihood of observing the actual rankings provided by the 50 experts. We employed a gradient descent optimization algorithm, as detailed in Algorithm~\ref{fig:plackett_luce_algo}, to iteratively adjust the worth scores to maximize this log-likelihood function. For numerical stability, the optimization is performed on the logarithm of the scores. The output of this algorithm is a 'worthiness score' for each of the four design concepts, normalized for each problem set. In the context of this study, this score represents the collective, latent preference of the expert panel. It is a powerful quantitative metric that captures not just which design was ranked first most often, but the overall strength of preference for each design relative to its competitors as perceived by the experts.

\subsection{Outcome Metrics}
\label{subsec:phase3_metrics}
The assessment of the experts of all the concepts presented to them was processed to gain insight into the quality of the final designs and the efforts associated with the different workflow paths.

\begin{itemize}
    \item \textbf{Time Efficiency:} The effort put in by the individual designers was measured in terms of the total time taken by them to complete the task for each of the four paths. The time for individual stages was measured using a stopwatch.

    \item \textbf{Design Quality: Worthiness Score:} To convert the ordinal ranking data from the experts into a quantitative measure of quality, the Inverse Plackett-Luce model, which is describe in the previous section was employed. Using this approach, we calculate a cardinal 'worthiness' score for each design by modelling the probability of the observed rankings using an gradient descent based optimization strategy.

    \item \textbf{Design Effectiveness:} To assess how well each design performed against the set of criteria, a 'Design Effectiveness' score was calculated from the 5-point Likert scale ratings. The mean expert ratings for the eight criteria were plotted on a spider chart for each rendering. The Design Effectiveness score is defined as the ratio of the area of the polygon formed by the design's actual mean ratings to the maximum possible area (i.e., if all criteria were rated a perfect '5'). This metric provides a holistic measure of a design's performance across all quality criteria; a score closer to 100\% indicates higher overall effectiveness.
\end{itemize}

\subsection{Outcomes}
\label{subsec:outcomes_phase3}
A detailed visual summary of the end-to-end design process and the outputs for each of the four problem statements in the Phase 3 study is shown below. Each figure corresponds to a specific design problem and visually contrasts the four experimental paths (RA, RB, RC, and RD) as executed by the assigned designers according to the Latin Square protocol. (a) Path RA shows the fully manual conventional process from a digital moodboard to a final rendering. (b) Path RB demonstrates a hybrid approach, using EUPHORIA for feature extraction to guide manual sketching. (c) Path RC utilizes EUPHORIA and the RETINA AI pipeline to generate reference sketches for a final manual concept. (d) Path RD showcases the end-to-end automated pipeline where the designer's role is to curate the final AI-generated rendering.

\subsubsection{Problem S1: Music Player}
\begin{figure*}[htbp]
    \centering
    \includegraphics[width=0.7\linewidth]{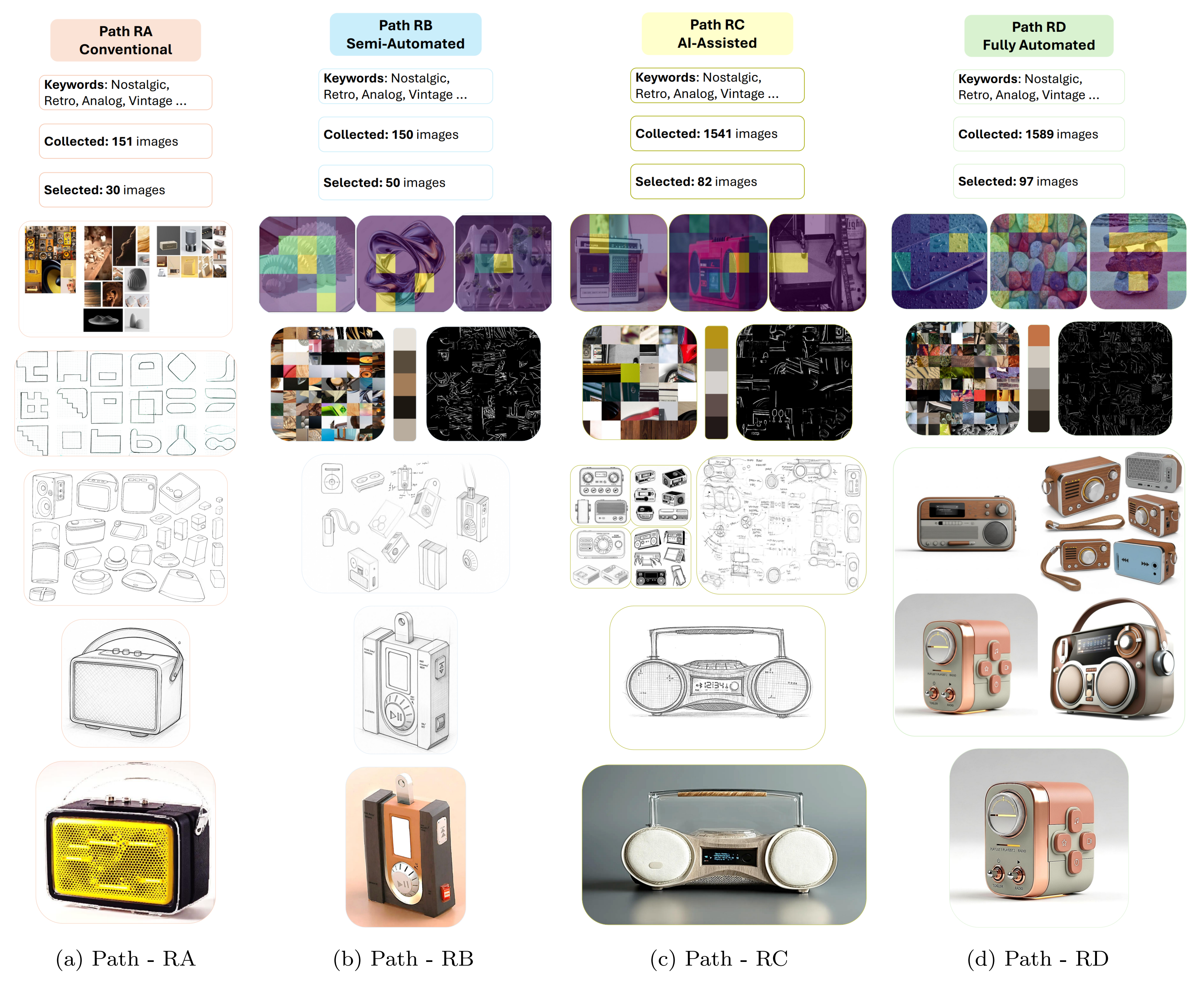}
    \caption{The complete design workflows and outputs for the four experimental paths (RA-RD) for the 'Portable Music Player' (S1) design problem. }
    \label{fig:music_player_workflow_four_paths}
\end{figure*}

The first design challenge (S1) required participants to design a Portable Music Player that fused retro (1960s-80s) and futuristic aesthetic keywords. Figure~\ref{fig:music_player_workflow_four_paths} provides a comprehensive visual summary of this task, documenting the end-to-end workflow and final outputs for each of the four experimental paths.

\subsubsection{Problem S2: Study Lamp}
\begin{figure*}[htbp]
    \centering
    \includegraphics[width=0.7\linewidth]{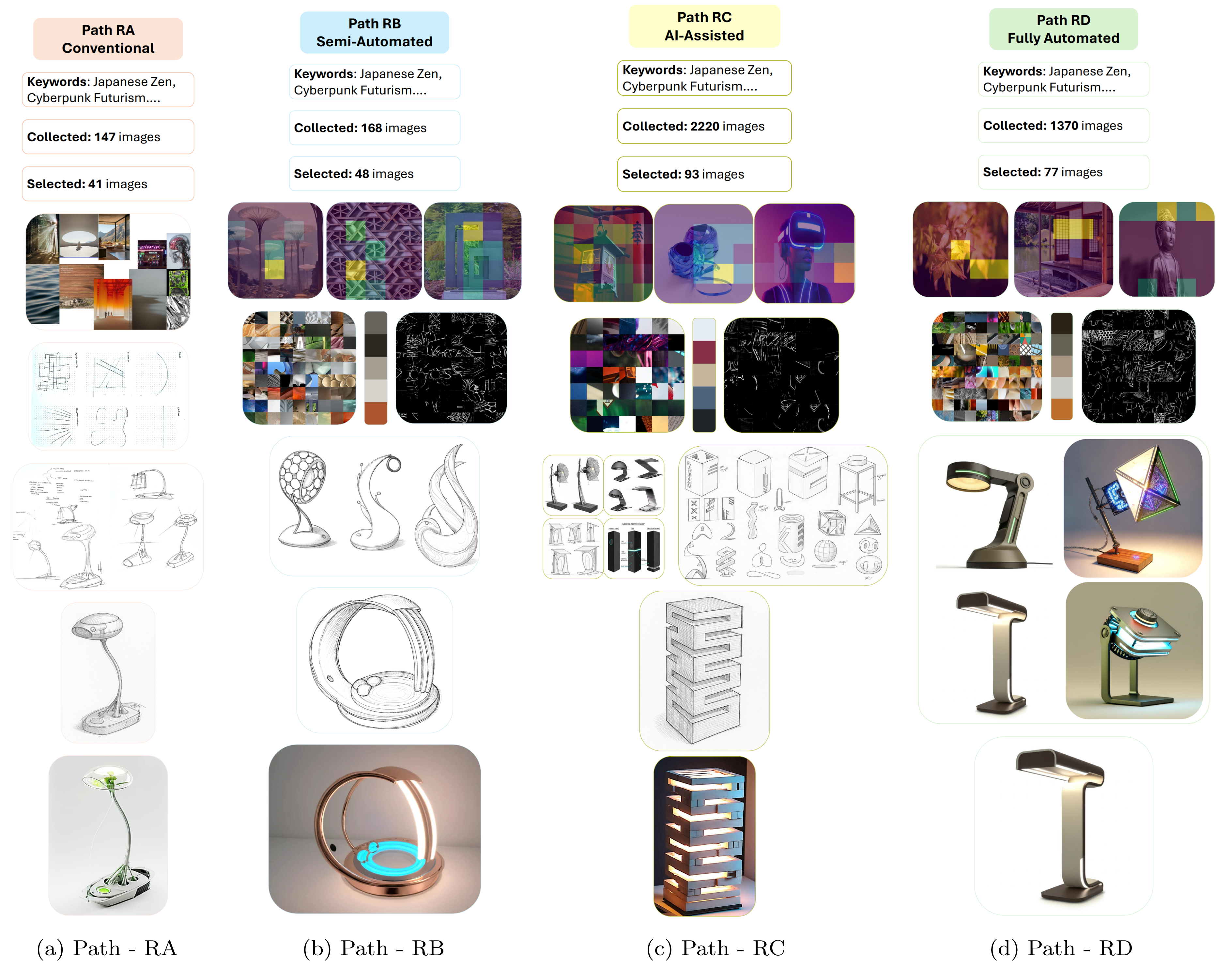}
    \caption{The complete design workflows and outputs for the four experimental paths (RA-RD) for the 'Study Lamp' (S2) design problem.}
    \label{fig:study_lamp_workflow_four_paths}
\end{figure*}

For the second problem statement (S2), designers were tasked with creating a Study Lamp that reconciled the contradictory styles of Japanese Zen and cyberpunk futurism. The complete procedural flow and resulting designs for this challenge are detailed in Figure~\ref{fig:study_lamp_workflow_four_paths}.

\subsubsection{Problem S3: Flash Light}
\begin{figure*}[htbp]
    \centering
    \includegraphics[width=0.7\linewidth]{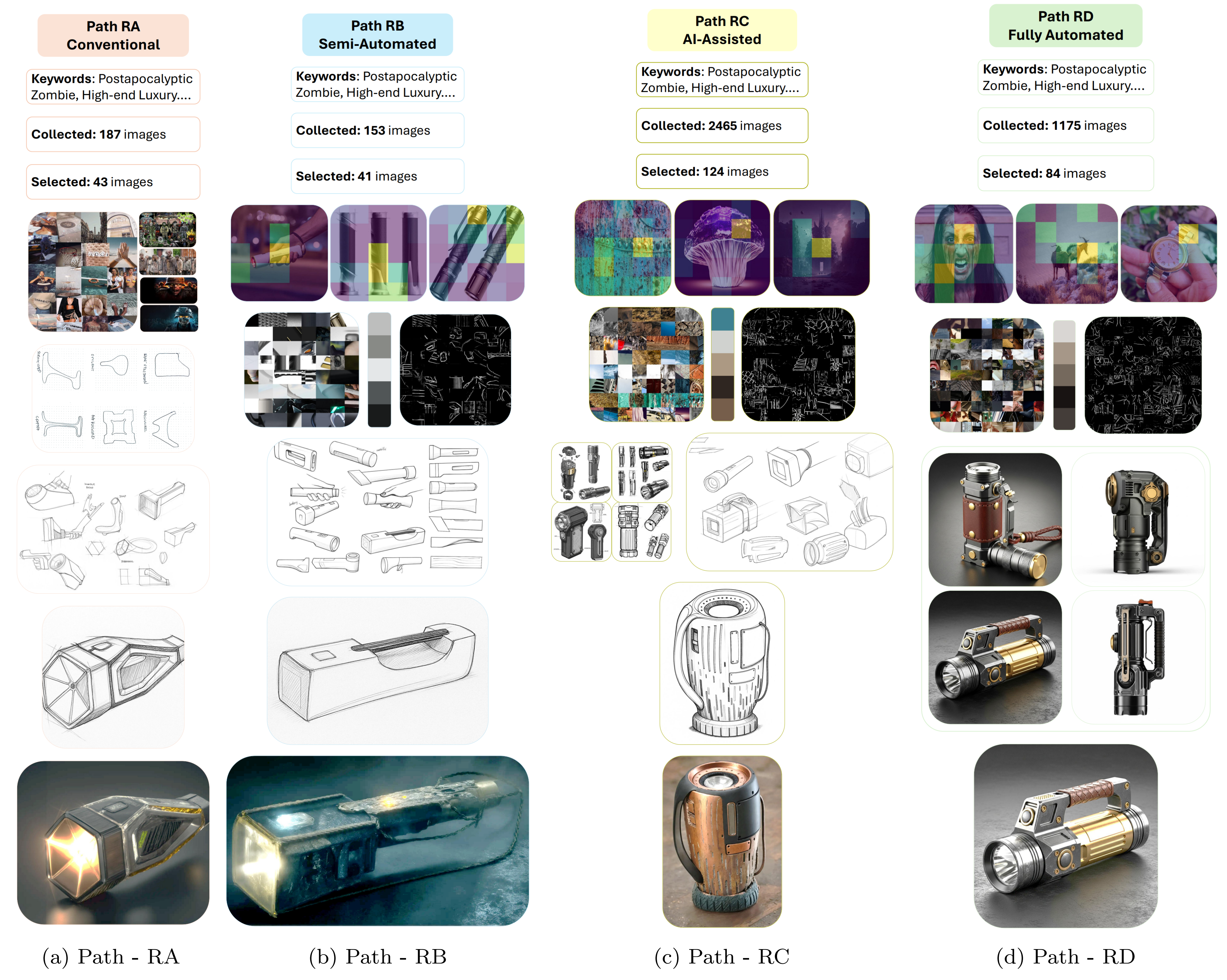}
    \caption{The complete design workflows and outputs for the four experimental paths (RA-RD) for the 'Handheld Flashlight' (S3) design problem.}
    \label{fig:flash_light_workflow_four_paths}
\end{figure*}

The third design task (S3) challenged participants to design a Handheld Flashlight combining post-apocalyptic ruggedness with the elegance of a luxury accessory. Figure~\ref{fig:flash_light_workflow_four_paths} documents the tangible outcomes of this task.

\subsubsection{Problem S4: Candle Holder}
\begin{figure*}[htbp]
    \centering
    \includegraphics[width=0.7\linewidth]{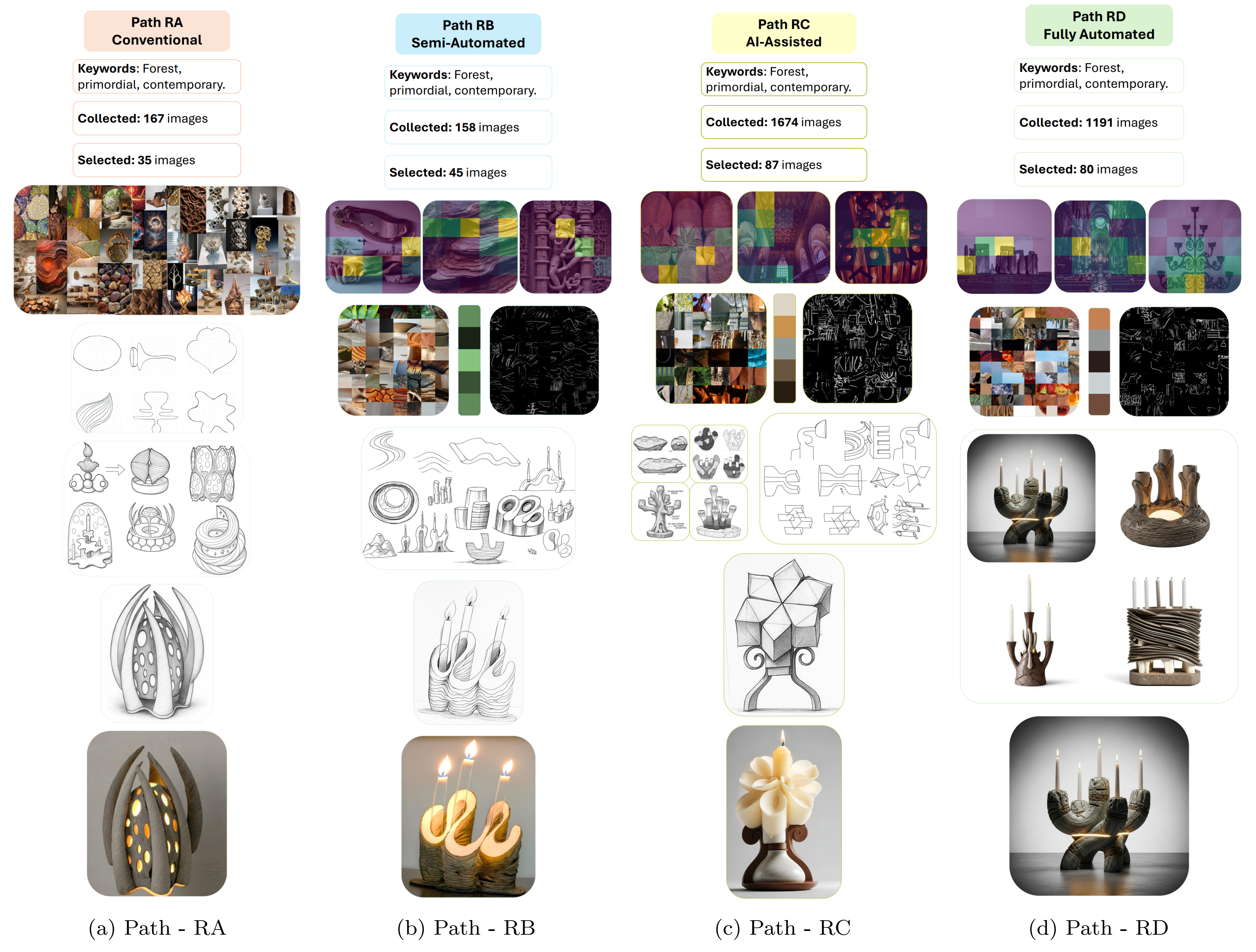}
    \caption{The complete design workflows and outputs for the four experimental paths (RA-RD) for the 'Candlestick Holder' (S4) design problem.}
    \label{fig:candle_holder_workflow_four_paths}
\end{figure*}

The final problem statement (S4) involved the design of a Candlestick Holder inspired by prehistoric natural forms for a contemporary setting. The visual record of the design process and outcomes for this task is presented in Figure~\ref{fig:candle_holder_workflow_four_paths}.


\subsection{Results}
\label{subsec:phase3_results}
\begin{figure*}[!ht]
    \centering
    \includegraphics[width=0.9\linewidth]{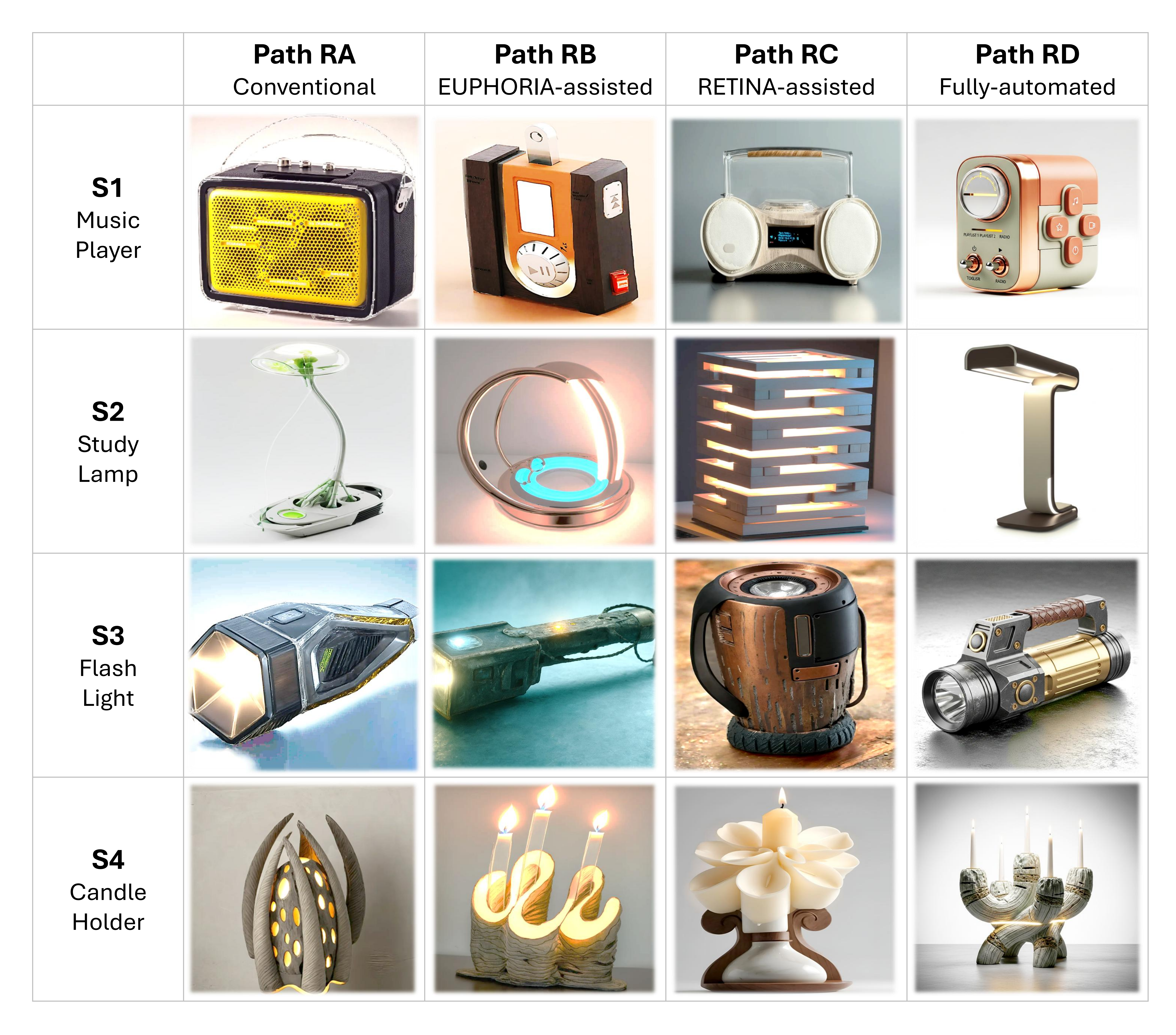}
    \caption{Complete matrix of all 16 final product renderings from the Phase 3 study. The figure presents the final design outputs in a 4x4 grid, where each row corresponds to one of the four problem statements (S1-S4) and each column corresponds to one of the four experimental workflow paths (Path RA-RD). Each cell displays the final rendering created by the assigned designer (D1-D4) for that specific task, in accordance with the Latin Square experimental design.}
    \label{fig:all_renderings_matrix}
\end{figure*}

The primary data from the Phase 3 study consisted of the time taken to complete each task and the 16 final product renderings (Figure~\ref{fig:all_renderings_matrix}). The analysis of this data was structured around the three core outcome metrics: Time Efficiency, Worthiness Score, and Design Effectiveness.

\subsubsection{Time Efficiency}
A primary objective of the study was to quantify the impact of the different workflows on the time required to progress from a problem statement to a final rendered concept. The average timings for each stage across all four problem statements are presented in Table~\ref{tab:study_times}.

\begin{table*}[!ht]
\centering
\caption{Approximate Average Time Taken (in HH:MM:SS) Across All Four Problem Statements for Each Workflow Path}
\label{tab:study_times}
\renewcommand{\arraystretch}{1.4}
\begin{tabular}{
    >{\raggedright\arraybackslash}p{6.0cm}
    >{\centering\arraybackslash}p{2.5cm}
    >{\centering\arraybackslash}p{2.5cm}
    >{\centering\arraybackslash}p{2.5cm}
    >{\centering\arraybackslash}p{2.5cm}
}
\toprule
\textbf{Stage / Sub-Process} & \textbf{Path RA \small{(Conventional)}} & \textbf{Path RB \small{(EUPHORIA-Assisted)}} & \textbf{Path RC \small{(RETINA-Assisted)}} & \textbf{Path RD \small{(Fully Automated)}} \\
\midrule
\textbf{Keyword Identification} & 00:04:30 & 00:04:30 & 00:05:15 & 00:06:30 \\
\midrule
\multicolumn{5}{l}{\textit{\textbf{Moodboarding Stages}}} \\
\quad Collection \& Selection & 01:32:30 & 01:22:30 & 00:15:00 & 00:15:00 \\
\quad Composition \& Reflection & 00:45:00 & 00:15:00 & 00:15:00 & 00:15:00 \\
\quad Extraction & 00:21:30 & 00:03:00 & 00:03:00 & 00:03:00 \\
\midrule
\multicolumn{5}{l}{\textit{\textbf{Sketching \& Ideation Stages}}} \\
\quad Shape Abstraction & 00:15:00 & 00:02:00 & 00:02:00 & 00:02:00 \\
\quad Rough Sketch Exploration & 00:40:00 & 00:38:15 & 00:05:00 & 00:00:00 \\
\quad Concept Sketch Creation & 00:35:15 & 00:32:15 & 00:27:30 & 00:05:00 \\
\midrule
\textbf{Rendering} & 00:09:00 & 00:08:30 & 00:08:15 & 00:05:00 \\
\midrule[\heavyrulewidth]
\textbf{Total Time} & \textbf{04:22:45} & \textbf{03:06:00} & \textbf{01:21:00} & \textbf{00:51:30} \\
\bottomrule
\end{tabular}
\end{table*}

The conventional Path RA, by a significant margin, was the most time-consuming, with an average total time of 04:22:45. Path RB, which automated the latter half of the moodboarding process, offered an improvement by reducing the average time to 03:06:00. A substantial leap in efficiency was observed with Path RC; the fully automated moodboarding and AI-assisted sketching brought the average time down to just 01:21:00. Finally, the fully automated Path RD was the most efficient by far, requiring an average of only 00:51:30 for the designer to arrive at a final, curated design. This represents a more than fourfold reduction in time from the conventional process to the end-to-end EUPHORIA-RETINA pipeline, confirming that the proposed system offers a vastly more time-efficient approach to form design. The results reveal a consistent reduction in the time required to complete the design task as the level of automation and computer assistance increases. 

\subsubsection{Worthiness Score (Based on Expert Rankings)}
\begin{figure*}[htbp]
    \centering
    \begin{subfigure}[b]{0.24\textwidth}
        \centering
        \includegraphics[width=\textwidth]{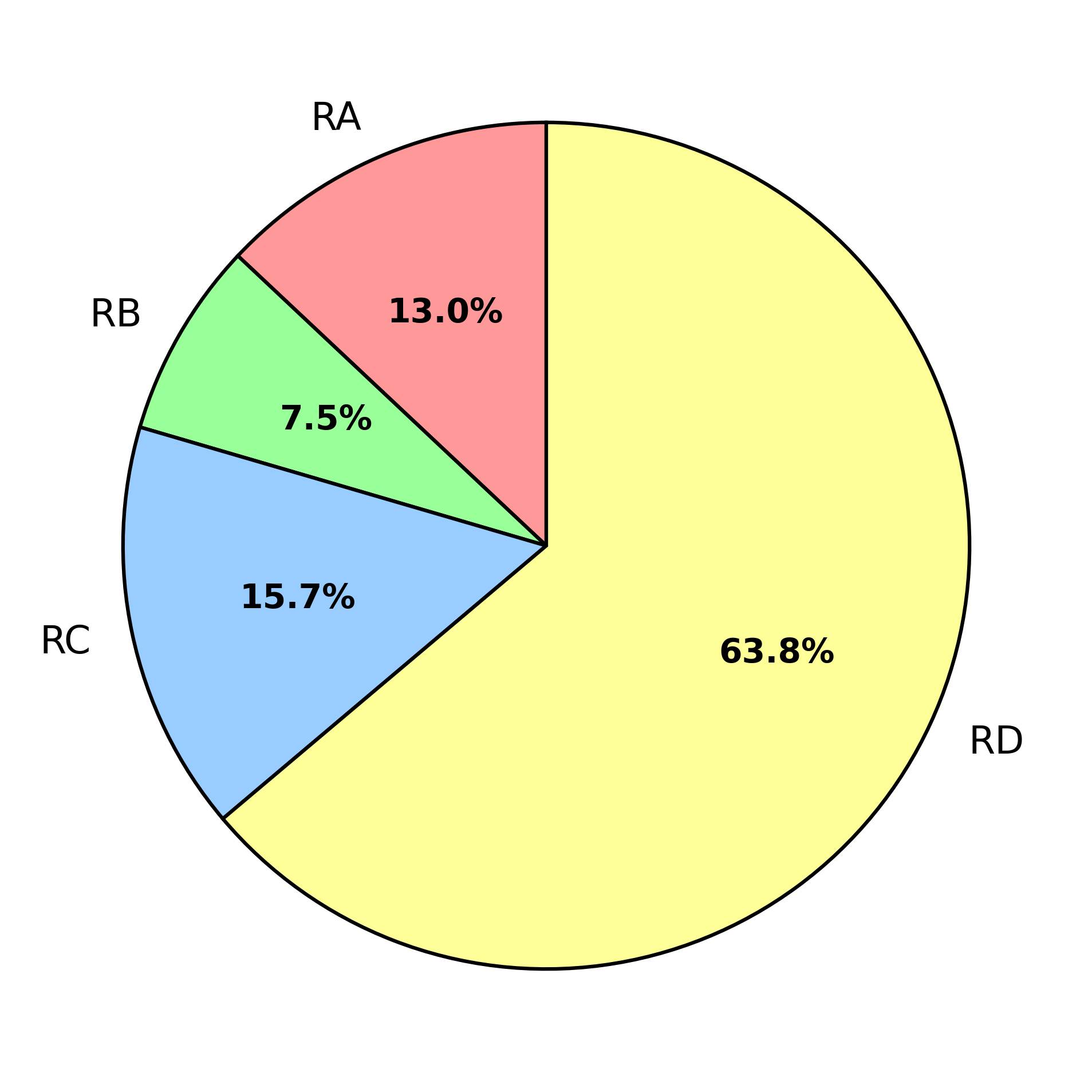}
        \caption{Music Player}
        \label{fig:worth_music_player}
    \end{subfigure}
    \hfill 
    \begin{subfigure}[b]{0.24\textwidth}
        \centering
        \includegraphics[width=\textwidth]{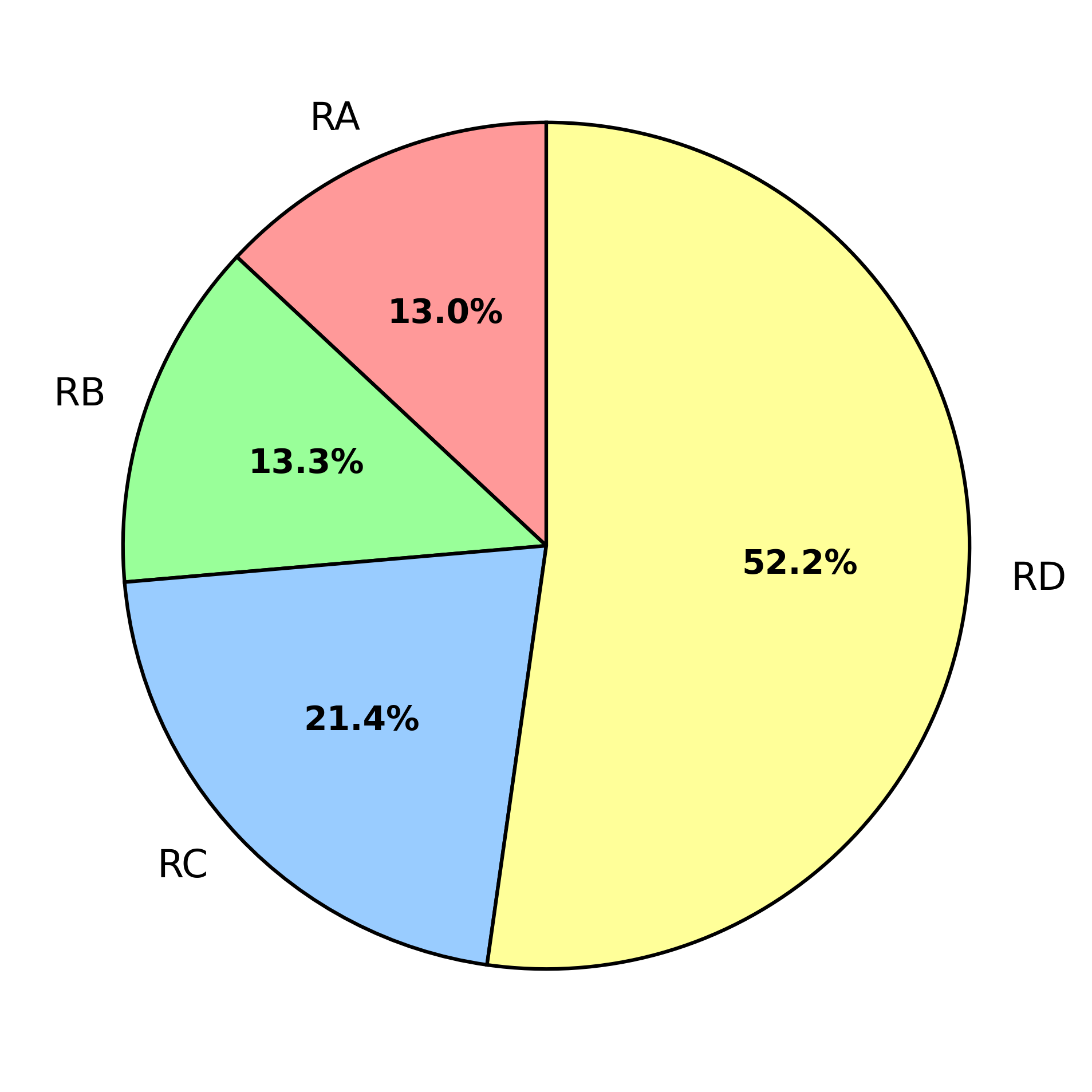}
        \caption{Study Lamp}
        \label{fig:worth_study_lamp}
    \end{subfigure}
    \hfill
    \begin{subfigure}[b]{0.24\textwidth}
        \centering
        \includegraphics[width=\textwidth]{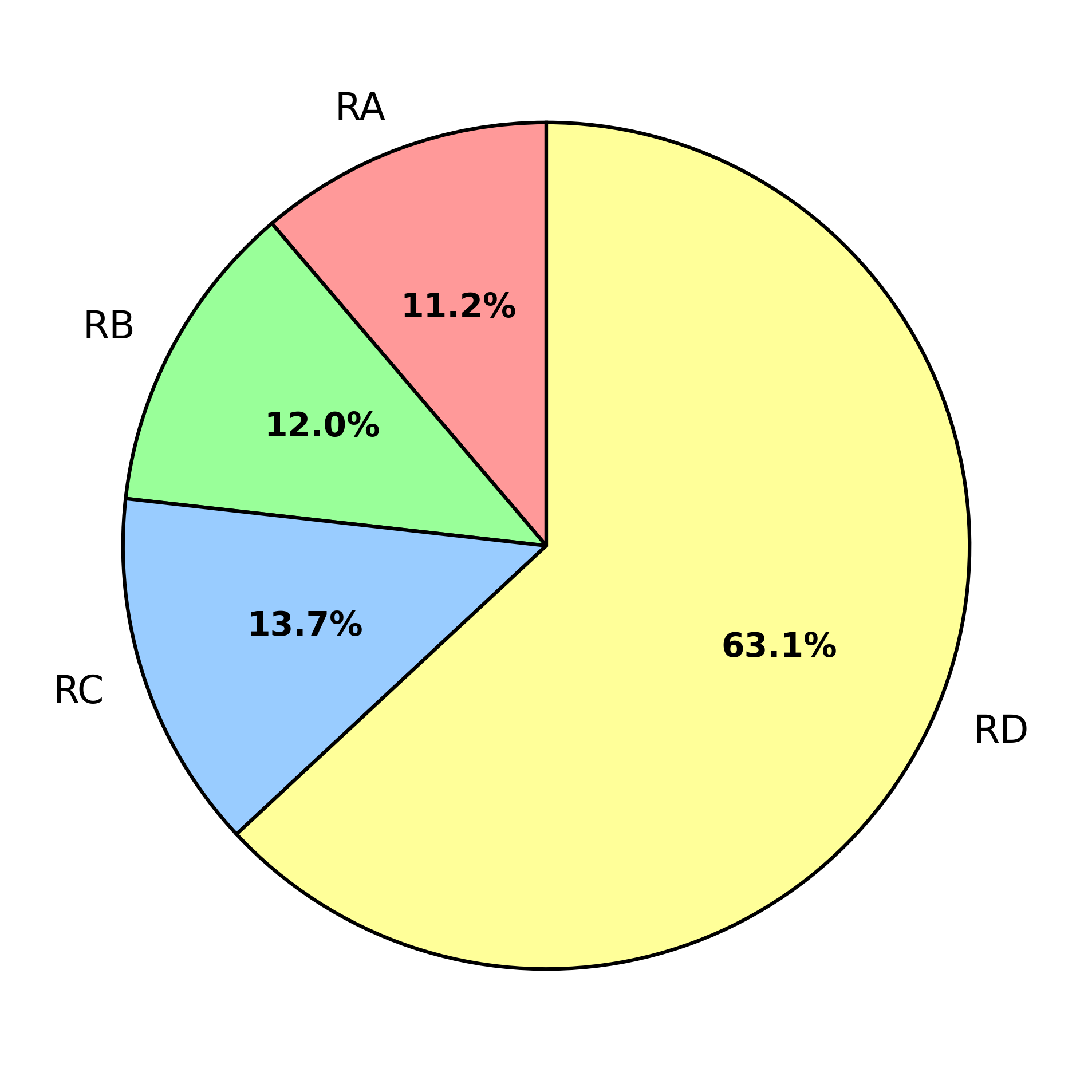}
        \caption{Flashlight}
        \label{fig:worth_flashlight}
    \end{subfigure}
    \hfill 
    \begin{subfigure}[b]{0.24\textwidth}
        \centering
        \includegraphics[width=\textwidth]{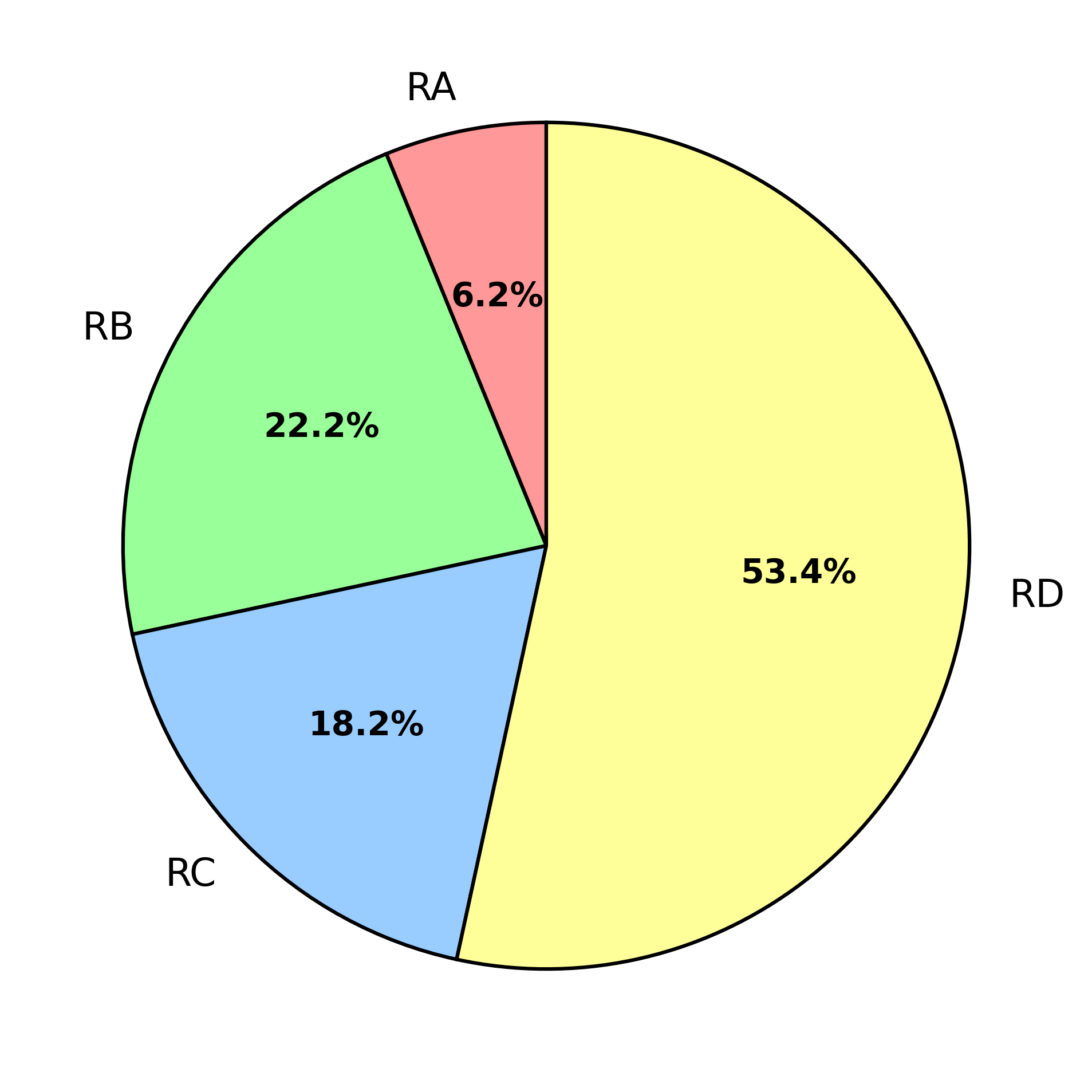}
        \caption{Candle Holder}
        \label{fig:worth_candle}
    \end{subfigure}
    \caption{Plackett-Luce worthiness scores for the four workflow path concepts (RA, RB, RC, RD) across each of the four product design problems. The percentage in each slice represents the calculated worth of the designs as perceived by the raters.}
    \label{fig:worthiness_scores_all}
\end{figure*}

To assess the overall perceived quality of the designs from an expert's comparative viewpoint, the ranking data were converted into cardinal Worthiness Scores using the Plackett-Luce model (Algorithm~\ref{fig:plackett_luce_algo}). These scores represent the collective preference of the expert panel. Figure~\ref{fig:worthiness_scores_all} presents the calculated worthiness scores for the four paths for each of the four problem statements.

The designs generated by the fully automated Path RD (EUPHORIA-RETINA pipeline) were consistently perceived as having the highest worth by the expert panel. For all the problems, the concept generated from Path RD captured over 50\% of the total worth of the experts. Across all problems, the Path RD designs were the unambiguous successor. In contrast, the worthiness scores for concepts generated by Paths RA, RB, and RC were clustered together with no significant differences between them, indicating that the experts perceived their overall quality as comparable, despite the differences in their underlying processes.

\subsubsection{Design Effectiveness (Based on Expert Ratings)}
\begin{figure}[htbp]
    \centering
    \includegraphics[width=\linewidth]{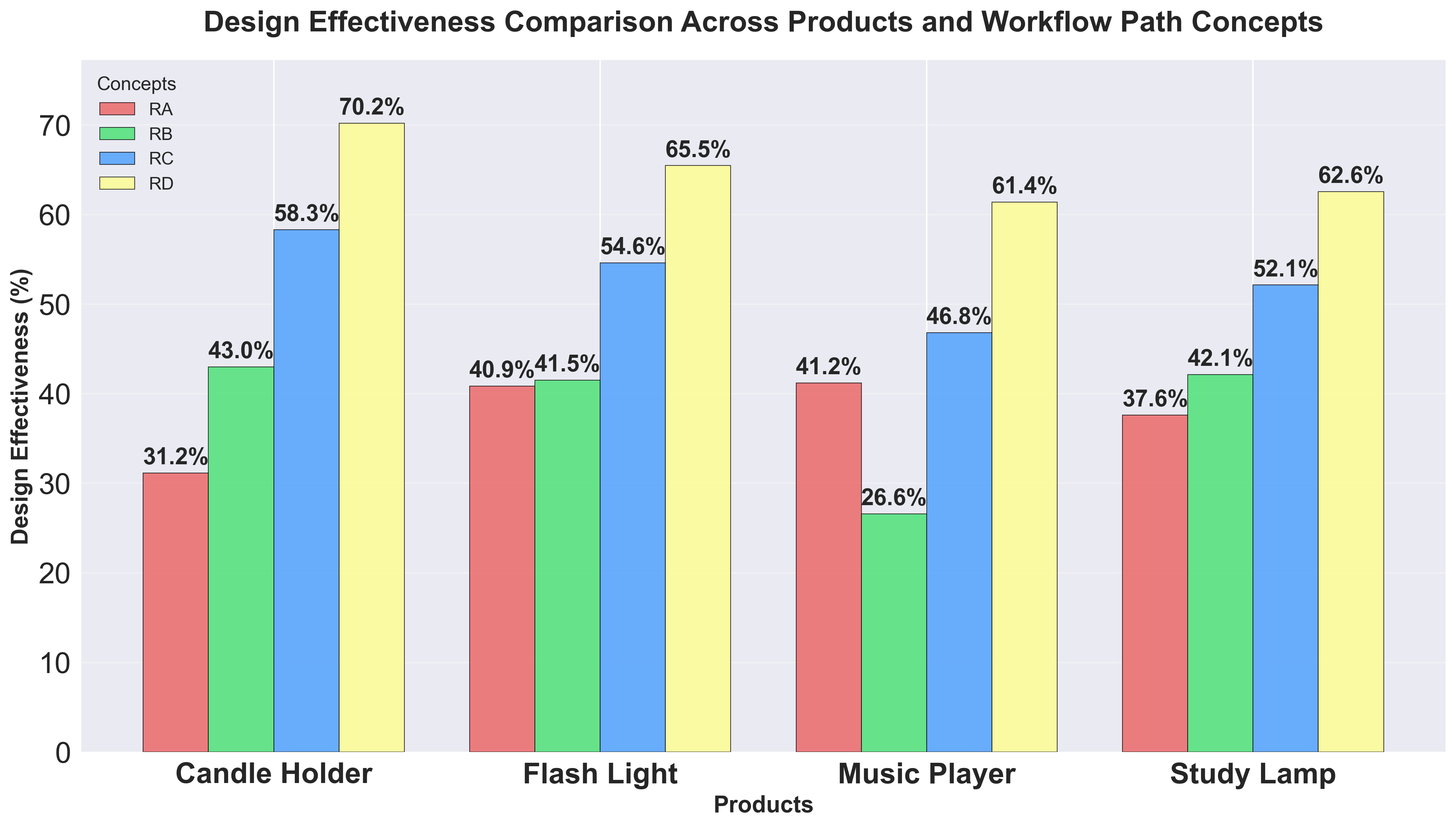}
    \caption{Comparison of design effectiveness scores across the four products (Candle Holder, Flashlight, Music Player, Study Lamp) and the four workflow path concepts (RA, RB, RC, RD). The bars represent the percentage of design effectiveness, showing how each concept performed for each product.}
    \label{fig:design_effectiveness_comparison}
\end{figure}

\begin{figure*}[!ht]
    \centering
    \begin{subfigure}[b]{0.48\textwidth}
        \centering
        \includegraphics[width=\textwidth]{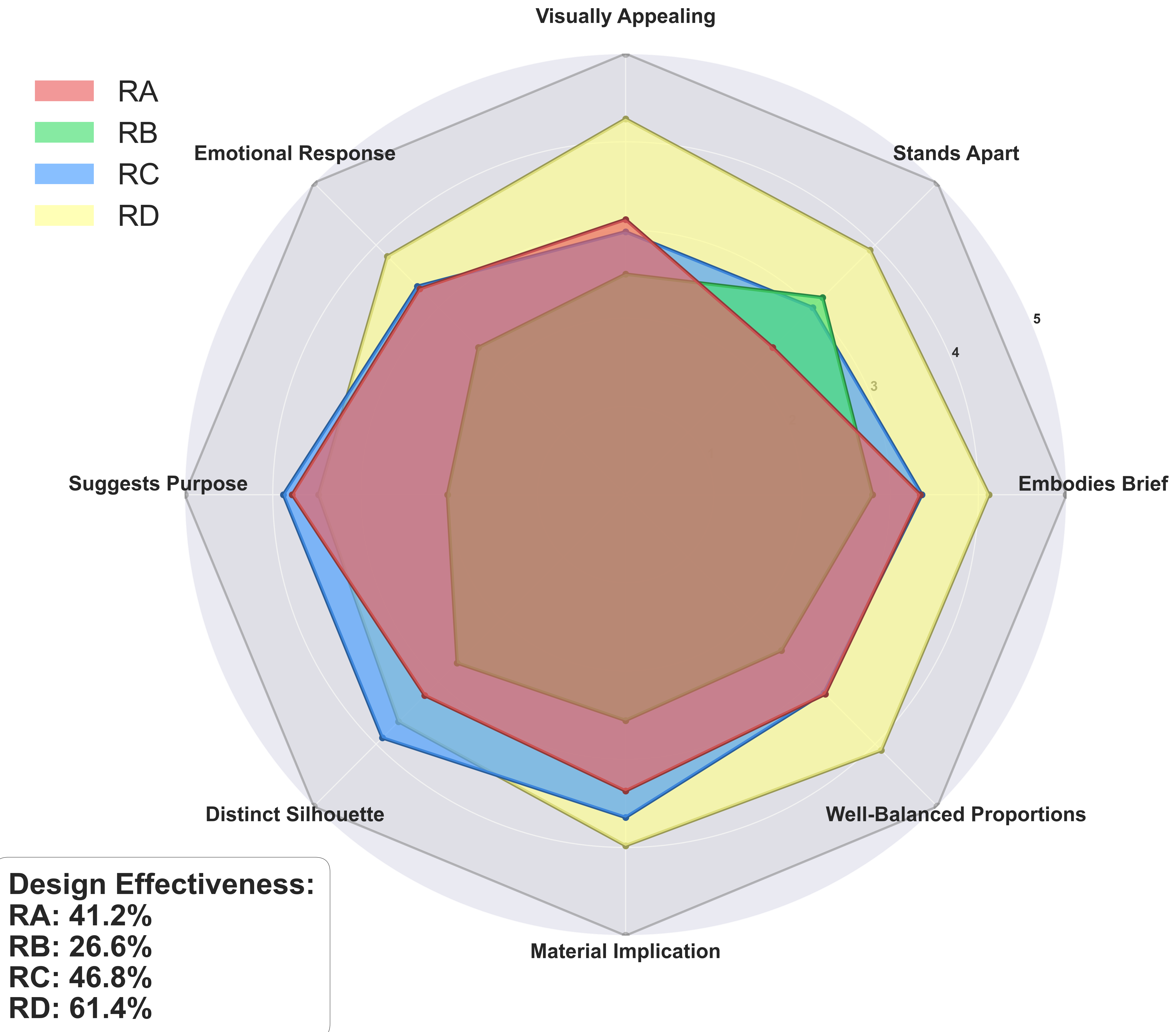}
        \caption{Music Player}
        \label{fig:spider_music_player}
    \end{subfigure}
    \hfill 
    \begin{subfigure}[b]{0.48\textwidth}
        \centering
        \includegraphics[width=\textwidth]{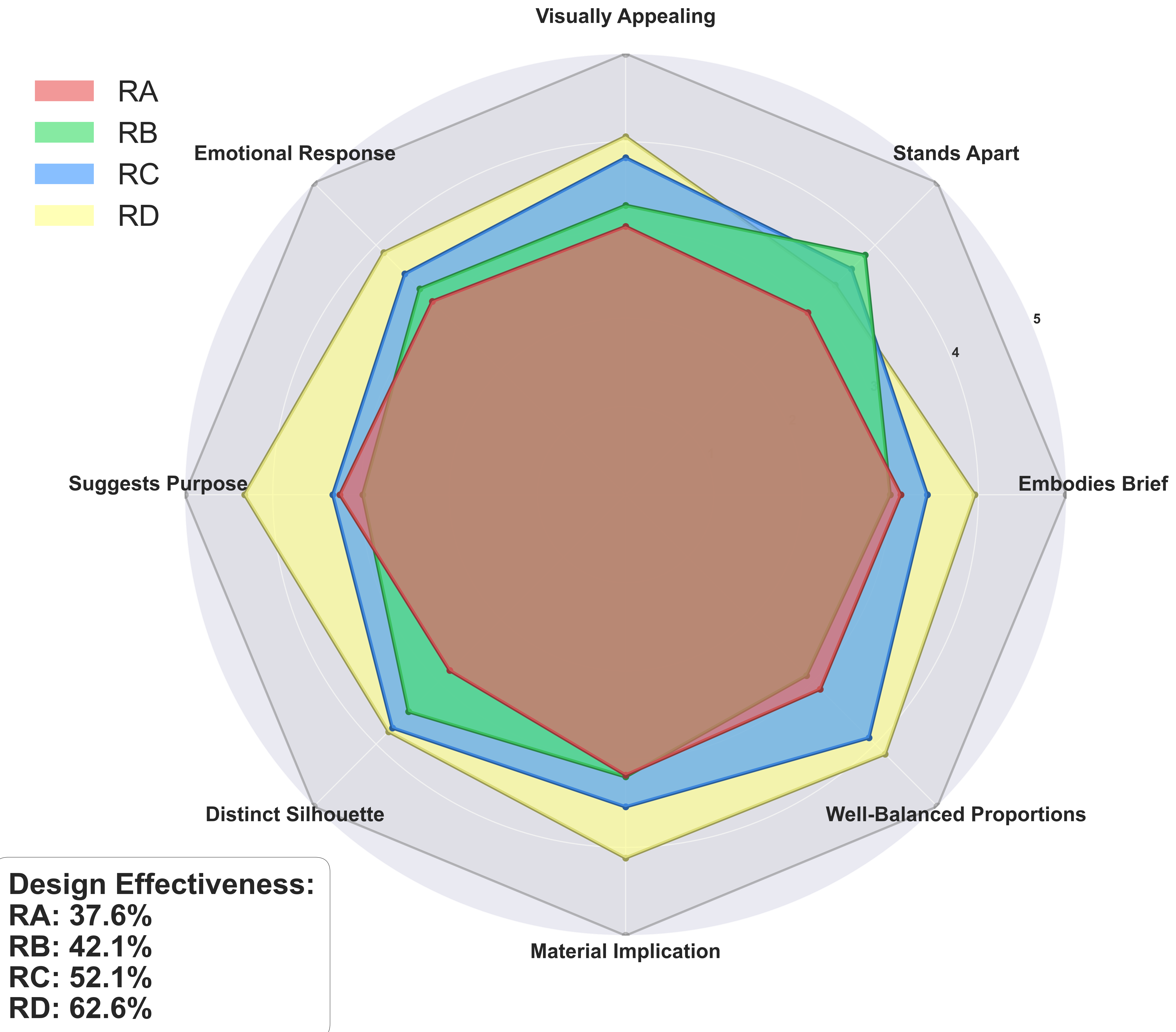}
        \caption{Study Lamp}
        \label{fig:spider_study_lamp}
    \end{subfigure}
    \vfill 
    \begin{subfigure}[b]{0.48\textwidth}
        \centering
        \includegraphics[width=\textwidth]{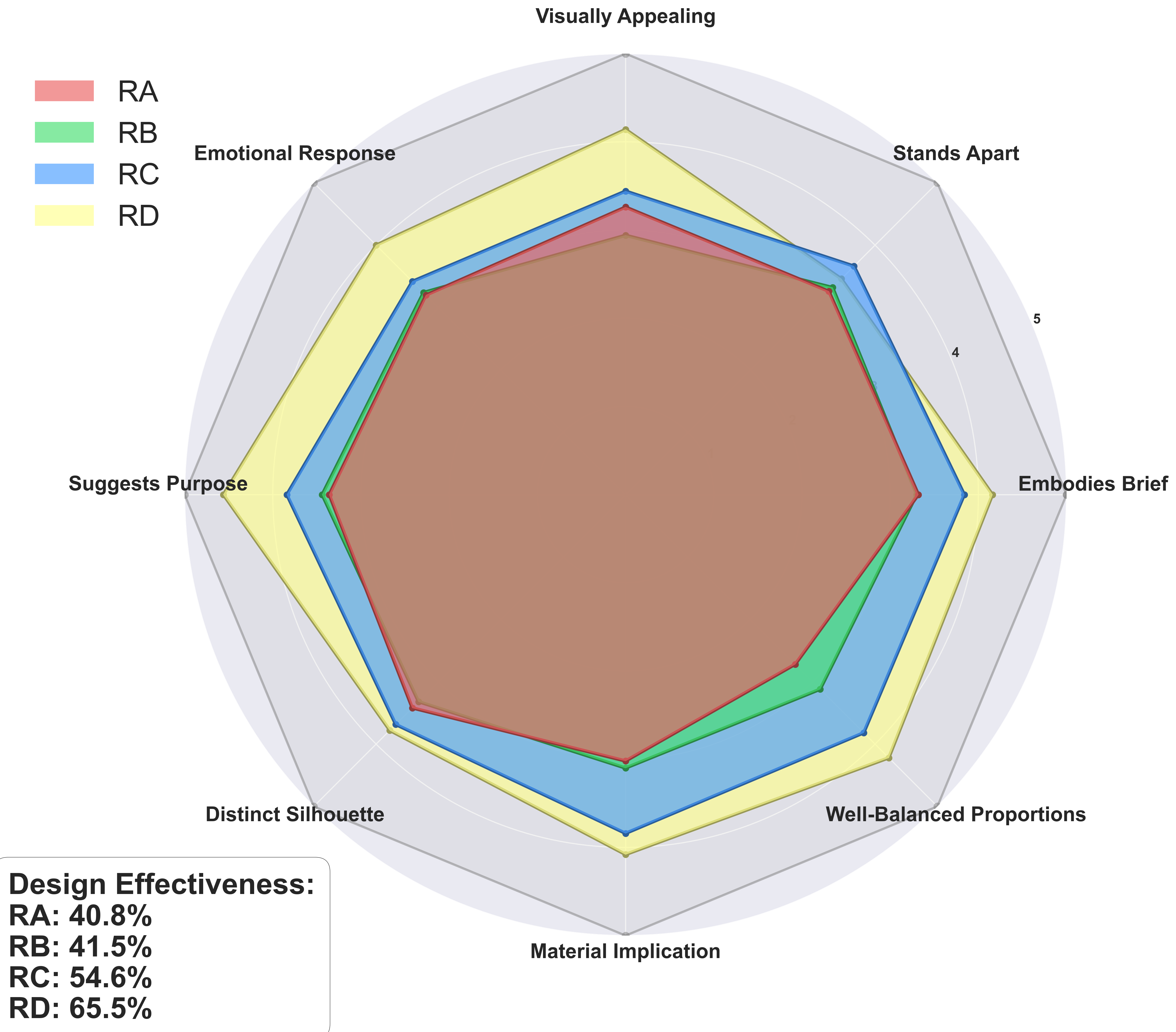}
        \caption{Flash Light}
        \label{fig:spider_flashlight}
    \end{subfigure}
    \hfill 
    \begin{subfigure}[b]{0.48\textwidth}
        \centering
        \includegraphics[width=\textwidth]{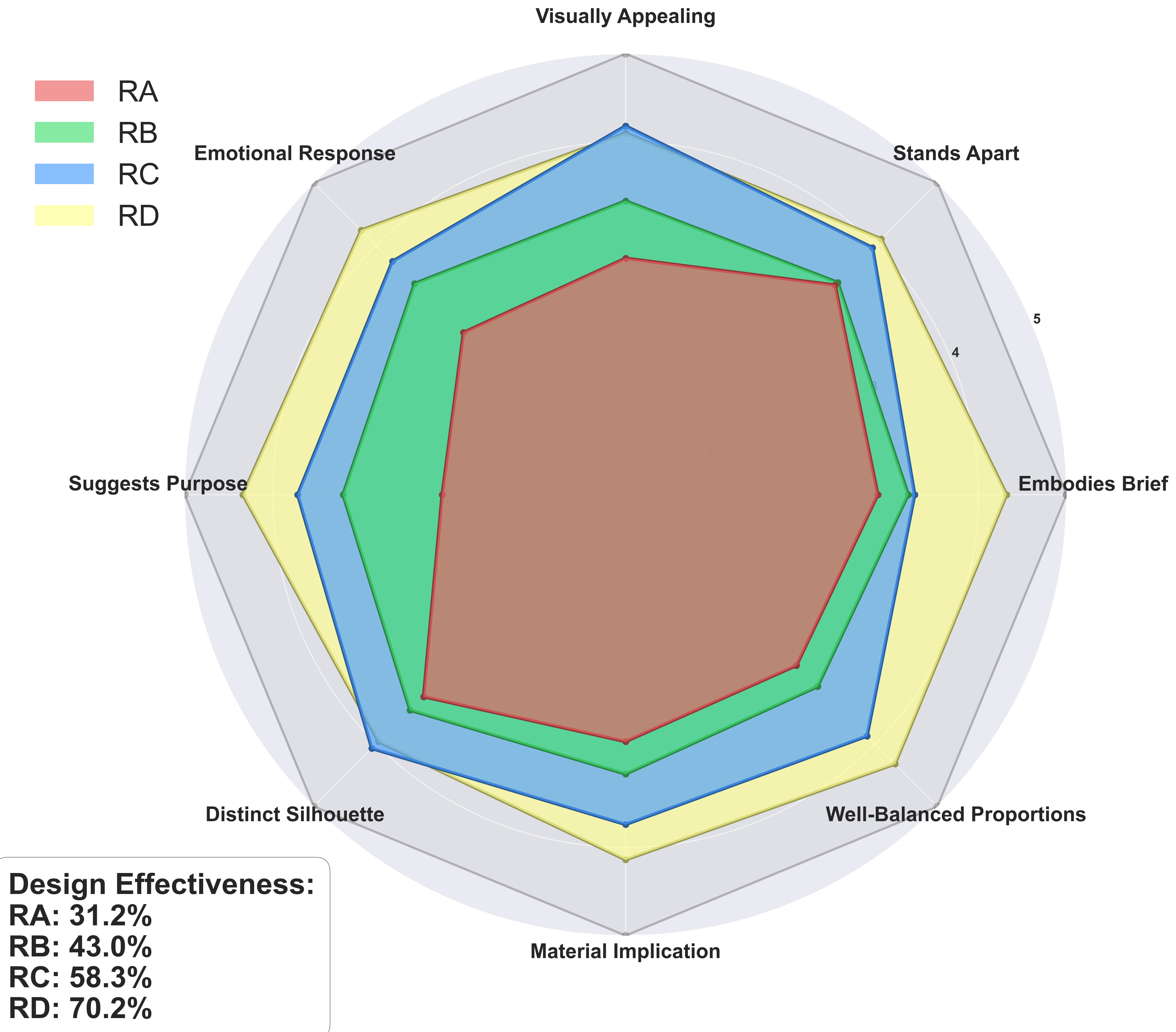}
        \caption{Candle Holder}
        \label{fig:spider_candle}
    \end{subfigure}
    \caption{Cumulative mean ratings from expert designers for the four design problems. Each spider plot shows the evaluation across eight criteria for the four different workflow paths (RA, RB, RC, RD).}
    \label{fig:spider_plots_all}
\end{figure*}

The expert panel rated each of the 16 renderings against eight specific criteria. The mean ratings for each design are visualized in the spider plots in Figure~\ref{fig:spider_plots_all}, and the final calculated Design Effectiveness scores are compared in the bar chart in Figure~\ref{fig:design_effectiveness_comparison}.

The results for Design Effectiveness mirror and reinforce the findings from the worthiness scores. The designs from Paths RC and RD consistently and significantly outperformed those from Paths RA and RB. As shown in Figure~\ref{fig:design_effectiveness_comparison}, the designs of RD (fully-automated) achieved the highest effectiveness scores across all four problem statements, followed closely by Path RC (RETINA-Assisted). The spider plots in Figure~\ref{fig:spider_plots_all} provide a visual explanation for this result; the polygons for Paths RC and RD are visibly larger and more complete close to the idea, indicating higher average ratings across nearly all criteria. The scores for the conventional Path RA and the semi-automated Path RB were again comparable to each other and markedly lower than those for the more automated paths.

\subsection{Discussion}
\label{subsec:phase3_discussion}
The results from the study provide strong evidence that the integrated EUPHORIA-RETINA system not only accelerates the form design process but also leads to form design outcomes that are judged by experts to be of a higher quality. The interpretation of these findings suggests a potential paradigm shift in workflow for creative design.

\subsubsection{Interpreting the Efficiency and Quality Outcomes}
The dramatic four-fold reduction in time is a clear indicator of the system's efficiency. However, this is not merely a matter of speed. The time saved in Paths RC and RD was primarily from the elimination of logistical and manual tasks (e.g., hours of image searching, manual sketching iterations) that dominate Path RA. This reallocated the designer's limited cognitive resources away from laborious execution and towards higher-level creative thinking and aesthetic curation.

The superior performance of Path RD in all the outcome metrics suggests that the end-to-end agentic pipeline was more successful at the "creative leap", translating abstract features into a coherent and compelling form, than the human designers were in Paths RA and RB. A key reason for this may be the inspiration source; designers in Paths RC and RD were exposed to a much larger set of visual stimuli within EUPHORIA (averaging over 1200 images viewed) compared to the manually collected sets (typically around 100 images). The Agentic AI, therefore, had a richer, more diverse set of implicitly-selected features from which to generate concepts.

The relatively stagnant performance of Path RB compared to Path RA suggests that even when a designer is provided with explicit, data-driven feature maps from EUPHORIA, the manual and highly subjective task of synthesizing these features into a final form remains a significant bottleneck. This indicates that simply assisting with feature extraction is not enough; true improvement requires support during the generative synthesis stage as well.

\subsubsection{From Computer-Assisted Design (CAD) to Designer-Assisting Computers (DAC)}
The comparative results of the four paths illustrate a fundamental paradigm shift, such as Paths RA and RB represent a model of \textbf{Computer-Assisted Design (CAD)}, where the computer is a passive tool that the designer uses to execute their own manually-driven ideas. The designer is the sole source of creative synthesis.

Paths RC and RD, however, exemplify a new paradigm of \textbf{Designer-Assisting Computers (DAC)}. In this model, the computer is an active, agentic partner in the creative process. Designer assists the EUPHORIA by providing their implicit intent, and RETINA proactively generates novel solutions based on that intent. The designer's role evolves from that of a manual creator and labourer to a high-level creative director and strategic curator. They guide the system's inspiration, and then apply their expertise and taste to select and refine the AI's output. This DAC paradigm does not replace the designer but elevates their role, freeing them to focus on the most essential aspects of creative decision-making. The superior results of Path RD suggest that this collaborative DAC model, where the designer's implicit taste guides a powerful generative engine, will lead to better designs than the traditional CAD model.

\greyline
\section{Summary of Findings}
\label{subsec:summary}
This research addressed the persistent challenges in conventional product form design—a process often hindered by its time-consuming nature, reliance on subjective designer experience, and the opaque, internal 'creative leap' required to translate inspiration into form. To overcome these limitations, this study introduced and validated a novel, attention-aware workflow. This workflow is facilitated by two integrated systems: {EUPHORIA}, an immersive VR-based system for capturing a designer's implicit visual preferences, and {RETINA}, an agentic AI framework for translating those preferences into tangible design concepts.

The research began by establishing the foundational principles of this approach through a systematic validation process. The first study confirmed a statistically significant positive correlation ($r = 0.3819$, $p < 0.0001$) between a user's implicit visual attention (measured by gaze duration) and their explicit preference for an image. It also validated that unguided preferences are highly individualistic. This demonstrated that eye-tracking within an immersive moodspace is a viable method for objectively capturing subjective aesthetic taste. The following study demonstrated that this attentional process could be effectively guided. By pre-conditioning participants with emotional stimuli, their visual selections and qualitative thoughts were shown to converge, confirming that the system can be used for goal-directed inspiration gathering.

The main contribution of this work was presented by conducting a comparative study of four distinct design workflows on four challenging design problems. The results were conclusive across three key metrics. Firstly, the integrated EUPHORIA-RETINA pipeline was found to be over four times more time-efficient than the conventional manual process (Path RA), reducing the average design time from over four hours to under one hour. Secondly, expert evaluation of the form renderings revealed that the designs from the fully automated Path RD were consistently ranked as having the highest Worthiness Score. Finally, these designs also achieved the highest Design Effectiveness scores, indicating superior performance across eight specific design criteria, including novelty, emotional resonance, and visual appeal.

In essence, this research successfully demonstrated a potential paradigm shift from traditional Computer-Assisted Design (CAD), where the computer is a passive tool, to a more collaborative active model of Designer-Assisting Computers (DAC). The proposed workflow not only streamlines the form design process but can also lead to objectively higher-quality creative outputs by synergizing the designer's implicit taste with the generative power of agentic AI computing systems.

\subsection{Limitations and Future Work}
\label{subsec:limitations}
Despite the promising results, this study has a few limitations that provide clear directions for future research.

\begin{itemize}
    \item \textbf{Sample Size and Diversity:} The comparative study, while structurally robust due to its Latin Square design, was conducted with four designers. Although their expertise was varied, this small sample size limits the statistical generalizability of the findings. Future studies should involve a larger and more culturally diverse cohort of designers to validate these results on a broader scale.

    \item \textbf{Scope of Design Problems:} The four design problems were intentionally challenging but were limited to relatively simple consumer products. The efficacy of the EUPHORIA-RETINA workflow for more complex systems with stringent engineering, ergonomic, or systemic constraints (e.g., medical devices, automotive interiors, service design) has not yet been tested. Future work should apply this agentic pipeline to more complex design challenges to explore its scalability.

    \item \textbf{Longitudinal Effects:} The study was conducted in single, isolated sessions. The long-term effects of using a Designer-Assisting Computer (DAC) system on a designer's creative skills, cognitive processes, and potential over-reliance on AI are unknown. A longitudinal study, where designers use the system over an extended period, would be helpful for understanding its impact on the evolution of their creative practice.

    \item \textbf{Evaluation Scope:} While expert evaluation is a standard in design research, it remains an inherently subjective measure of quality. The study concluded with digital renderings and did not proceed to physical prototyping. Future evaluations could be powerfully supplemented by user testing of physical prototypes derived from the AI-generated forms, which would add a layer of validation regarding their real-world appeal, usability, and emotional impact.
\end{itemize}

\greyline
\section{Conclusion}
\label{sec:conclusion}
This research set out to address the fundamental challenges that have long defined the practice of product form design: its significant time investment, its deep reliance on the subjective and often opaque expertise of the designer, and the inherent difficulty in translating abstract inspiration into tangible, high-quality forms. The conventional workflow, while creatively valuable, erects considerable barriers for novices and imposes a heavy cognitive and logistical burden on experts. In response, this paper has presented and rigorously validated a novel, attention-aware methodology that systematically restructures the creative process through an integrated framework of two systems: \textbf{EUPHORIA}, for immersive, implicit preference capture, and \textbf{RETINA}, for agentic, AI-driven form generation.

What we have achieved is a comprehensive, end-to-end validation of a new design paradigm. The research began by establishing the foundational principles of our approach. Through initial empirical studies, we first demonstrated that a designer's implicit visual attention, as measured by their gaze within an immersive environment, serves as a robust and objective proxy for their subjective aesthetic preferences. We then proved that this attention mechanism could be effectively channelled; by pre-conditioning the designer's mind with thematic goals, their creative exploration became convergent and thematically coherent. These foundational findings confirmed that a designer's subconscious gaze could be harnessed as a reliable input signal for a computational creative system.

The central contribution of this work was the comparative validation of this new workflow against conventional methods. The results were unequivocal. The integrated EUPHORIA-RETINA pipeline was shown to be profoundly more efficient, reducing the time required to generate a final design concept by a factor of more than four. More importantly, this acceleration did not come at the cost of quality. On the contrary, the final design outputs from the fully automated workflow were consistently judged by a large panel of independent experts to be of a higher quality, achieving superior 'Worthiness' and 'Design Effectiveness' scores across all tested problem statements.

The key takeaway for the reader is the presentation of a viable and powerful new paradigm for creative work: a shift from traditional \textbf{Computer-Assisted Design (CAD)}, where the computer is a passive tool for manual execution, to a more collaborative model of \textbf{Designer-Assisting Computers (DAC)}. In this new model, the designer is elevated from a labourer to a strategic director. Their unique, intuitive taste—captured effortlessly through their gaze—becomes the guiding signal for an intelligent, agentic system that handles the laborious tasks of feature analysis and form generation. This research, therefore, offers more than just a new tool; it presents a systematic, data-driven methodology that democratizes the design process, amplifies the designer's creative intent, and provides a compelling blueprint for the future of human-AI partnership in the creative industries.


\greyline
\bibliographystyle{elsarticle-harv} 
\bibliography{2_bibliography}

\end{document}